\documentclass[useAMS,usenatbib]{mn2e}
\usepackage{graphicx}
\usepackage{natbib}
\usepackage{enumerate}  
\usepackage{amssymb}
\usepackage{longtable,lscape}
\usepackage{rotating}
\usepackage{textcomp}
\usepackage{balance}
\usepackage{placeins}
\usepackage{amsmath}
\usepackage{txfonts}
\usepackage{multicol}
\usepackage{latexsym}
\usepackage{lineno}
\usepackage[format=hang,font=small,labelfont=bf]{caption}
\usepackage{enumerate,letltxmacro}

\LetLtxMacro\itemold\item
\renewcommand{\item}{\itemindent0.8cm\itemold}

\newcommand{\av}{$A_V$~}
\newcommand{\cloudy}{{\sc{cloudy}}~}
\newcommand{\galfit}{{\sc galfit}}
\newcommand{\hb}{H$\beta$~}
\newcommand{\ha}{H$\alpha$~}
\newcommand{\hgamma}{H$\gamma$~}
\newcommand{\hdelta}{H$\delta$~}
\newcommand{\iraf}{{\sc iraf}}
\newcommand{\msun}{M$_{\odot}$~}
\newcommand{\neiii}{[Ne{\scriptsize{III}}]}
\newcommand{\twospace}{$\!\!$}
\newcommand{\onespace}{\hspace{-3pt}}
\newcommand{\starlight}{{\sc{starlight}}~}
\newcommand{\popstar}{{\sc{popstar}}~}
\newcommand{\reff}{$r_{\mathrm{eff}}$~}

\DeclareRobustCommand{\ion}[2]{%
\relax\ifmmode
\ifx\testbx\f@series
{\mathbf{#1\,\mathsc{#2}}}\else
{\mathrm{#1\,\mathsc{#2}}}\fi
\else\textup{#1\,{\mdseries\textsc{#2}}}%
\fi}

\newcommand{\hii}{\ion{H}{ii}~}

\newcommand{\hei}{\ion{He}{i}}
\newcommand{\heii}{\ion{He}{ii}}
\newcommand{\nii}{[\ion{N}{ii}]}

\newcommand{\oii}{[\ion{O}{ii}]}
\newcommand{\oiir}{\ion{O}{ii}}
\newcommand{\oiii}{[\ion{O}{iii}]}
\newcommand{\sii}{[\ion{S}{ii}]}
\newcommand{\siii}{[\ion{S}{iii}]}

\newcommand{\hiiexplorer}{HII{\sc{explorer}}~}

\makeatletter
\newcounter{subsubsubsection}[subsubsection]
\renewcommand\thesubsubsubsection{\thesubsubsection .\@arabic\c@subsubsubsection}
\newcommand\subsubsubsection{\@startsection{subsubsubsection}{4}{\z@}%
                                     {-3.25ex\@plus -1ex \@minus -.2ex}%
                                     {1.5ex \@plus .2ex}%
                                     {\normalfont\normalsize}}
\newcommand*\l@subsubsubsection{\@dottedtocline{3}{10.0em}{4.1em}}
\newcommand*{\subsubsubsectionmark}[1]{}
\makeatother

\title[Ionizing stellar population in NGC 3310]{Ionizing stellar population in the disc of NGC 3310 -- I. The impact of a minor merger on galaxy evolution\thanks{Based on observations collected at the Centro Astronómico Hispano-Alemán (CAHA) at Calar Alto, operated jointly by the Max-Planck Institut für Astronomie and the Instituto de Astrofísica de Andalucía (CSIC).}}

\author[Miralles-Caballero et al.]{\parbox{\textwidth}
{D. Miralles-Caballero$^{1}$\thanks{E-mail:
daniel.miralles@uam.es},  A. I. D\'iaz$^{1}$, F. F. Rosales-Ortega$^{2}$, E. P\'erez-Montero$^{3}$ and S. F. S\'anchez$^{3,4}$}\vspace{0.4cm}\\
$^{1}$Departamento de F\'isica Te\'orica, Universidad Aut\'onoma de Madrid, 28049 Madrid, Spain\\
$^{2}$Instituto Nacional de Astrof\'isica, \'Optica y Electr\'onica, Luis E. Erro 1, 72840 Tonantzintla, Puebla, Mexico\\
$^{3}$Instituto de Astrof\'isica de Andaluc\'ia, CSIC, Apdo. 3004, 18080, Granada, Spain\\
$^{4}$Centro Astron\'omico Hispano Alem\'an, 04004 Almer\'ia, Spain}
\begin{document}

\date{Accepted 2014 March 5. Received 2014 February 28; in original form 2013 December 19}

\pagerange{\pageref{firstpage}--\pageref{lastpage}} \pubyear{...}

\maketitle  

\label{firstpage}

\begin{abstract}
Numerical simulations of minor mergers predict little enhancement in the global star formation activity. However, it is still unclear the impact they have on the chemical state of the whole galaxy and on the mass build-up in the galaxy bulge and disc. We present a two-dimensional analysis of NCG 3310, currently undergoing an intense starburst likely caused by a recent minor interaction, using data from the PPAK Integral Field Spectroscopy (IFS) Nearby Galaxies Survey (PINGS). With data from a large sample of about a hundred \hii regions identified throughout the disc and spiral arms we derive, using strong-line metallicity indicators and direct derivations, a rather flat gaseous abundance gradient. Thus, metal mixing processes occurred, as in observed galaxy interactions.

Spectra from PINGS data and additional multiwavelength imaging were used to perform a spectral energy distribution fitting to the stellar emission and a photoionization modelling of the nebulae. The ionizing stellar population is characterized by single populations with a narrow age range \mbox{(2.5-5 Myr)} and a broad range of masses \mbox{(10$^4$--6$\times 10^{6}$ \msun\twospace)}. The effect of dust grains in the nebulae is important, indicating that 25--70\% of the ultraviolet photons can be absorbed by dust. The ionizing stellar population within the \hii regions represents typically a few percent of the total stellar mass. This ratio, a proxy to the specific star formation rate, presents a flat or negative radial gradient. Therefore, minor interactions may indeed play an important role in the mass build-up of the bulge.

\end{abstract}


\section{Introduction}

The friction between gas and dust in a merger event can have an important impact on the evolution of the galaxies involved. Minor mergers are usually defined as the collision between two galaxies with a mass ratio smaller than 1:3-1:4. Enhanced star formation when compared with isolated objects has been observed in samples of interacting/merging galaxies (e.g.~\citealt{Kennicutt87,Barton03,Geller06,Woods07}). Numerical simulations of minor mergers indicate that they can trigger nuclear activity~\citep{Mihos94c,Hernquist95,Eliche-Moral11}, alter the morphologies of galaxies ~\citep{Robertson06} and activate bars~\citep{Laine99,Romano-Diaz08}. Contrary to major mergers, minor mergers are much less violent dynamical processes since they do not destroy the disc of the main progenitor. Nevertheless, they have been increasingly recognized as important players in galaxy evolution and, in particular, in the formation and assembly of bulges especially in lower mass systems (\citealt{Guo07},~\citealt{Hopkins10}, and 
references therein). Yet, whether they are also important for the total stellar mass build-up in galaxies in general is unclear and controversial~\citep{Bournaud07,Lopez-Sanjuan11,Newman12,Xu13}.

The induced star formation associated with the gas motions created by an interaction or the presence of bars is also expected to have an impact in the chemical distribution of the galaxies. Inflows of metal poor gas from the outer parts of the galaxy can decrease the metallicity in inner regions and modify the radial abundance gradients across spiral discs~\citep{Rupke10}. In fact, some studies have found that interacting galaxies do not follow the well established correlation between luminosity and metallicity found in normal disc galaxies. Shallower or null metallicity gradients and relatively high metallicity \hii\onespace-like regions have been observed in the outer part of the galaxies in major merger events (~\citealt{Kewley06,Ellison08,Michel-Dansac08,Rupke08,Peeples09,Miralles-Caballero12,Sanchez14}). A similar effect can have the action of inward and outward radial flows of interstellar gas induced by the non-axisymmetrical potential of the bars (e.g.,~\citealt{Friedli94,Roy97}
).

There is extensive literature on the impact of a major merger event on the evolution of the involved galaxies. However, only a handful recent works have been focused on minor mergers (e.g.~\citealt{Krabbe11,Alonso-Herrero12}). Therefore, more observational studies are needed to provide a deeper insight on the effect of less violent dynamical phenomena on galaxy evolution. 

NGC 3310 is a very distorted spiral galaxy classified as an SAB(r)bc by~\cite{deVaucouleurs91}, with strong star formation. The star formation activity, especially a vigorous circumnuclear star-forming ring, has been studied over a wide range of wavelengths from X-rays to radio (e.g.~\citealt{Balick81,Pastoriza93,Zezas98,Diaz00a,Elmegreen02,Hagele10b,Hagele13,Mineo12}). Several studies support that this galaxy collided with a poor metal dwarf galaxy, which caused a burst of star formation~\citep{Balick81,Schweizer88,Smith96}. It has also been suggested that NGC 3310 has actually experienced several interactions with small galaxies~\citep{Wehner06}. The starburst activity of this galaxy began some 100 Myr ago~\citep{Meurer00}, although some of the clusters are rather young, indicating that starburst galaxies may remain in the starburst mode for quite some time.~\cite{Pastoriza93} reported that the circumnuclear regions in NGC 3310 present low metal abundances (\mbox{0.2-0.4 Z$_\odot$}), in contrast to what 
is generally found in early-type spirals~\citep{Diaz07}. 

In order to provide a better insight on the effects of the past interaction in the disc of NGC 3310, we need to investigate the properties of the hot ionized gas and the stellar population distributed along the whole disc. Studies based on NGC 3310 have mainly been focused on the properties of the young massive population (i.e.~\mbox{$\tau < 10$ Myr} and stellar mass \mbox{m$_\star$ = 10$^5$-10$^6$ \msun\onespace}) in the circumnuclear region. These studies made use of photometric images and long-slit spectroscopy, which are limited to either spectral range and/or sampling biases (i.e. area coverage, only the most luminous clusters and \hii -like regions with a very limited range in ionization conditions or only a unique aperture in large sample of galaxies, etc.). With the advent of the Integral Field Spectroscopy (IFS) such limitations can be overcome or at least significantly diminished. Nowadays, the use of IFS to perform wide-field 2D analyses of galaxies is well established and continuously growing. 
Large fields of view (FoV) can now be observed with simultaneous spatial and spectral coverage. We thus devised the PPAK Integral-field-spectroscopy Nearby Galaxies Survey (PINGS;~\citealt{Rosales-Ortega10}) in order to solve the limitations just mentioned. This is a survey specially designed to obtain complete emission-line maps, stellar populations and extinction using an IFS mosaicking imaging for nearby (\mbox{$D_\mathrm{L} \leq $ 100 Mpc}) well-resolved spiral galaxies. It takes the advantage of one of the world's widest FoV integral field unit (IFU).   

This paper focuses thus on the characterization of the ionizing population in the stellar disc of NGC 3310 so as to study the impact of the minor merger on the star formation properties of the remnant. The specific goals of this study are: (i) to obtain gaseous abundance determinations of not a handful but a non-biased luminosity hundred or so \hii regions distributed all along the disc, which are necessary to better trace radial abundance gradients; (ii) to characterize the age, mass and other star formation properties of the ionizing stellar population and (iii) to investigate the extent at which the interaction could have affected the mass growth in the disc. The PINGS data, together with retrieved multiwavelength images covering from the near-$UV$ to the near-$IR$ spectral range, are well suited to deal with these issues with unprecedented statistics. In a companion paper we will focus on the study of the Wolf--Rayet (WR) population present in the circumnuclear regions and arms in NGC 3310. The presence 
of this population, which has been detected in the PINGS spectra, sets important constraints on the age and nature of the most massive ionizing stellar population (\mbox{i.e.~M $\geq$ 40 \msun\onespace}).

The paper is organized as follows. We present the data set used in Sect.~\ref{sec:obs}. In Sect.~\ref{sec:analysis_results}, we describe the analysis techniques used (i.e. to identify the \hii regions, decouple the gaseous and the continuum stellar emission, to obtain the chemical abundances of the gas and to estimate the age and the mass of the ionizing stellar population) and present our abundance and stellar population property derivations for \hii regions identified in the disc of NGC 3310. We first discuss in Sect.~\ref{sec:discussion} the effects due to biases and systematics of the analysis techniques. We then proceed with the scientific discussion on the radial abundance gradient in the galaxy and on the star formation properties in the disc of NGC 3310 and how they might have been affected by the past minor interaction. We finally draw our conclusions in Sect.~\ref{sec:conclusions}. Throughout this paper, the luminosity distance to NGC 3310 is assumed to be 16.1 Mpc (taken from the NASA 
Extragalactic Database). With an adopted cosmology of \mbox{H$_0$ = 73 km s$^{-1}$ Mpc$^{-1}$} an angle of 1 arcsec corresponds to a linear size of 78 pc.

\section[]{Observations and data acquisition}
\label{sec:obs}
\subsection{PINGS data}

NGC 3310 observations were carried out with the 3.5m telescope of the Calar Alto Observatory using the Postdam Multi-Aperture Spectrograph (PMAS;~ \citealt{Roth05}) in the PMAS fibre package mode (PPAK;~\citealt{Verheijen04},~\citealt{Kelz06}). This was part of the PINGS (\citealt{Rosales-Ortega10}). In brief, the V300 grating was used to cover the 3700-7100 \AA{} spectral range with a spectral resolution of \mbox{10 \AA{}}, corresponding to \mbox{600 km s$^{-1}$}, at \mbox{$\lambda$ = 5000 \AA{}}. We took three pointings with a dithered pattern (three dithered exposures per pointing), using a mean acquisition time per PPAK field in dithering mode (including set-up + integration time) of \mbox{2$\times$600 s} per dithering position. This observing strategy allowed us to re-sample the PPAK 2.7 arcsec-diameter fibre to a final mosaic with a 1 arcsec spaxel sampling and a FoV of about \mbox{148 $\times$ 130 arcsec$^2$}.

The data were reduced following~\cite{Sanchez06}, and can be summarized as follows: pre-reduction, identification of the location of the spectra on the detector, extraction of each individual spectrum, distortion correction of the extracted spectra, wavelength calibration, fibre-to-fibre transmission correction, flux calibration, allocation of the spectra to the sky position, dithered reconstruction and re-sampling. 

The prereduction processing was performed using standard \iraf\footnote{\iraf~is distributed by the National Optical Astronomy Observatories, which are operated by the Association of Universities for Research in Astronomy, Inc., under cooperative agreement with the National Science Foundation.} packages while the main reduction was performed using the {\small R}3{\small D} software for fibre-fed and IFS data~\citep{Sanchez06}. A thorough description of all the reduction and flux calibration procedures is described in~\cite{Rosales-Ortega10} and~\cite{Sanchez11}. 

The reduced IFS data were stored in row-stacked-spectra (RSS) files. RSS format is a 2D FITS image where the $X-$ and $Y-$ axes contain the spectral and spatial information respectively, regardless of their position in the sky. This format requires an additional file that stores the position of the different spatial elements on the sky. After reducing each individual pointing with a first-order flux calibration we built a single RSS file for the mosaic following an iterative procedure: (1) a master pointing that has the best possible flux calibration and sky extinction correction and signal-to-noise ratio (S/N)  is selected; (2) the mosaic is then constructed by adding consecutive pointings following the mosaic geometry; (3) overlapping spectra are replaced by the average between the previous pointing and the new rescaled spectra (obtained by using the average ratio of the brightest emission lines found in the overlapping spectra). (4) the resulting spectra are incorporated into the final RSS file, updating 
the corresponding position 
table.

In order to obtain the most accurate absolute spectrophotometric calibration, an additional correction was performed by comparing the IFS data with available broad-band photometry in $B$, $g$ and $r$ band (see next section). The estimated spectrophotometric accuracy of the IFS mosaic is of the order of 10-15\%. 
\begin{figure}
\centering
\includegraphics[width=0.95\columnwidth]{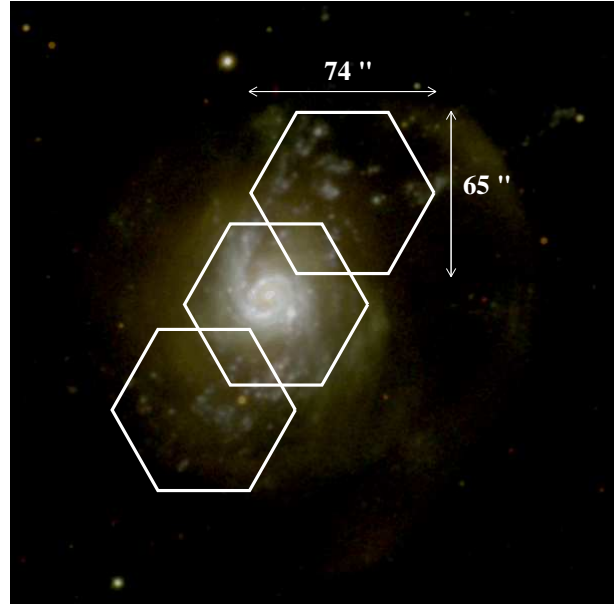}
  \caption{False colour image of NGC 3310, produced by the combination of the $u$ (in blue), $g$ (in green) and $r$ (in red) SDSS filters. Knots of star formation are clearly visible in the central region and extended outer regions. Overlaid apertures show the three pointings that were taken with PPAK, with a central position centred in the galaxy's nucleus and two offsets of (–35, 35) and (35, –35) arcsec in (RA, Dec.) in north-west and south-east directions respectively.
}
  \label{fig:combined_im}
\end{figure}
A total of 8705 spectra were finally produced, spatially resolved in spaxels of \mbox{1 $\times$ 1 arcsec$^2$}. 
\subsection{Multi-wavelength data}
We retrieved publicly available broad-band imaging of this galaxy in order to perform an absolute flux re-calibration. Specifically, we used the Sloan Digital Sky Survey (SDSS\footnote{http://www.sdss.org/}) broad-band $g-$ and $r-$ filter images (with a spatial resolution of about 1 arcsec) and an $HST$\footnote{http://www.stsci.edu/hst/} image taken with the Wide Field Planetary Camera 2 (WFPC2, with a spatial resolution of about 0.05 arcsec) using the $F439W$ filter (similar to $B$ Johnson). Although the latter does not cover the entire FoV of the galaxy, we could perform the calibration by obtaining the photometry using an aperture large enough to cover the central 30 arcsec. 

We present in Fig.~\ref{fig:combined_im} a false colour image of NGC 3310, produced using the $u$, $g$ and $r$ SDSS filters. From the figure, it can be clearly seen that NGC 3310 is a very distorted spiral galaxy with strong star formation, with a very bright central nucleus, surrounded by a ring of luminous HII regions. Given its morphology, a tailored mosaic pattern was constructed for this galaxy (overplotted hexagons in the figure). 

\begin{figure*}
\includegraphics[trim = -1cm 7.3cm -1cm 0cm,clip=true,width=0.85\textwidth]{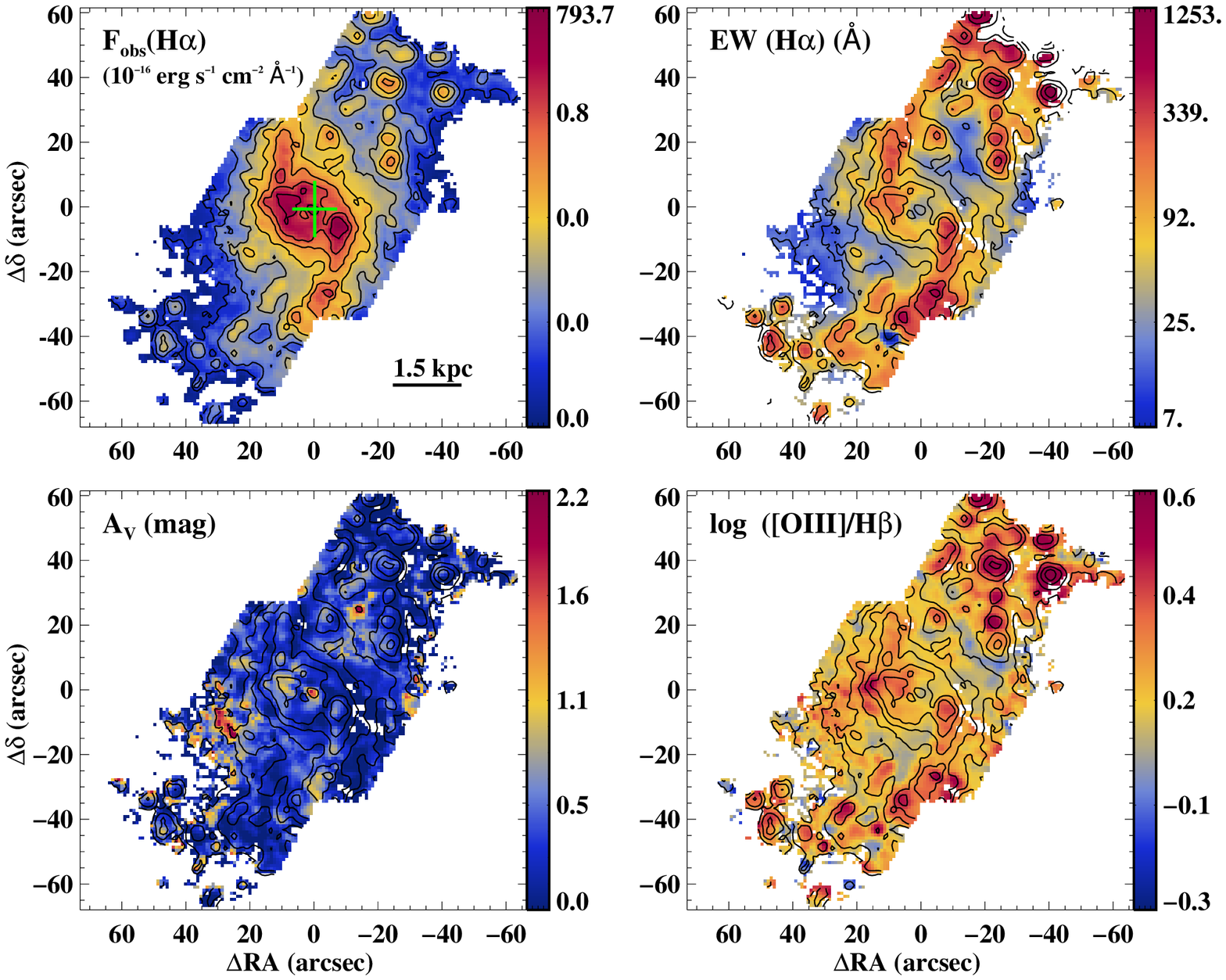}
  \caption{Maps for the observed \ha flux, \ha EW, $A_V$ extinction and \oiii$\lambda$5007 to \hb ratio. Contours with the \ha emission are overplotted in each case. The centre of the galaxy, which is enclosed by an \ha peaked contour, is marked with a green plus sign. The physical scale of 1.5 kpc at the distance of the galaxy is represented by the straight line at the bottom-right corner of the first map. Axes scales are in arcsec and orientation is as usual: north up, east to the left.}
  \label{fig:emission_maps}
\end{figure*}

We also obtained UV images of the galaxy. In particular, taken with the  $UVW2$ and $UVM2$ filters (with effective wavelenghts of 2087 and 2297 \AA{}, respectively), mounted on the OM camera onboard the \textit{XMM--Newton} Satellite\footnote{http://xmm.esac.esa.int/}. The observations, with ID 0556280201, were taken in 2008. NGC 3310 was observed for more than 38 and 8 ks with the $UVW2$ and $UVM2$ filters, respectively. We reduced the observation data files using {\small SAS} version 13.0.0. The {\small SAS}\footnote{http://xmm.esac.esa.int/sas/} script \textit{omichain} was used to produce calibrated images. 

The Sloan and OM ultraviolet images provide valuable information on the continuum emission of the young ionizing stellar population. 

\section[]{Analysis and results}
\label{sec:analysis_results}

We present here the results obtained from the analysis of the ionizing stellar population in NGC 3310 using optical IFS data together with available broad-band ancillary data. 

\subsection{2D general properties of the ionized gas}
\label{sub:2D_gas}

We study the spatially resolved distribution of the physical properties in the \hii regions of NGC 3310 by obtaining a complete 2D view of: (i) the main emission lines used in the optical range for typical abundance diagnostic methods; (ii) important spectral features useful for the analysis of the ionizing stellar populations.

In order to extract any physical information from the data set, we first need to identify the detected emission lines of the ionized gas and to decouple them from the stellar continuum. A preliminary fast decoupling was performed using an improved version of the {\small FIT3D} package \citep{Sanchez06,Sanchez07}. This software includes several routines to model and subtract the underlying stellar population of a spectral energy distribution (SED) using synthetic stellar spectra, and subsequently to fit and deblend the nebular emission lines. We refer the reader to \cite{Sanchez07} for further details.

\begin{figure*}
\includegraphics[trim = -2cm 18.2cm 0cm 0cm,clip=true,width=0.85\textwidth]{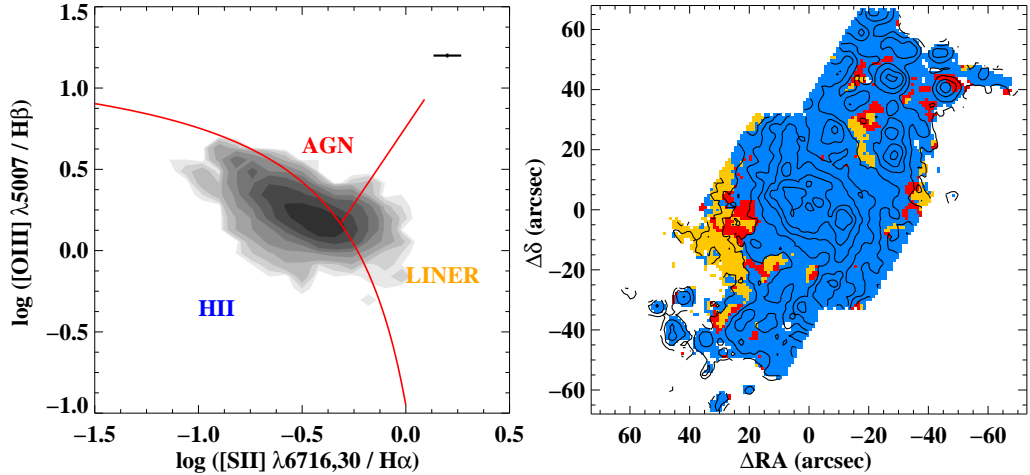}
  \caption{\textbf{Left:} diagnostic diagram (see text) for all the spaxels with good quality measurements. The contours show the density distribution of spaxels. The Kewley et al. (\citeyear{Kewley01a}) red demarcation line, which is usually invoked to distinguish between star-forming regions (\hii\onespace) and other source of ionization (AGN, LINER  or composite, C), is also plotted. The error-bars at the top-right indicate the typical mean errors for the considered line ratios. \textbf{Right:} colour-coded maps showing the 2D spatial distribution of the ionizing source of ionization: star-forming regions (in blue), LINER or composite (in yellow) and AGN (in red).  }
  \label{fig:bpt}
\end{figure*}
As a result of the line fitting procedure on the stellar-subtracted (i.e. gas) spectra, a set of measured emission-line intensities was obtained for each observed spectrum of the final clean mosaic. From these sets of emission-line intensities, emission-line maps were created by interpolating the intensities derived for each individual line in each individual spectrum, based on the position tables of the clean mosaics, and correcting for the dithering overlapping effects when appropriate. The observed \ha\onespace /\hb line ratio was used to correct the line maps for extinction, on a spaxel-by-spaxel basis. See~\cite{Rosales-Ortega10} for a full description.

We present a set of line emission maps and those with derived properties in Fig.~\ref{fig:emission_maps}. The top panel map on the left shows the 2D distribution of the \ha observed flux. The nuclear part of the galaxy is easily identified together with the two spiral arms extending to the north and south of the galaxy. Bright inner clumps are also clearly seen within the circumnuclear region and along the whole extent of the galaxy. The lowest level contours are likely to enclose diffuse-like emission. 

In practical, all regions with \ha emission peaks, the derived EWs \mbox{($ > 20$ \AA{})}, as shown on the top right map, are consistent with the presence of very young \mbox{(i.e. $\tau < 10$ Myr)} stellar populations. This is expected in \hii regions. In the nucleus, however, the relatively low values of the equivalent width (EW) distribution indicate the prevalence of the underlying old population. 

We used the reddening constants from the Balmer decrement between \ha and \hb as compared to the theoretical value for mean nebular conditions given by \cite{Osterbrock89} in order to estimate the reddening. We then used the law by \cite{Cardelli89}, assuming $R_V = 3.1$, in order to correct the emission-line fluxes. The distribution of dust extinction (lower left map) although somewhat clumpy, is typically low (\mbox{$A_V < 1$ mag)}. In fact, the average extinction derived from the map corresponds to \mbox{$\overline{A_V} = 0.35 \pm 0.31$ mag}. The extinction peaks up to about \mbox{$A_V > 2$ mag} only in the nucleus and along some diffuse-like emission areas.

We also show the 2D distribution of the emission-line ratio log (\oiii $\lambda 5007$/\hb), a parameter sensitive to the excitation, which correlates with the effective temperature of the exciting stars~\citep{Hunter92}. The range sampled generally lies between 0 and about 0.6, which shows that the ionized gas excitation in this galaxy is not very high. As mentioned in the introduction, NGC 3310 is a very distorted spiral galaxy with strong and rather complex star formation, very likely undergoing a merger-driven global starburst~\citep{Smith96,Kregel01,Wehner06}. It is widely known that massive stellar populations as young as about less than 10 Myr (i.e. hot OB stars) can well be responsible for the ionization observed in galaxies. The identification of the main ionizing mechanism in an observed object can be relatively well accomplished by the use of the so-called diagnostic diagrams (\citealt{Baldwin81},~\citealt{Veilleux87}). They usually involve two or three strong emission lines (SEL) that depend on 
the 
ionization degree and, to a lesser extent, on electron temperature or abundance. Fig.~\ref{fig:bpt} (left) shows a common  diagnostic diagram (log \oiii/\hb versus log \sii $\lambda \lambda$6717,6731/\ha\onespace) as derived for each spaxel. Colour-coded maps were also created (Fig.~\ref{fig:bpt}, right) so as to spatially identify the regions ionized by different physical processes. As shown in the figure, the vast contribution of the ionization in this galaxy comes from young massive stars (\hii\onespace-like ionization). There are only a few regions (mainly not associated with \ha peaks) associated with other ionization mechanisms, though with larger uncertainties in the determination of their emission-line ratios. This paper focuses on the ionizing stellar population. Thus, with these maps we can guarantee that our study on \hii regions in NGC 3310 is not perceptibly contaminated by other sources of ionization.  
\subsection{Identification of \hii regions and star--gas decoupling}
\begin{figure*}
\includegraphics[angle=90,trim = 2.7cm -1.5cm 3cm -2.0cm,clip=true,width=1.05\textwidth]{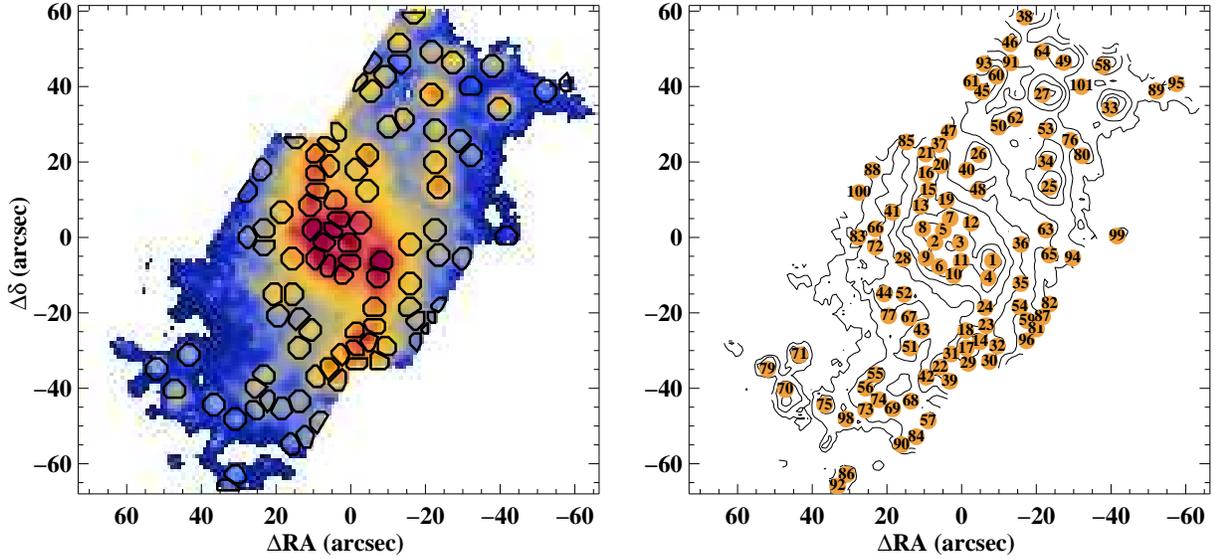}
 \caption{\textbf{Left:} \ha intensity map in arbitrary units with the apertures of the identified \hii regions from the \hiiexplorer routine. \textbf{Right:} \ha contour map showing the spatial 2D distribution of the identified \hii regions with their identification number (\mbox{HII ID} in tables).}
  \label{fig:hii_id}
\end{figure*}

The detection and segregation of \hii regions, and subsequent extraction of the spectra for each of them, was performed using the semi-automatic procedure \hiiexplorer (\citealt{Sanchez12b}). A segmentation map that identifies each detected \hii region is provided by the code, together with the corresponding extracted spectra. In our case, a total of 99 \hii regions were identified. Fig.~\ref{fig:hii_id} shows the segmentation map overplotted on the \ha map (left) and the identification of each ionizing region (right). The regions have typical radii of about 2--3 arcsec, similar to, or somewhat larger than the spatial resolution of the instrument. This translates into radii in the range of 150--250 pc, within the size range of extragalactic giant \hii regions (\citealt{Kennicutt84,Oey03,Hunt09,Lopez11}). Table~\ref{table:hii_catalogue} reports a catalogue with the main observed properties of the identified \hii regions in NGC 3310.

As mentioned before, particular care has always been taken in subtracting the stellar continuum so as to correctly decouple the stellar continuum from the nebular line emission. When the analysis was carried out in a spaxel-by-spaxel basis, an optimized fast analysis aimed at obtaining the best polynomial fit to the data (not necessarily giving a meaningful physical solution) was done. For the \hii regions we made use of the spectral synthesis code \starlight (\citealt{Cid-Fernades04b,Cid-Fernandes05}). This code mixes computational techniques originally developed for empirical population synthesis with ingredients of evolutionary synthesis models. Briefly, an observed spectrum is fitted with a combination of \textit{N} simple stellar populations from a set of evolutionary synthesis models. Extinction by foreground dust is also modelled by the V-band extinction \av\twospace.~Line-of-sight (LOS) stellar motions are modelled as well, but given the rather low spectral resolution of the PINGS 
data ($\sim$ 600 km s$^{-1}$), off-set velocities and stellar velocity dispersions were set to a fixed default value which in any case does not affect our results since we do not intend to study the kinematics of this galaxy. 

Basically, four inputs are needed for \starlight\onespace: the observed spectrum, a configuration file, a mask-file to mask emission lines and a file that allocates the base spectra (i.e. the set of evolutionary synthesis models). In our analysis we made use of a compilation of a few hundreds of synthesis models covering metallicities from Z$_\odot / 200$ to $1.5 \times  \mathrm{Z}_\odot$ and ages from 1 Myr to 17 Gyr. This set compiles models of \cite{Gonzalez-Delgado05} for \mbox{$\tau < $ 63 Myr} with the MILES library \cite[as updated by \citealt{Falcon-Barroso11}]{Vazdekis10} for larger ages. The models are based on the Salpeter initial mass function (IMF) and the evolutionary tracks by \cite{Girardi00}, except for the youngest ages (i.e. \mbox{$\tau =$ 1-3 Myr}), which are based on Geneva tracks (\citealt{Schaller92,Schaller93,Charbonnel93}).

\begin{figure*}
\centering
\includegraphics[trim = 2cm 0cm 2cm -1cm,clip=true,angle=90,width=1.02\columnwidth]{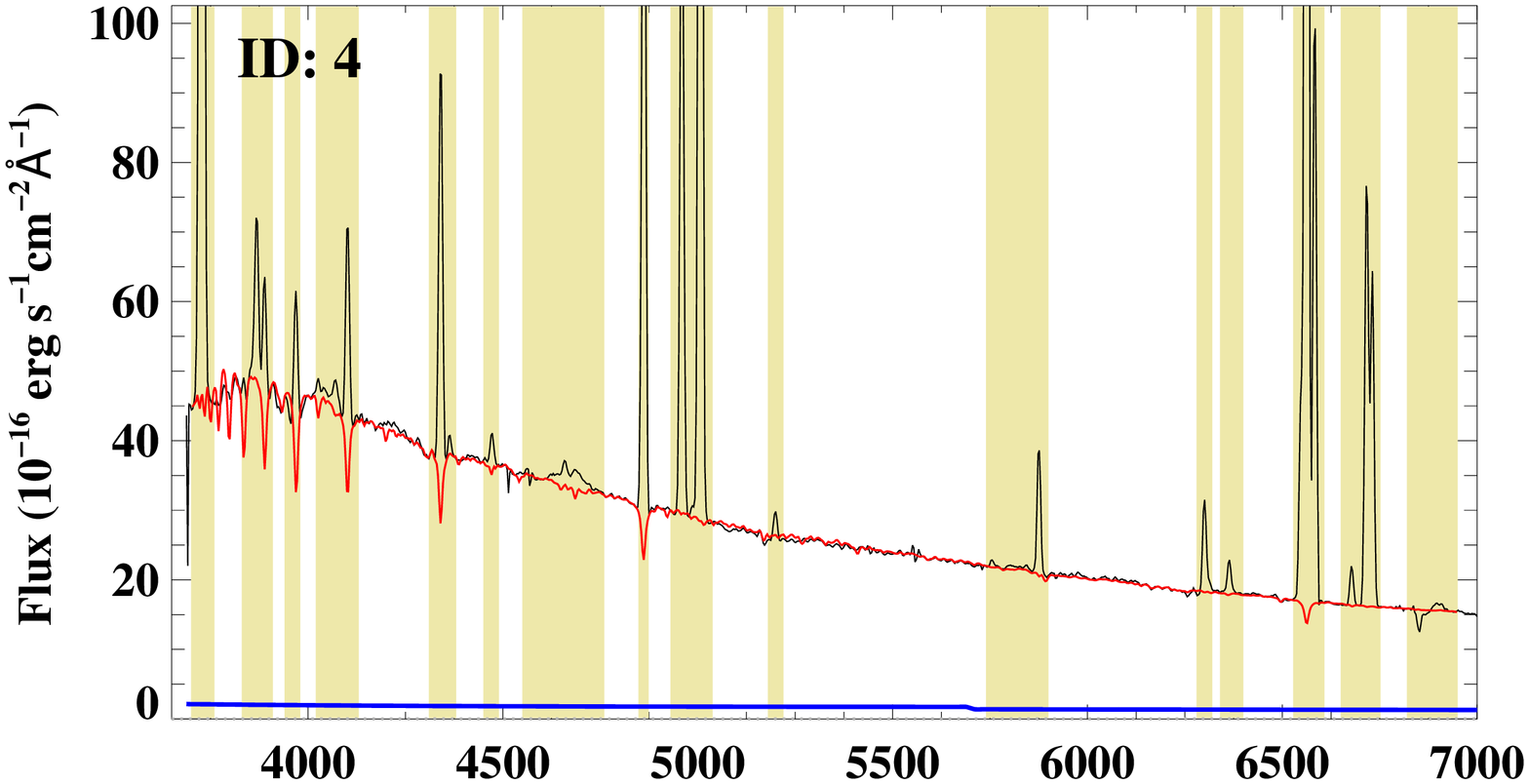}
\includegraphics[trim = 2cm 0cm 2cm 0cm,clip=true,angle=90,width=1.02\columnwidth]{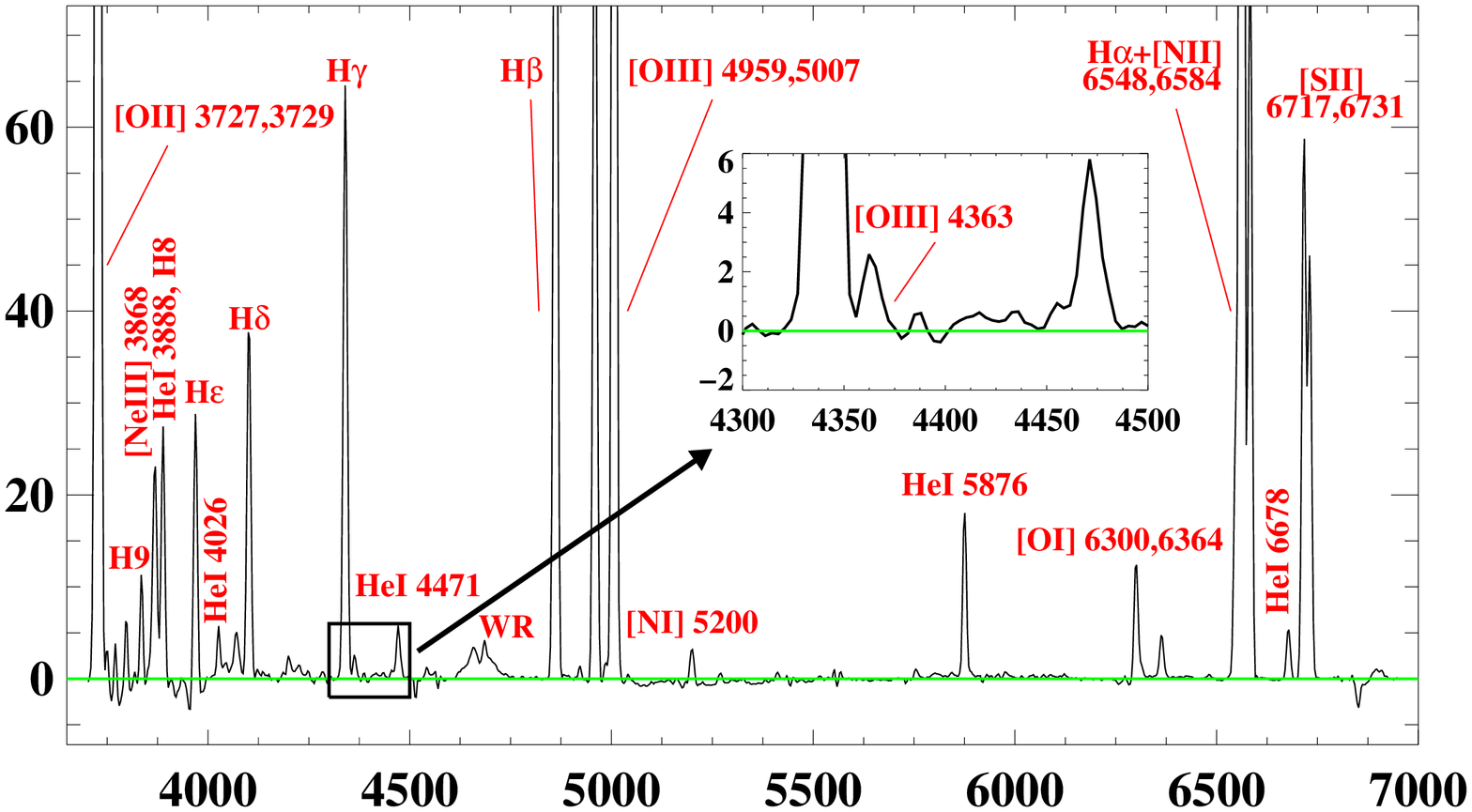}

\includegraphics[trim = 2cm 0cm 2cm -1cm,clip=true,angle=90,width=1.02\columnwidth]{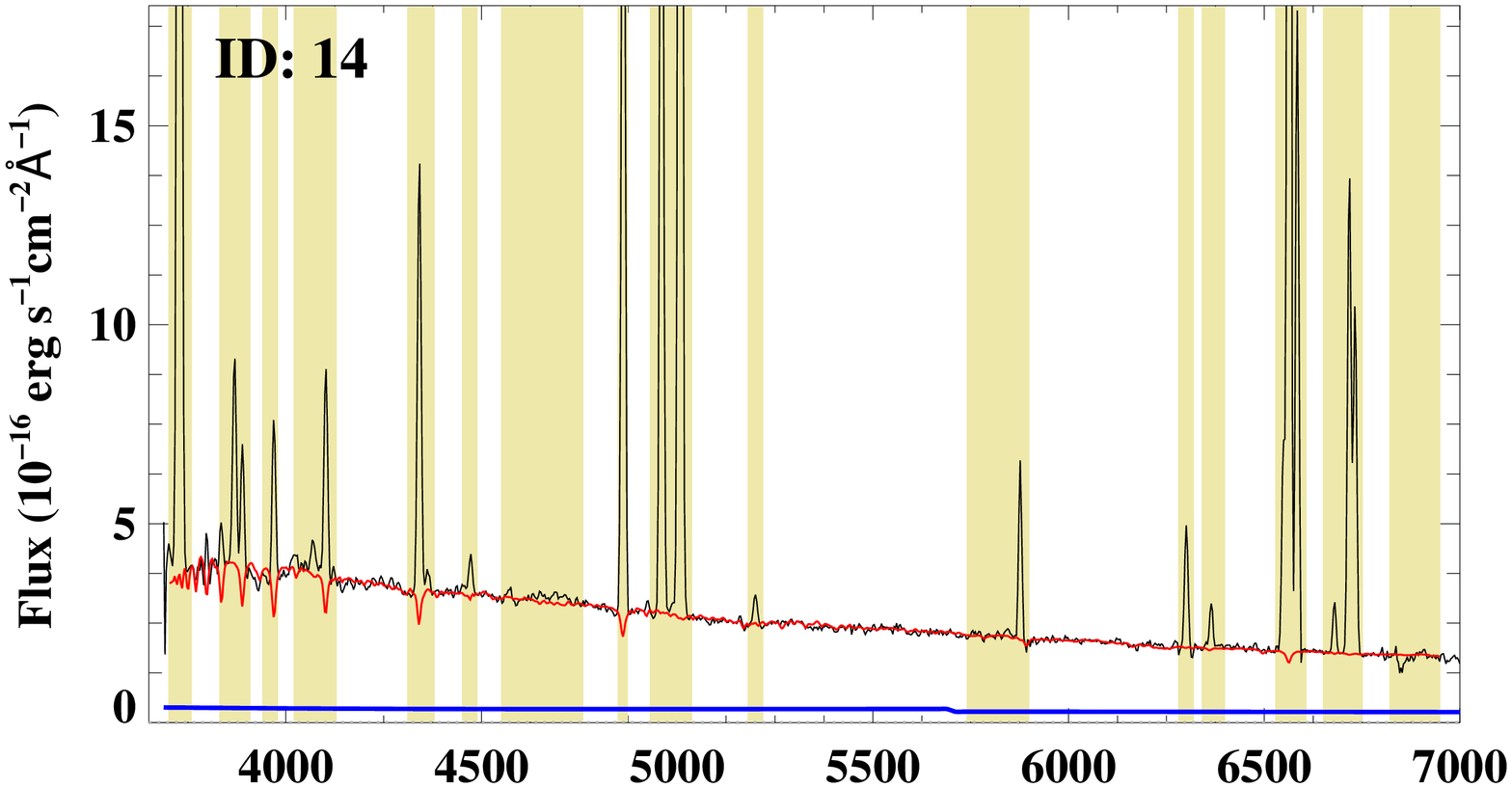}
\includegraphics[trim = 2cm 0cm 2cm 0cm,clip=true,angle=90,width=1.02\columnwidth]{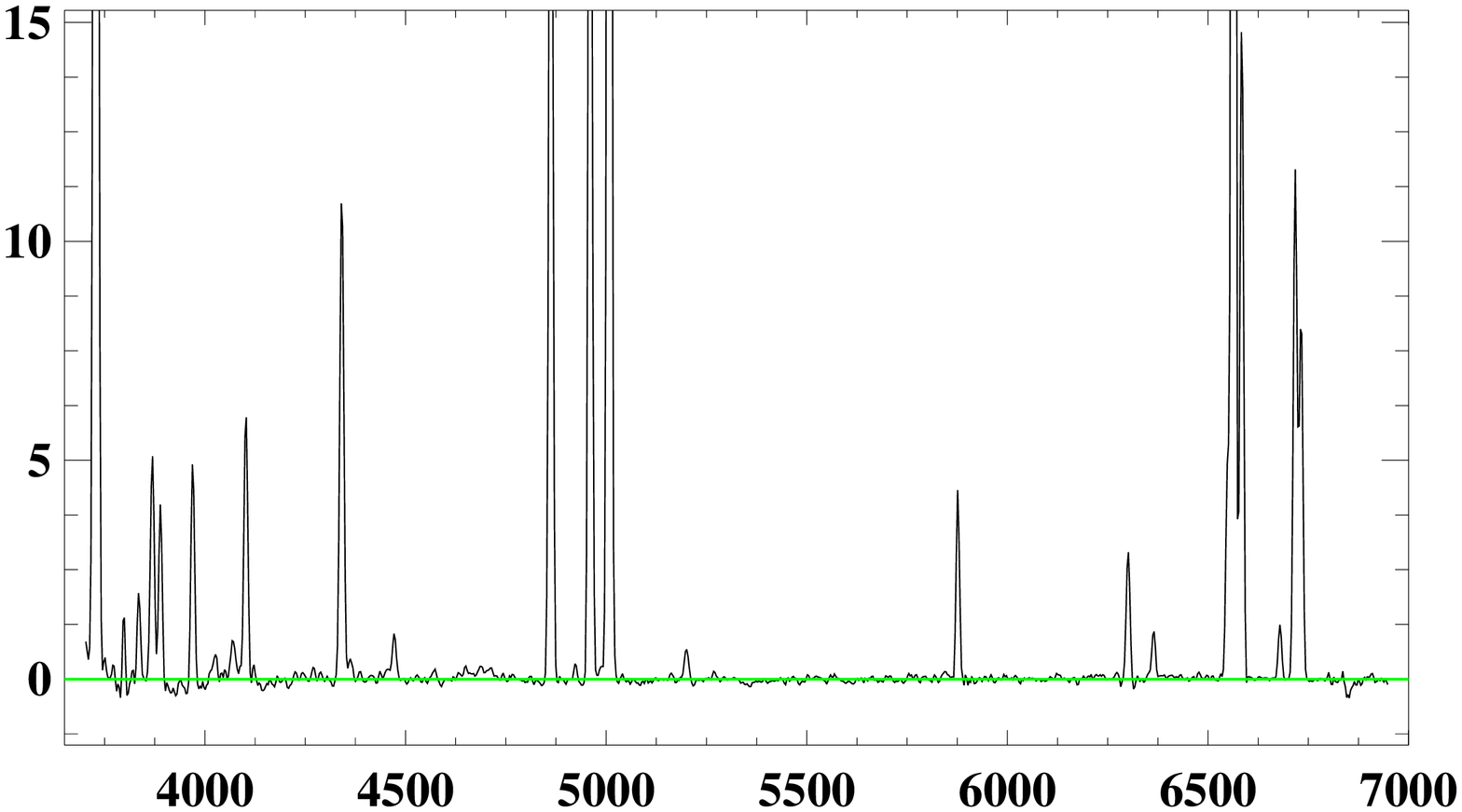}

\includegraphics[trim = 1cm 0cm 2cm -1cm,clip=true,angle=90,width=1.02\columnwidth]{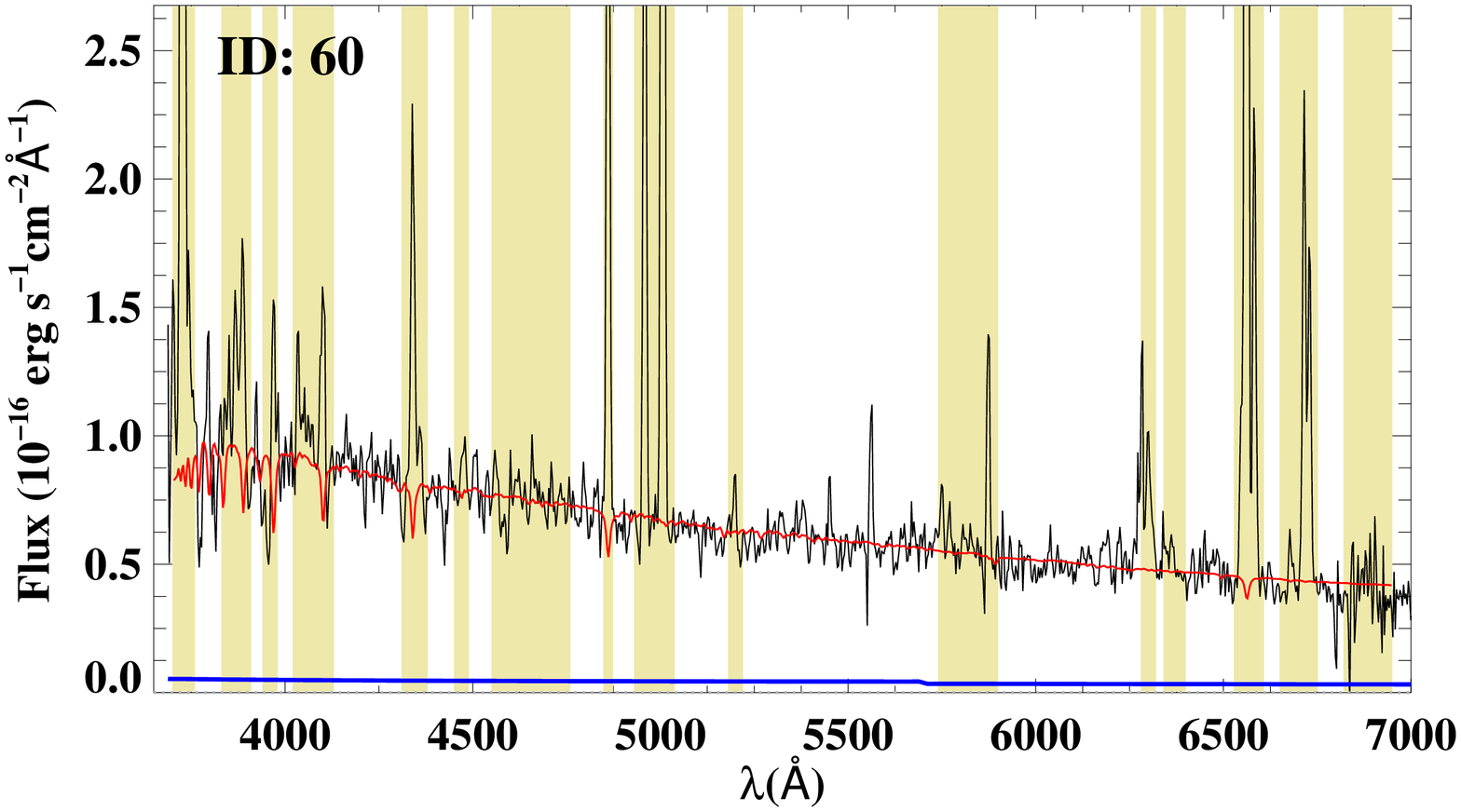}
\includegraphics[trim = 1cm 0cm 2cm 0cm,clip=true,angle=90,width=1.02\columnwidth]{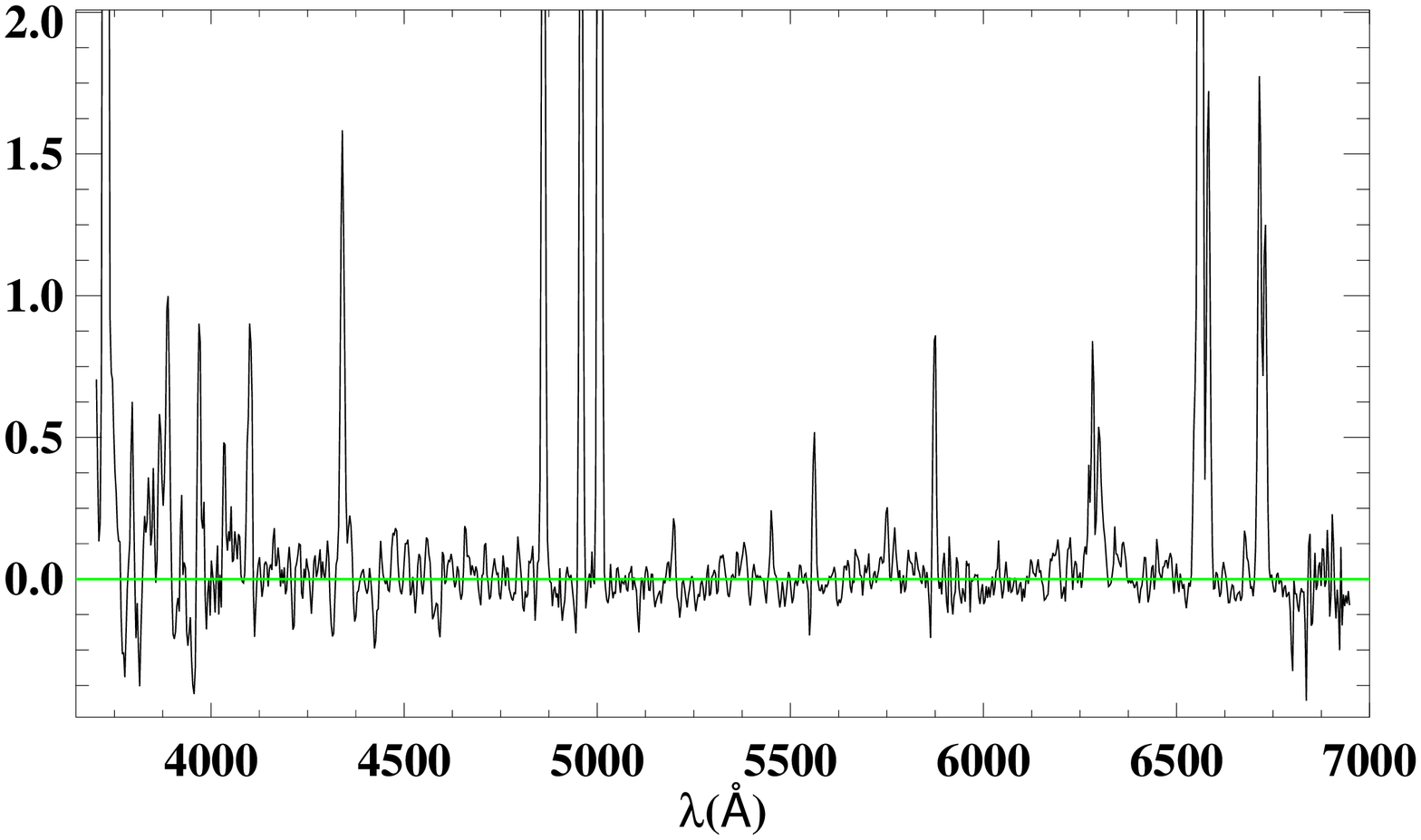}

 \caption{\textbf{Left-hand panel:} example of rest-frame observed spectra of three different \hii regions (left). Blue line shows the derived nebular continuum (subtracted before \starlight is run) and red line shows the continuum \starlight fit (C$_{\mathrm{fit}}$). Yellowish shaded areas correspond to masked spectral intervals  where typical nebular  and stellar (i.e. WR) features are not included in the model libraries (i.e. blue WR bump). Masked intervals are not used in the fit. \textbf{Right-hand panel:} corresponding residual gas spectrum (\mbox{= observed -- C$_{\mathrm{fit}}$ -- nebular spectrum}) for each \hii region. Typical emission lines and the WR bump are labelled on the plot at the top of the panel. Inset plot there shows a zoom of the \oiii$\lambda$4363 auroral line whereabouts in the spectrum.}
  \label{fig:SL_examples}
\end{figure*}

The observed spectrum was not used as an input for \starlight\twospace. Although the model templates used have enough spectral resolution to deal with our spectra, those corresponding to the ionizing population (i.e. $\tau \leq$ 10 Myr) lack the nebular emission component. The contribution of this component can be significant in massive \hii regions. We thus first subtracted a ``nebular continuum spectrum'' from each observed spectrum. This nebular spectrum was computed using the derived \ha luminosity for each region, assuming a metallicity of Z$_\odot$/3, following the procedure explained in~\cite{Molla09} and~\cite{Martin-Manjon10}. That particular metallicity was chosen because it is consistent with the gaseous abundances reported in~\cite{Pastoriza93}. It is also compatible with our derived metallicities (see Sect.~\ref{sec:radial_gradients}). Note that this continuum can be underestimated if the \ha luminosity is underestimated (i.e. if photon leakage or absorption by UV photons by dust 
grains is important). 

Fig.~\ref{fig:SL_examples} shows three examples of the fitted continuum spectrum (in red, left-hand panels) and the \textit{residual} gas spectrum (right-hand panels) for \hii regions with high and low S/N. As  can be seen, the nebular continuum spectrum (in blue, left-hand panels) can contribute significantly to the measured emission (in region ID 14, this continuum represents about 25\% of the observed light at the reddest wavelengths). Although the residual spectrum has been identified as emission from ionized gas, sometimes a  spectral feature centred around 4680 \AA{} can be observed. This stellar spectral feature, easily recognizable in the figure for the \hii region with ID 4, corresponds to the blue WR bump and indicates the presence of WR stars.   

\begin{figure*}
\centering
\includegraphics[trim = -2cm 16cm 0cm 0.2cm,clip=true,width=0.9\textwidth]{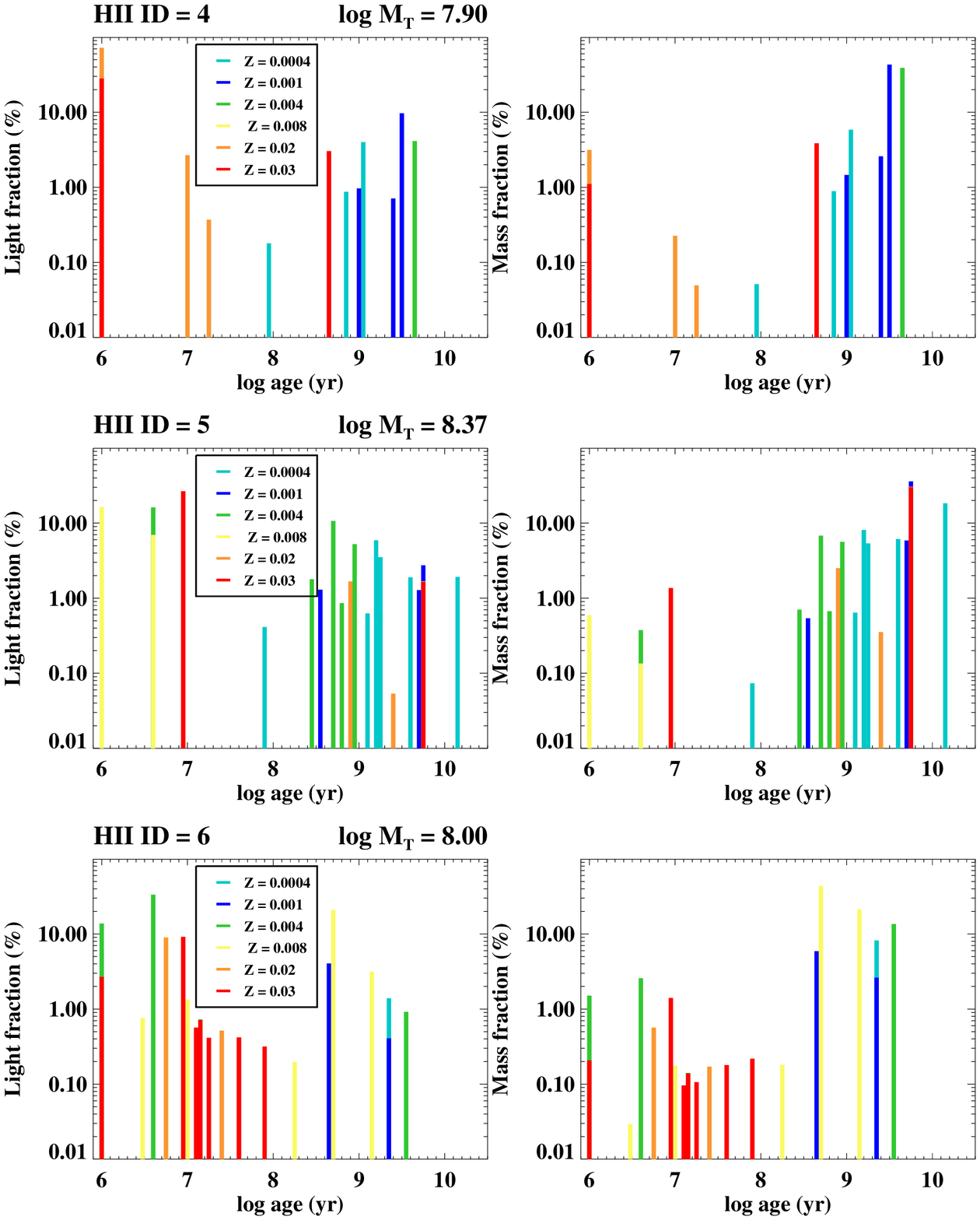}
\includegraphics[trim = -2cm 0cm 0cm 16.2cm,clip=true,width=0.9\textwidth]{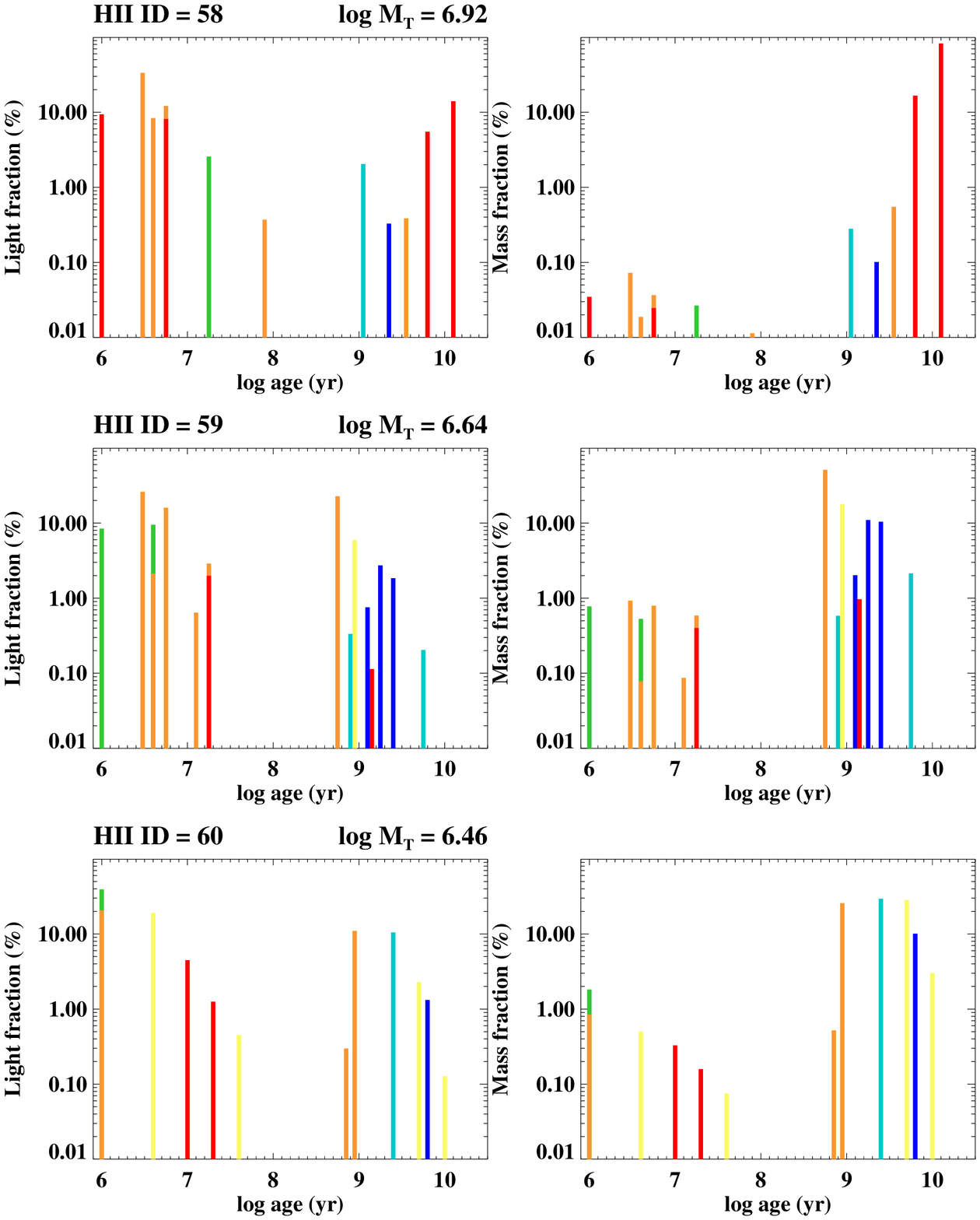}
 \caption{Examples of light- (left) and mass- (right) weighted populations obtained with on run of the STARLIGHT fitting procedure for two identified \hii regions in NGC 3310. Different colours represent different stellar metallicities ($Z$). The derived total mass, in log and solar units (\mbox{log M$_\mathrm{T}$}), is also shown.}
  \label{fig:SL_pop_examples}
\end{figure*}

Once \starlight is run, the contribution in light and in mass of the base spectrum that best fits the input spectrum is provided as a result. In general, less than 20-30 populations are needed to fit our spectra, where normally a handful of young populations dominates the luminosity and a handful of old populations dominates the mass (see examples in Fig.~\ref{fig:SL_pop_examples}). In order to assess the accuracy of the continuum subtraction, the code was run a hundred times for each \hii region spectrum. 

Once the analysis with \starlight is done, the continuum and gas spectra are decoupled, as in Sect.~\ref{sub:2D_gas}, though this time for the stacked spectra of the identified \hii regions, rather than for spectra of individual spaxels. Individual emission-line fluxes were then measured by considering spectral window regions of \mbox{$\sim$ 200 \AA{}}. We produced {\sc idl} routines to perform a simultaneous fitting of several emission lines within the spectral window with Gaussian functions. Only one spectral component is observed for each line (if different components exist, the difference in velocity must be below the spectral resolution of the data,  $\sim$ 600 km s$^{-1}$). This was done for each of the 100 cleaned gas spectra for each \hii region. The statistical errors associated with the observed emission were computed taking into account: (1) the measuring method, given by the fitting and (2) the following expression: 
\begin{equation}
\sigma_\mathrm{l} = \sigma_c N^{1/2} \left[1 + \frac{\mathrm{EW}}{\mathrm{N}\Delta} \right]^{1/2} 
\end{equation}
where $\sigma_\mathrm{l}$ is the error in the observed line flux, $\sigma_c$ refers to the standard deviation in a box near the measured emission line, $N$ is the number of pixels used in the measurement of the line flux, EW is the line EW and $\Delta$ is the wavelength dispersion in \mbox{\AA{} pixel$^{-1}$}~\citep{Gonzalez-Delgado94}. The first term represents the error introduced in placing the continuum level and the second term scales the S/N of the continuum to the line. As a conservative approach, the maximum value between both error estimates was considered. For a given line, we took the median of the computed error and flux line of the 100 spectra as the adopted values for each \hii region, respectively.  Adopted line intensities of the \hii regions are reported in Table~\ref{table:hii_catalogue}.

We performed a sanity test to ensure that no overcorrection was done on the absorption stellar features. For instance, if the fitted old population is such that the \hb absorption is too strong, the obtained ratio \ha\onespace/\hb may well be (within errors) below its theoretical value in case of no dust extinction (i.e. \ha\onespace/\hb = 2.86, assuming case B recombination;~\citealt{Osterbrock89}). This test is described in Sect.~\ref{sec:tests_SL}.

\begin{table*}
\hspace{0.2cm}
\begin{minipage}{0.6\textwidth}
\renewcommand{\footnoterule}{\kern -0.65cm}  
\begin{footnotesize}
\caption{Identified circumnuclear \hii regions in NGC 3310 in previous studies.}
\label{table:comparison_hagele}
\begin{center}
\begin{tabular}{@{\hspace{0.20cm}}l@{\hspace{0.20cm}}c@{\hspace{0.20cm}}@{\hspace{0.20cm}}c@{\hspace{0.15cm}}@{\hspace{0.15cm}}c@{\hspace{0.15cm}}@{\hspace{0.15cm}}c@{\hspace{0.15cm}}@{\hspace{0.15cm}}c@{\hspace{0.15cm}}@{\hspace{0.15cm}}c@{\hspace{0.15cm}}@{\hspace{0.15cm}}c@{\hspace{0.15cm}}@{\hspace{0.15cm}}c@{\hspace{0.15cm}}}
\hline \hline
   \noalign{\smallskip}
H{\tiny II} & ID & $L$ (H$\alpha$) & $L$ (H$\alpha$) & EW (H$\beta$) &  EW (H$\beta$) & $c$ (H$\beta$) & $c$ (H$\beta$) \\
 ID& H10 & & H10 & & H10 & & H10\\
 \hline
   \noalign{\smallskip}
1+4 & J & 554 $\pm$ 55 & 573 & 55.6 $\pm$ 0.9 & 82.5 & 0.10 $\pm$ 0.05 & 0.23 \\ 
3 & N & 161 $\pm$ 16 & 113 & 10.3 $\pm$ 0.2 & 11.0 & 0.14 $\pm$ 0.03 & 0.42 \\ 
5 & R4 & 137 $\pm$ 13 & 144 & 26.1 $\pm$ 0.3 & 32.4 & 0.08 $\pm$ 0.03 & 0.23 \\ 
7 & R5+R6+S6 & 130 $\pm$ 13 & 193 & 22.5 $\pm$ 0.2 & 16.7 & 0.09 $\pm$ 0.03 & 0.24 \\ 
8 & R1+R2 & 219\footnote{Actually, our regions ID8+ID2 match those of R1+R2+R3+R16 in \cite{Diaz00a}, but cannot be individually separated with our observations. Adding up all fluxes we obtain $L$(H$\alpha$) = 535 for ID8+ID2 and about 415 for R1+R2+R3+R16.} $\pm$ 21 & 102 & 43.4 $\pm$ 0.3 & 28.6 & 0.28 $\pm$ 0.02 & 0.23 \\ 
11 & R10+R11 & 108 $\pm$ 10 & 102 & 15.6 $\pm$ 0.2 & 9.7 & 0.05 $\pm$ 0.03 & 0.23 \\ 
12 & R7 & 95\footnote{Large discrepancy due to aperture effects. The radius measured in ~\cite{Diaz00a} for this region corresponds to 1.5 arcsec, smaller than the 3 arcsec radius of the corresponding region detected with our automatic software.} $\pm$ 9 & 45 & 13.8 $\pm$ 0.2 & 19.4 & 0.00 $\pm$ 0.01 & 0.17 \\ 
\hline \noalign{\smallskip}
\multicolumn{9}{@{} p{\columnwidth} @{}}{{\footnotesize \textbf{Notes.} \ha luminosities are in 10$^{38}$ erg s$^{-1}$ and EWs in \AA{}.}}
\end{tabular}
\end{center}
\end{footnotesize}
\end{minipage}
\end{table*}

\hii regions detected in previous studies (\citealt{Hagele10b}; hereafter, H10) have been identified and the \ha extinction-corrected luminosities, \hb EWs and extinction coefficients c(H$\beta$) are compared in Table~\ref{table:comparison_hagele}. Some discrepancies are found, very likely due to the different angular resolution and aperture among the different studies (e.g.,~\citealt{Pastoriza93,Diaz00a}; see footnotes in the table). 
\subsection{Chemical abundance properties of the \hii regions in NGC 3310}
\label{sec:abundance_properties}
With our sample of \hii regions we can investigate the radial distribution of the oxygen abundance in NGC 3310 with better statistics than works based on samples of a few number of \hii regions, basically due to the limitations on the use of slits. 
\subsubsection{Direct oxygen abundance determination}
\label{sec:abundance_direct}
In principle, recombination lines (i.e. \oiir~$\lambda$4649, $\lambda$4089)  would provide the most accurate determination of the abundance, due to their weak dependence on nebular temperature. However, these lines are very faint and most of the observed emission in nebulae correspond to collisionally excited lines (CEL), whose intensities depend exponentially on the temperature. The electron temperature of the gas is useful as an abundance indicator since higher chemical abundances increase nebular cooling, leading to lower \hii region temperatures. This temperature can be determined from the ratios of auroral lines like \oiii $\lambda$4363 to lower excitation lines such as $\lambda\lambda$4959,5007, or \nii $\lambda$5755 to $\lambda\lambda$6548,6584. Methods that rely on the measurements of these lines  in order to obtain the abundance are the so-called \textit{direct methods}.
\begin{table*}
\begin{minipage}{\textwidth}
\renewcommand{\footnoterule}{}  
\begin{small}
\caption{Direct abundance estimates of \hii regions}
\label{table:abundances}
\begin{center}
\begin{tabular}{@{\hspace{0.16cm}}l@{\hspace{0.16cm}}c@{\hspace{0.16cm}}@{\hspace{0.16cm}}c@{\hspace{0.16cm}}@{\hspace{0.16cm}}c@{\hspace{0.16cm}}@{\hspace{0.16cm}}c@{\hspace{0.16cm}}@{\hspace{0.16cm}}c@{\hspace{0.16cm}}@{\hspace{0.16cm}}c@{\hspace{0.16cm}}@{\hspace{0.16cm}}c@{\hspace{0.16cm}}@{\hspace{0.16cm}}c@{\hspace{0.16cm}}@{\hspace{0.16cm}}c@{\hspace{0.16cm}}@{\hspace{0.16cm}}c@{\hspace{0.16cm}}@{\hspace{0.16cm}}}
\hline \hline
   \noalign{\smallskip}
\hii& $I (\lambda$4636) & $t_{\mathrm{e}}$ \oiii $\lambda$4363 $\equiv t_3$ & $n_{\mathrm{e}}$ & $t_{2}~(t_{3}) $  & 12+ log (O/H) &  $I (\lambda$5755) & t$_{\mathrm{e}}$  \nii $\lambda$5755 & O$^{+}$/H$^+$  & O$^{2+}$/H$^+$ & 12+ log (O/H) \\
 ID & &  ($\times$10$^4$) K & (cm$^{-3}$)  & ($\times$10$^4$ K) & &  & ($\times$10$^4$ K) &  & &  ($t_{3}$,$t_{\mathrm{e}}$  \nii $\lambda$5755)  \\
 (1) & (2) & (3) & (4)  & (5) & (6) & (7) & (8) & (9) & (10) & (11)  \\
 \hline
   \noalign{\smallskip}
1& 1.37$\pm$ 0.30& 0.97$\pm$ 0.06& 108$\pm$ 8& 1.00$\pm$ 0.06& 8.30$\pm$ 0.10& 0.74$\pm$ 0.13& 1.04$\pm$ 0.08& 7.96& 7.96&
8.26$\pm$ 0.08\\
4& 1.52$\pm$ 0.24& 1.03$\pm$ 0.05& 115$\pm$ 8& 1.03$\pm$ 0.06& 8.23$\pm$ 0.08& 0.64$\pm$ 0.12& 0.95$\pm$ 0.06& 8.19& 7.82&
8.34$\pm$ 0.09\\
5& 1.93$\pm$ 0.29& 1.12$\pm$ 0.05& 123$\pm$ 17& 1.09$\pm$ 0.05& 8.09$\pm$ 0.07&  \ldots &  \ldots &  \ldots &  \ldots &
 \ldots \\
7& 1.70$\pm$ 0.33& 1.13$\pm$ 0.07& 90$\pm$ 21& 1.13$\pm$ 0.06& 8.03$\pm$ 0.08& 0.89$\pm$ 0.18& 1.01$\pm$ 0.08& 8.02& 7.60&
8.16$\pm$ 0.12\\
8& 1.84$\pm$ 0.42& 1.08$\pm$ 0.07& 112$\pm$ 18& 1.07$\pm$ 0.06& 8.17$\pm$ 0.10& 0.79$\pm$ 0.11& 1.00$\pm$ 0.05& 8.08& 7.76&
8.25$\pm$ 0.08\\
13& 1.51$\pm$ 0.35& 1.07$\pm$ 0.07& 77$\pm$ 16& 1.10$\pm$ 0.06& 8.12$\pm$ 0.10& 1.19$\pm$ 0.11& 1.17$\pm$ 0.05& 7.81& 7.69&
8.06$\pm$ 0.06\\
14& 1.64$\pm$ 0.36& 0.96$\pm$ 0.06& 37$\pm$ 6& 1.07$\pm$ 0.07& 8.30$\pm$ 0.10&  \ldots &  \ldots &  \ldots &  \ldots &
 \ldots \\
15& 1.98$\pm$ 0.33& 1.14$\pm$ 0.06& 52$\pm$ 20& 1.19$\pm$ 0.05& 8.02$\pm$ 0.06& 0.89$\pm$ 0.14& 1.04$\pm$ 0.07& 8.03& 7.64&
8.18$\pm$ 0.09\\
16& 2.03$\pm$ 0.34& 1.17$\pm$ 0.06& 108$\pm$ 24& 1.13$\pm$ 0.05& 8.07$\pm$ 0.07& 1.10$\pm$ 0.22& 1.19$\pm$ 0.12& 7.81& 7.59&
8.02$\pm$ 0.11\\
19& 2.23$\pm$ 0.46& 1.35$\pm$ 0.11& 174$\pm$ 64& 1.16$\pm$ 0.05& 7.96$\pm$ 0.06&  \ldots &  \ldots &  \ldots &  \ldots &
 \ldots \\
20& 2.61$\pm$ 0.45& 1.29$\pm$ 0.08& 109$\pm$ 48& 1.21$\pm$ 0.05& 7.95$\pm$ 0.05&  \ldots &  \ldots &  \ldots &  \ldots &
 \ldots \\
21& 2.46$\pm$ 0.35& 1.27$\pm$ 0.07& 116$\pm$ 38& 1.19$\pm$ 0.05& 7.99$\pm$ 0.05& 1.26$\pm$ 0.25& 1.27$\pm$ 0.13& 7.74& 7.46&
7.93$\pm$ 0.12\\
22& 2.18$\pm$ 0.48& 1.05$\pm$ 0.07& 34$\pm$ 15& 1.15$\pm$ 0.07& 8.16$\pm$ 0.09&  \ldots &  \ldots &  \ldots &  \ldots &
 \ldots \\
26& 1.63$\pm$ 0.34& 1.06$\pm$ 0.06& 52$\pm$ 14& 1.12$\pm$ 0.06& 8.12$\pm$ 0.09&  \ldots &  \ldots &  \ldots &  \ldots &
 \ldots \\
27& 3.68$\pm$ 0.47& 1.17$\pm$ 0.05& 92$\pm$ 25& 1.15$\pm$ 0.05& 8.12$\pm$ 0.05&  \ldots &  \ldots &  \ldots &  \ldots &
 \ldots \\
32& 2.11$\pm$ 0.47& 1.05$\pm$ 0.07& 14: & 1.21$\pm$ 0.10& 8.12$\pm$ 0.09&  \ldots &  \ldots &  \ldots &  \ldots &
 \ldots \\
\hline \noalign{\smallskip}
\multicolumn{11}{@{} p{\textwidth} @{}}{{\footnotesize \textbf{Notes.} Col (1): \hii region identification number. Col (2): reddening-corrected flux of the \oiii $\lambda$4363 auroral line with respect to \hb\onespace, normalized to  \mbox{\hb = 100}.~Col(3): direct determination of the electron temperature of \oiii. Col (4): derived electron density. Col (5): derived electron temperature of \oii, assuming that \mbox{$t_\mathrm{e}$ (\oii) $\sim$ $t_\mathrm{e}$ (\nii)} and the theoretical relation between $t_\mathrm{e}$ (\oii) and $t_\mathrm{e}$ (\oiii) as prescribed in~\cite{Perez-Montero03} (Eq.~\ref{eq:t2_t3}). Col (6): derived oxygen abundance with a direct-temperature determination of \oiii. Col (7): reddening-corrected flux of the \nii $\lambda$5755 auroral line with respect to \hb, normalized to  \mbox{\hb = 100}. Col (8): direct determination of the electron temperature for \nii. Col (9): O$^+$ ionic abundance, derived using both direct determinations of the electron temperature of \oiii~and \nii. Col 
(10): same as in 
col. (9), but for the O$^{2+}$ ion. Col. (11): derived oxygen abundance using both direct determinations of the electron temperature of \oiii~and \nii.}}
\end{tabular}
\end{center}
\end{small}
\end{minipage}
\end{table*}

We measured the auroral line \oiii $\lambda$4363 with a detection level of \mbox{S/N $>$ 4} from the ``residual'' gas spectra of 16 \hii regions. We refer the reader to Sect.~\ref{sec:t_obs_res}, where we argue about the systematics introduced when measuring this line on the observed spectrum. The determination of the electron temperature ($t_\mathrm{e}$\footnote{In what follows $t_\mathrm{e}$ denotes electron temperature in units of $10^4$ K.}) and other physical conditions of the gas such as the electron density ($n_\mathrm{e}$, in cm$^{-3}$), and the ionic abundances were obtained using the procedures outlined in~\cite{Perez-Montero07} and~\cite{Hagele08b}. We have assumed two distinct layers of ionization for each \hii region: a high ionization zone (\heii, \oiii, \neiii, etc.) and a low ionization zone (\oii, \nii, \sii, etc.), whose electron temperatures are given by $t_3$ and $t_2$, respectively.

Different prescriptions can be found in the literature that relate the $t_3$ and the $t_2$ temperatures. In our case, the $t_2$ temperature was determined using photoionization models that take into account the dependence on the electron density~\citep{Perez-Montero03,Perez-Montero09}:

\begin{equation}
 \label{eq:t2_t3}
t_2 \equiv t_\mathrm{e}(\textrm{\oii}) = \frac{1.2 + 0.002n_\mathrm{e} + 4.2/n_\mathrm{e}}{t_3^{-1} + 0.08 +0.003n_\mathrm{e} + 2.5/n_\mathrm{e}},
\end{equation}
where the electron density was determined from the \sii $\lambda\lambda$6717,6731 doublet.

In a few cases the \nii$\lambda$5755 line was also detected and measured. In those cases, we could derive the temperature of \nii, the temperature of the low ionization zone.

The ionic abundances were calculated based on the functional forms provided by~\cite{Hagele08b}, who published a set of equations for the determination of oxygen abundances in \hii regions based on a five-level atom model:
\begin{flalign}
 & 12 + \textrm{log(O}^{2+}/\textrm{H}^+) & \mspace{-20.0mu}=~&   \textrm{log}(\textrm{\oiii}\lambda\lambda4959,5007/\textrm{H}\beta) + 6.144~+ \\
 & & &\mspace{-90.0mu} +1.251/t_3 - 0.55\textrm{log}t_3  \notag  \\\notag\\
 & 12 + \textrm{log(O}^+/\textrm{H}^+) & \mspace{-20.0mu}=~& \textrm{log}(\textrm{\oii}\lambda\lambda3727,3729/\textrm{H}\beta) + 5.992~+    \\
 & & & \mspace{-90.0mu}+ 1.583/t_2 - 0.681\textrm{log}t_2 + \textrm{log}(1 + 0.00023n_\mathrm{e})  \notag
\end{flalign}
Finally, the total oxygen abundance was obtained assuming that

\begin{equation}
\frac{\textrm{O}}{\textrm{H}} = \frac{\textrm{O}^+}{\textrm{H}^+} + \frac{\textrm{O}^{2+}}{\textrm{H}^{+}}
\end{equation}

Auroral line measurements, electron density and temperatures, along with the derived oxygen abundances using the methodology described in this section are reported in Table~\ref{table:abundances}. With typical electron densities of the order of \mbox{100 cm$^{-3}$} and typical electron temperatures of the order of \mbox{10000 K}, the derived direct abundances are subsolar, in the range \mbox{8 $\lesssim$ 12 + log(O/H) $\lesssim$ 8.3}.

\subsubsection{Strong-line methods}
\label{sec:empirical_methods}

The ratios used in direct methods involve the detection and measurement of at least one intrinsically weak line, which in objects of low excitation and/or low surface brightness, often result too faint to be observed. Given these limitations, \textit{strong-line methods} based on the use of strong, easily observable, optical lines have been developed throughout the years. Several abundance calibrators have been proposed involving different emission-line ratios and have been applied to determine oxygen abundances in objects as different as individual HII regions in spiral galaxies, dwarf irregular galaxies, nuclear starbursts and emission-line galaxies.  By far, the most commonly strong-line calibrator used  is the ratio (\oii $\lambda$3727 + \oii $\lambda\lambda$4959, 5007)/\hb, known as the $R_{23}$ method \citep{Pagel79}. However, it is known to present many problems basically because it is double valued with a wide transition/turn-over region (\mbox{12+log(O/H) $\sim$ 8.0--8.3}). Furthermore, a few 
observed circumnuclear \hii regions in NGC 3310 are known to have a moderately low metallicity (\mbox{12+log(O/H) $\sim$ 8--8.3};~\citealt{Pastoriza93}), just in the middle of the turn-over region of the calibrator. The spectral range covered by the PINGS data allows us, nevertheless, to employ more up-to-date strong-line calibrations that make use of more information via strong nitrogen and/or sulphur lines. This are as follows.
\begin{itemize}

\begin{figure*}
\centering
\includegraphics[trim = 0cm 1cm 0cm 0cm,clip=true,width=0.80\textwidth]{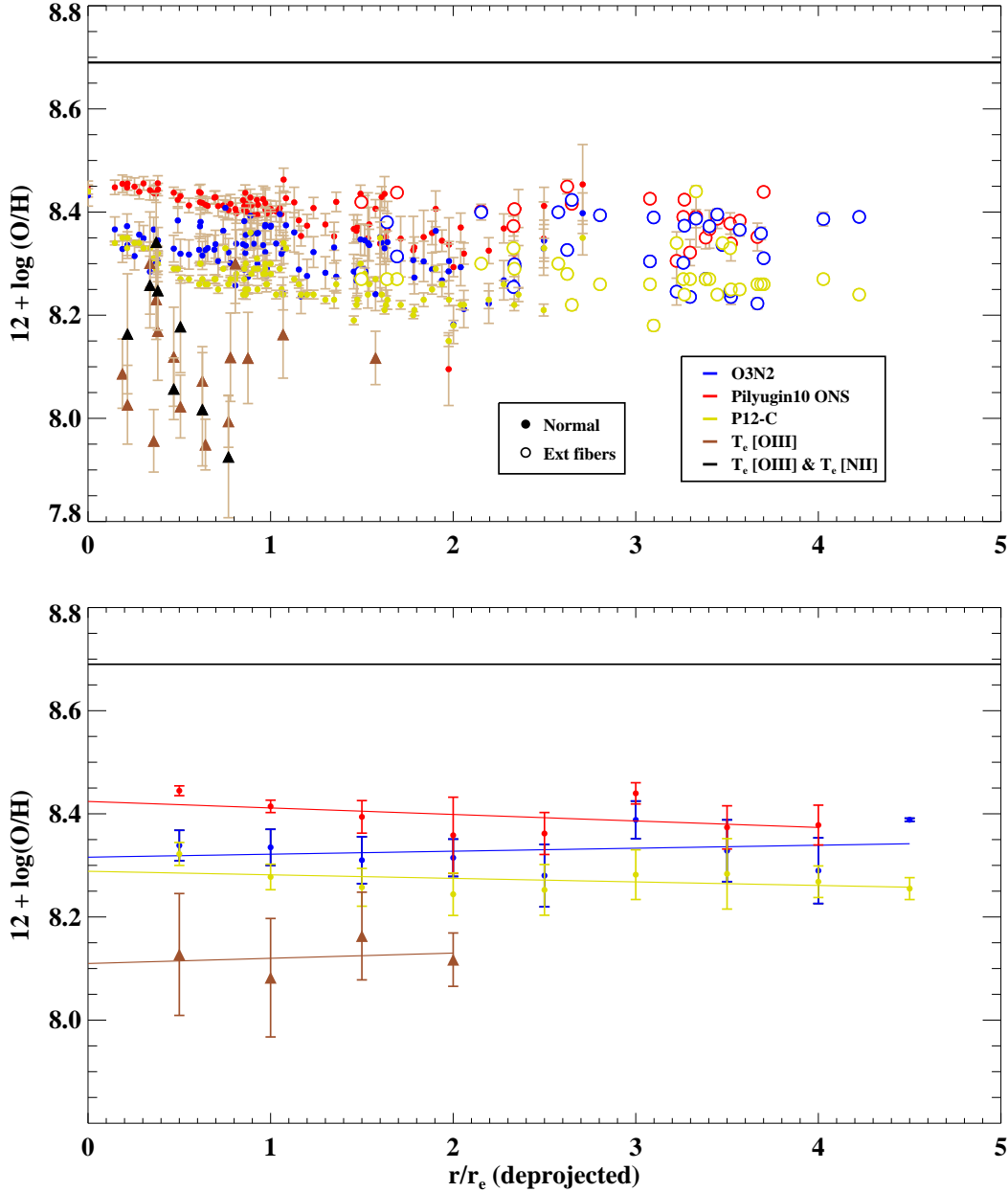}

 \caption{\textbf{Top:} radial distribution of oxygen abundance in NGC 3310. $X-$ axis shows the deprojected radial distance. Different colours refer to different methods use to derive the oxygen abundance (see legend). Estimates from spectra obtained with data within the hexagonal PPAK FoV are plotted in solid symbols (either circles or triangles, the latter referring to $t_\mathrm{e}$-based estimates), whereas an open symbol means that the spectra were obtained from external fibres (see text). The horizontal line highlights the position of the solar abundance of \mbox{12 + log(O/H) = 8.69}~\citep{Asplund09} in the diagram. \textbf{Bottom:} data have been grouped in distance bins of 0.5$r$\reff and median values have been plotted. The error bars represent the dispersion within a given bin. The lines show a straight line fitted to the median values. We have not grouped the estimates derived from T$_\mathrm{e}$-based methods that use both the auroral \oiii~ and \nii~lines due to the lack of data.}
  \label{fig:metal_radial_gradients_a}
\end{figure*}

\item \textit{$O3N2$-parameter calibrations} -- The $O3N2$ parameter, first introduced by \cite{Alloin79}, depends on two SEL intensity ratios: 

\begin{equation}
  O3N2 = \textrm{log} \left ( \frac{\textrm{\oiii}\lambda5007}{\textrm{H}\beta} \times \frac{\textrm{H}\alpha}{\textrm{\nii}\lambda6584} \right )
\end{equation}

This parameter is almost independent of either reddening correction or flux calibration. Several calibrations using this parameter have been defined (e.g.,~\citealt{Pettini04,Perez-Montero09}). Recently,~\cite{Marino13} have provided an improved calibration using the largest compilation so far of temperature-based abundance determinations, and has proved its validity for \mbox{12 + log(O/H) $> 8.1-8.2$} with a dispersion somewhat lower than 0.2 dex.

\item \textit{ONS calibration} -- This calibration~\citep{Pilyugin10} is based on several  strong-line-intensity ratios to \hb from several species, including oxygen, nitrogen and sulphur as
\begin{flalign}
\label{eq:indeces}
&R_2 = \textrm{\oii}\lambda\lambda3727,3729/\textrm{H}\beta & \notag\\
&R_3 = \textrm{\oiii}\lambda\lambda4959,5007/\textrm{H}\beta & \\
&N_2 = \textrm{\nii}\lambda\lambda6548,6854/\textrm{H}\beta  & \notag\\
&S_2 = \textrm{\sii}\lambda\lambda6717,6731/\textrm{H}\beta & \notag
\end{flalign}
It uses the excitation parameter, ($P = R_3/(R_3 + R_2)$), that takes into account the effect of the ionization parameter. This calibration was derived using a set of \hii regions with measured electron temperatures. 

\item \textit{C-method} -- The `counterpart' method, P12-C~\citep{Pilyugin12b}, is based on the standard assumption that \hii regions with similar intensities of SEL have similar physical properties and abundances. Given all the problems that different calibrations have (e.g., different branches, different applicability ranges, the no one-to-one correspondence between oxygen and nitrogen abundances) the authors propose a method that does not know a priori in which metallicity interval (or on which of the two branches) the \hii region is located. This calibration basically selects a number of reference (well-measured abundances) \hii regions and then the abundances in the target \hii region are estimated through extra-/interpolation. According to the authors, if the errors in the line measurements are within 10\%, then one can expect that the uncertainty in the \mbox{C-based} abundances is not in excess of 0.1 dex (although it can reach \mbox{0.15--0.2 dex} in the interval \mbox{7.8 $<$ 12 + log(O/H) 
$<$ 8.2}). 
\end{itemize}

We made use of the calibrations outlined in order to search for any radial abundance gradient trend in \mbox{NGC 3310} within dispersions of \mbox{0.1-0.2 dex}, though larger systematic offsets can be expected from the results obtained from one calibration to another.

\subsubsection{Radial abundance gradients}
\label{sec:radial_gradients}

We have only a limited sub-sample of \hii regions with reliable $t_\mathrm{e}$-based abundance determinations. We thus derived abundances for almost all identified \hii regions using the strong-line calibrations described in previous sections. Fig.~\ref{fig:metal_radial_gradients_a} shows the radial distribution of the oxygen abundance in NGC 3310 up to about four effective radii (\reff\onespace). At the assumed distance of this galaxy ($D_\mathrm{L}$ = 16.1 Mpc), this distance corresponds to about 11 kpc. We identify \reff as the half-light radius, which was determined using \galfit\footnote{http://users.obs.carnegiescience.edu/peng/work/galfit/galfit.html}, version 3.0~\citep{Peng10}, on the $g$-band SDSS image. Either  assuming a typical S\'ersic (letting the index $n$ vary) + an exponential disc profiles or a rotating S\'ersic (simulating the spiral arms) + an exponential disc profiles, the resulting \reff is around 35 arcsec~(i.e.~\mbox{$\sim 2.5$ kpc}).

For this part of the study, we have included in our catalogue spectra corresponding to the external fibres of the PPAK module (normally used to determine the local sky, grouped in bundles at $\sim$75 arcsec from the centre of the PPAK module) which show evident \hii\onespace-like emission with coincident redshift of that of NGC3310, i.e. fortuitous observations of \hii regions at very large galactocentric distances. Their 2D spatial distribution is shown in Fig.~\ref{fig:external_fibers}. In many cases, given the dithering used, the flux of 2 or 3 fibres could be integrated to obtain each spectrum of these `external' regions. Although the S/N of the continuum is generally very low, the emission lines are clearly detected. The extinction corrected line intensities  are reported in Table~\ref{table:ext_fibers_catalogue}. These additional regions allow us to investigate if there is any abundance radial gradient in this galaxy up to more than 10 kpc. If this data set is not included we still 
can sample \hii regions within about 5 kpc or about 2.8\reff. Several interesting aspects can be inferred from the plot as follows.

\begin{enumerate}
 \item All computed gaseous abundances are sub-solar, spanning the range \mbox{7.95 $<$ 12 + log(O/H) $<$ 8.45} (between a half and a fifth solar).

 \item All computed abundances using strong-line calibrations agree within 0.1--0.15 dex, covering a typical range \mbox{12 + log(O/H) = 8.2--8.4}. This is expected, since the accuracy of strong-line calibrations (\mbox{0.1--0.2 dex}), not included in the error bars, is of the order of this range.

 \item In practice, we have obtained $t_\mathrm{e}$-based estimates within one \reff. Their range span oxygen abundances between 8 and 8.3 with a larger dispersion than those computed with strong-line calibrations. In general, the $t_\mathrm{e}$-based estimates are also lower by about \mbox{0.2--0.3 dex}. With a sample of 16 \hii regions with direct abundance estimates and over 100 regions with strong-line estimates we can assure that the offset is not due to lack of statistics. In Sect.~\ref{sec:t_obs_res}, we discuss on this disagreement.

\begin{figure}
\centering
\includegraphics[angle=90,trim = 0.5cm 1cm 2cm 3.5cm,clip=true,width=1.1\columnwidth]{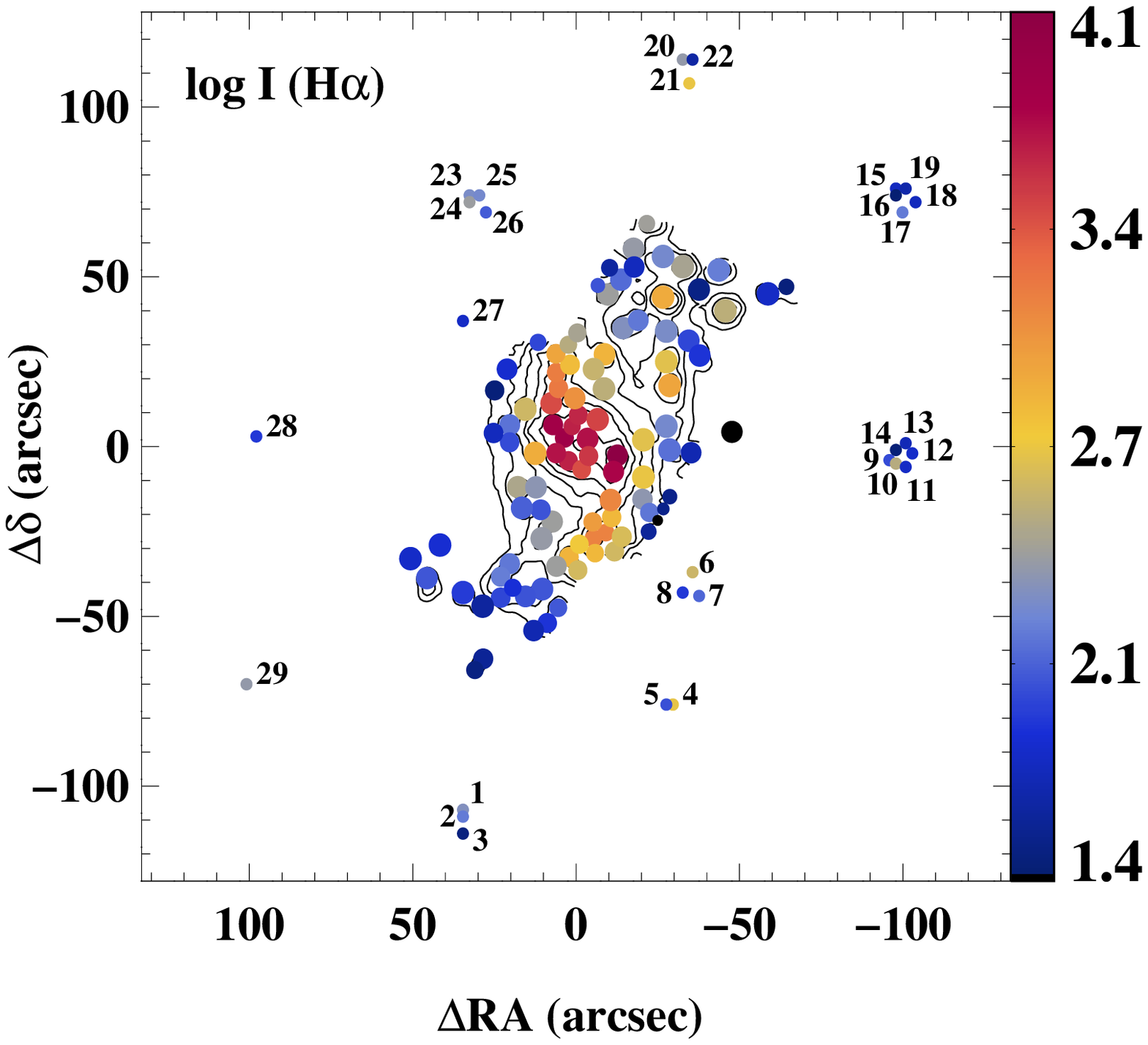}
 \caption{\ha contour map showing the spatial 2D distribution of the identified \hii regions and the external regions (with their identification number). While the size of the circles for the identified \hii regions represents approximately their actual size, that for the external regions corresponds to their average size. The circles are colour coded following the logarithm of the \ha intensity in units of \mbox{10$^{-16}$ erg s$^{-1}$ cm$^{-2}$}.}
  \label{fig:external_fibers}
\end{figure}

 \item With a sample of over 100 \hii regions we do not see a clear abundance gradient in NGC 3310, further than 10 kpc away from the nucleus. A weak gradient might be present from the centre and up to 2\reff (5 kpc). We have computed the Spearman correlation coefficient for $r < 2$\reff and have obtained a possible correlation for abundances that were derived using the ONS and the \mbox{P-12C}~calibrations ($r = 0.63,0.72$, respectively). The derived slopes of the fit are 0.06 and 0.05 dex/\reff (\mbox{$\sim$ 0.02 dex/kpc}), respectively. Should an abundant gradient exist, it is significantly flatter than the observed gradients in galactic discs (i.e. \mbox{$\sim$ -0.08 dex/kpc};~\citealt{Vilchez88,Kennicutt96,Rosolowsky08,Costa10}) and the universal gradient proposed in \cite{Sanchez14} (i.e. \mbox{ $\sim$ -0.10 dex/\reff}). For abundances obtained using the $O3N2$ calibration and the direct method (using the \oiii $\lambda$4363 emission line), the data are not correlated, according to the Spearman's rank-
order 
correlation  test. Besides, the intrinsic dispersion of strong-line calibrations is typically \mbox{0.1--0.2} dex. All facts considered, we conclude that either the abundance gradient of the gas in the disc of NGC 3310 is flat or at least significantly flatter than gradients observed in spirals.   

\end{enumerate}

\subsection{Characterization of the ionizing stellar population}

With the spectral information provided with the IFU data and available public wide-band imaging we can perform a detailed study on the ionizing stellar populations in NGC 3310. In particular, the line flux ratios, the EW measurements and broad-band UV and optical images help us to tightly constrain the  mass and the age of the ionizing stellar populations present in the \hii regions. 

\begin{figure*}
\centering
\includegraphics[trim = 8.5cm 0cm 0cm 0cm,clip=true,angle=90,width=0.9\textwidth]{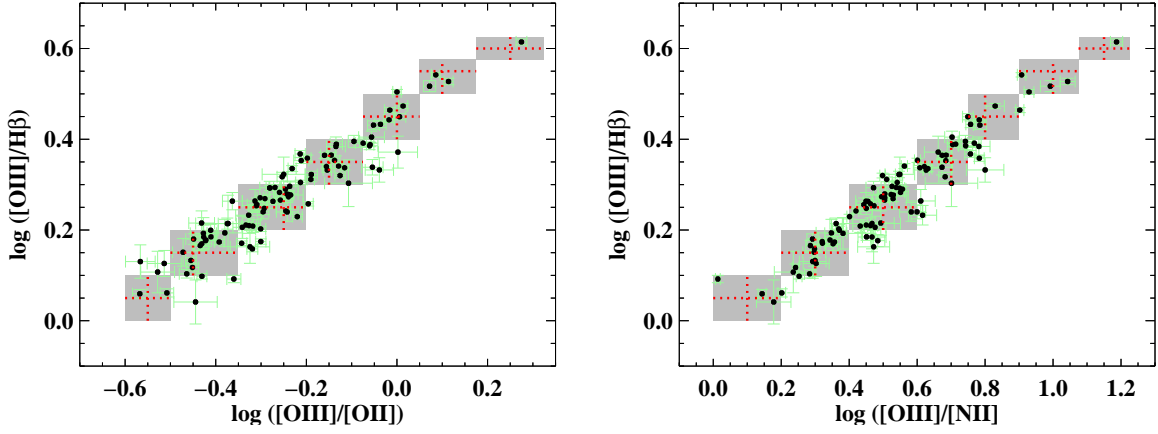}

 \caption{\oiii/\hb line intensity ratio versus several intensity ratios sensitive to the ionization structure of the nebulae for the identified \hii regions in NGC 3310. Seven interval classes have been defined as the shaded regions in the different plots. All ratios for a given \hii region lies on the same class or in-between two classes. A typical average point value for each class is drawn as a red cross.}
  \label{fig:ionization_ratios1}
\end{figure*}

The main advantage in subtracting the stellar continuum is that the end product of such procedure is the percentage of each contributing stellar population to the light and mass. That is, we can estimate the mass of each stellar population responsible for the continuum emission. We may note, however, that any analysis technique has its own advantages and limitations. In particular, given the limited spectral range fitted (i.e. part of the optical), \starlight gives a consistent description of the age of the `young' population within 0.15-0.20 dex independently of the stellar templates used for an average age of \mbox{$\tau \sim 100$ Myr}~\citep{Cid-Fernades13}. Constraining the age of the population at \mbox{$\tau \leq 10 $Myr} (expected for ionizing population) is highly uncertain using this technique alone. Therefore, including data from other spectral bands, especially at bluer wavelengths, helps to better constrain the properties of young ionizing populations. 

Since this population is ionizing the gas, the emission lines we observe in the gas spectra give us useful information on its properties. For that reason, in the following we first compare the measured emission-line fluxes with those computed in photoionization models. This comparison gives us a first rough estimate of the age of the ionizing population and of the presence of dust grains. Next, we perform a multiwavelength analysis, covering a wider spectral range using SDSS and \textit{XMM}-OM images. Finally, we combine both techniques to better constrain the age and the mass of the ionizing population.

\begin{figure*}
\centering
\includegraphics[trim = -0.5cm 0cm 0cm -0.5cm,clip=true,angle=90,width=0.9\textwidth]{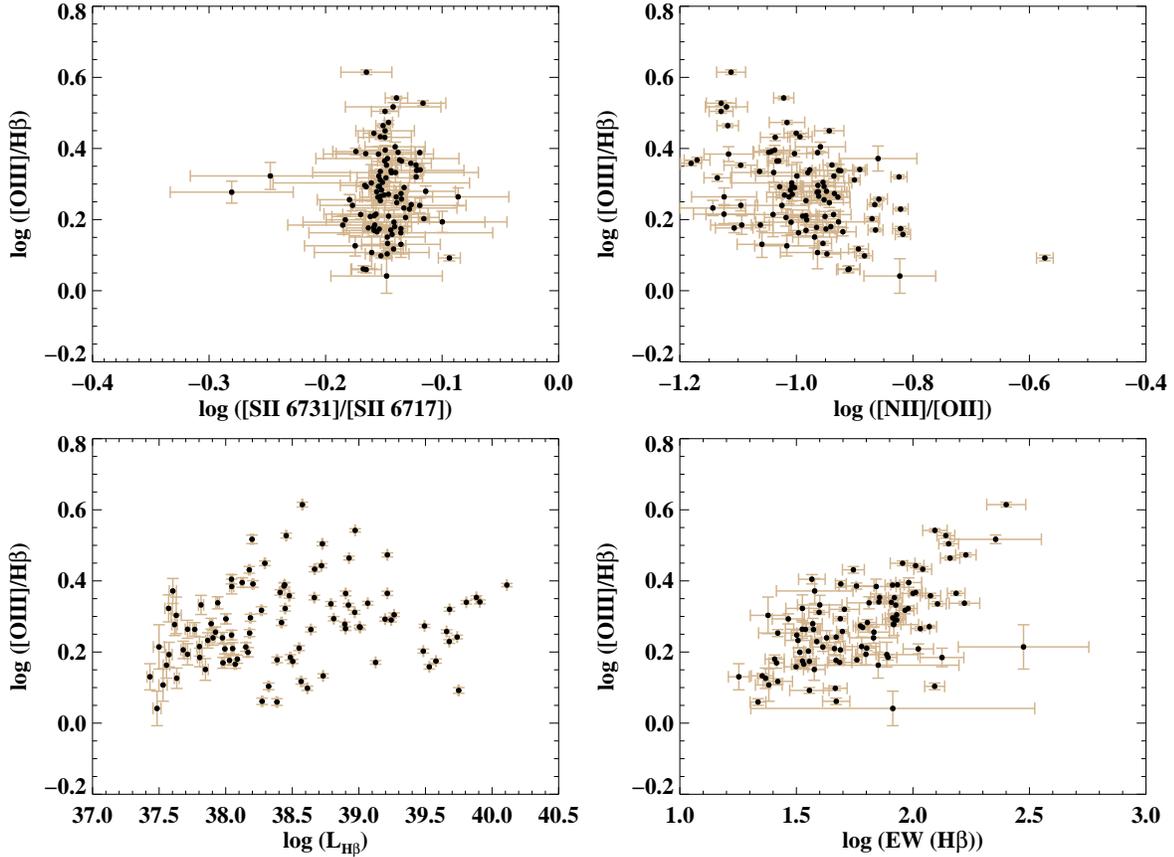}

 \caption{Same as in Fig.~\ref{fig:ionization_ratios1}, but versus other sensitive intensity and intensity ratios. In this case no class has been defined since correlation are either weak (i.e. with \nii/\oii, proxy to the abundance ratio N/O) or non-existent. The derived \hb luminosity is given in units of \mbox{erg s$^{-1}$}. The EW has been corrected by the continuum of the non-ionizing population.}
  \label{fig:ionization_ratios2}
\end{figure*}

\subsubsection{Photoionization models}
\label{sec:photo_models}

First of all, we took advantage of the knowledge of the ionization conditions of the gas within an \hii region. The strength of different line flux ratios depends on the shape of the ionizing continuum (i.e. age and metallicity of the young stellar population) and on the conditions and geometry of the cloud (i.e. electron density, temperature, clumpiness, absorption by dust grains, etc.). Based on this conception, we show in Figs.~\ref{fig:ionization_ratios1} and~\ref{fig:ionization_ratios2} the dependence of the \oiii$\lambda$5007/\hb line intensity ratio (in log units) on other line ratios that provide information on the degree of ionization (e.g., the \oiii/\oii~ratio), the electron density of the cloud (ratios of the sulphur lines), the abundance, the electron temperature,   the number of ionizing photons and the age of the burst (i.e. EW(\hb\onespace)). 

As Fig.~\ref{fig:ionization_ratios1} illustrates, there is a good correlation between the \oiii$\lambda$5007/\hb ratio and other line intensity ratios sensitive to the ionization structure of the cloud. We defined seven classes with different typical ratios (red dashed lines in the plot) and some width, which define the shaded areas, in such a way that in practically all cases the ratios for a given \hii region lies in the same class or in-between two classes. Class 1 would correspond to ratios on the bottom shaded area on the left with a typical log(\oiii/\hb\onespace) ratio of 0.05. Moving right and up in the plot, the other classes are defined by the area of the shaded rectangles up to class seven where the typical log(\oiii/\hb\onespace) ratio equals 0.6. On the other hand, as Fig.~\ref{fig:ionization_ratios2} shows, the ratio between the sulphur lines (sensitive to the electron density) and the \hb luminosities do not really depend on the \oiii/\oii~ratio. There is a mild dependence on the \nii/\
oii~ratio,
sensitive to the abundance ratio N/O~\citep{Perez-Montero05,Perez-Montero09} and on the EW(\hb\onespace), which has been corrected for the contribution to the continuum by non-ionizing population as derived with \starlight (see next section). Finally, although for classes 1--4 the logarithm of the \hb luminosity spans the whole range sampled (i.e. 37.5--40), for classes 5--7 the range is more restricted, from somewhat lower than 38.5 to somewhat higher than 39.0.

\begin{table*}
\begin{minipage}{0.85\textwidth}
\renewcommand{\footnoterule}{}  
\begin{center}
\caption{\cloudy simulations. Input line intensity ratio ranges, derived ages ($\tau$) and dust absorption factors ($f_{\mathrm{d}}$).}
\label{table:cloudy_tramos}
\begin{tabular}{cccccccc|cc}
\hline \hline
\multicolumn{1}{c}{Class} \vline &
\multicolumn{7}{c}{Line ratio and EW ranges} \vline &
\multicolumn{2}{c}{Result of the fit} \\

\multicolumn{1}{c}{} \vline &
\multicolumn{1}{c}{\oii/\hb}  & 
\multicolumn{1}{c}{\oiii/\hb}  & 
\multicolumn{1}{c}{\nii/\hb}  & 
\multicolumn{1}{c}{\sii$^{1}$/\hb} & 
\multicolumn{1}{c}{\sii$^{2}$/\hb} &
\multicolumn{1}{c}{log L(\hb)} &
\multicolumn{1}{c}{log EW (\hb)} \vline &
\multicolumn{1}{c}{log ($\tau$)} & 
\multicolumn{1}{c}{$f_\mathrm{d}$} \\

\multicolumn{1}{c}{} \vline &
\multicolumn{1}{c}{}  & 
\multicolumn{1}{c}{}  & 
\multicolumn{1}{c}{}  & 
\multicolumn{1}{c}{} & 
\multicolumn{1}{c}{}  &
\multicolumn{1}{c}{(erg s$^{-1}$)} &
\multicolumn{1}{c}{(\AA{})} \vline &
\multicolumn{1}{c}{(Myr)} & 
\multicolumn{1}{c}{} \\\hline
1	&	3.16--5.00	&	1.00--1.26	&	0.63--1.26	&	0.63--1.26	&	0.53--0.84	&	37.5-38.5	&	1.35-1.65	&	6.71 $^{+0.02}_{-0.03}$	&	2.1 $^{+0.8}_{-0.4}$	\\
\noalign{\smallskip}																			
	&		&		&		&		&		&	38.8-39.8	&		&	6.71 $^{+0.01}_{-0.03}$	&	2.3 $^{+0.6}_{-0.4}$	\\
\noalign{\smallskip}																			
2	&	2.82--5.00	&	1.26--1.58	&	0.63--1.00	&	0.50--1.00	&	0.35--0.67	&	37.5-38.5	&	1.35-1.65	&	6.71 $^{+0.01}_{-0.03}$	&	2.4 $^{+0.9}_{-0.6}$	\\
\noalign{\smallskip}																			
	&		&		&		&		&		&	38.8-39.8	&		&	6.71 $^{+0.02}_{-0.03}$	&	1.9 $^{+0.6}_{-0.4}$	\\
\noalign{\smallskip}																			
3	&	2.50--4.47	&	1.58--2.00	&	0.50--0.80	&	0.47--0.79	&	0.28--0.56	&	37.5-38.5	&	1.6-2.1	&	6.65 $^{+0.02}_{-0.15}$	&	2.9 $\pm$ 1.2	\\
\noalign{\smallskip}																			
	&		&		&		&		&		&	38.8-39.8	&		&	6.65 $^{+0.02}_{-0.05}$	&	3.0 $^{+1.3}_{-0.9}$	\\
\noalign{\smallskip}																			
4	&	2.37--3.76	&	2.00--2.51	&	0.45--0.63	&	0.30--0.60	&	0.24--0.45	&	37.5-38.5	&	1.6-2.1	&	6.63 $^{+0.06}_{-0.04}$	&	1.3 $^{+0.4}_{-0.2}$	\\
\noalign{\smallskip}																			
	&		&		&		&		&		&	38.8-39.8	&		&	6.64 $^{+0.03}_{-0.06}$	&	1.6 $\pm$ 0.4	\\
\noalign{\smallskip}																			
5	&	2.24--3.76	&	2.51--3.16	&	0.32--0.56	&	0.28--0.47	&	0.21--0.38	&	38.5-39.5	&	1.6-2.1	&	6.63 $^{+0.04}_{-0.05}$	&	1.4 $^{+0.4}_{-0.3}$	\\
\noalign{\smallskip}																			
6	&	2.11--3.35	&	3.16--3.76	&	0.32--0.47	&	0.27--0.42	&	0.21--0.32	&	38.5-39.5	&	2.05-2.35	&	6.58 $^{+0.03}_{-0.06}$	&	1.2 $^{+0.5}_{-0.1}$	\\
\noalign{\smallskip}																			
7	&	1.99--2.82	&	3.76--4.22	&	0.22--0.35	&	0.22--0.35	&	0.13--0.24	&	38.5-39.5	&	2.05-2.35	&	6.55 $^{+0.06}_{-0.09}$	&	1.8 $^{+1.0}_{-0.6}$	\\
\hline \noalign{\smallskip}
\multicolumn{10}{@{} p{\textwidth} @{}}{\textbf{Notes.} \oiii refers to \oiii$\lambda$5007 \AA{}, \nii to \nii$\lambda$6584 \AA{}, \sii$^{1}$ to \sii$\lambda$6717 \AA{}  and \sii$^{2}$ to \sii$\lambda$6731 \AA{}.}
\end{tabular}
\end{center}
\end{minipage}
\end{table*}

With this information, we have simulated the properties of the ionized gas and the stellar ionizing population for each class with the use of the photoionization code \cloudy (version 10.0;~\citealt{Ferland98}). We took as ionizing sources the SEDs of ionizing star clusters spanning ages from 1 to \mbox{10 Myr} using evolutionary synthesis techniques (\popstar\onespace;~\citealt{Molla09,Martin-Manjon10}). We used models with a Salpeter IMF, with masses between 0.15 and 100 \msun\onespace. We assumed in all the models a radiation-bounded spherical geometry, a constant density of 100 particles cm$^{-3}$  and a standard fraction of dust grains in the interstellar medium (ISM). The presence of solid grains within the ionized gas  can have some consequences on the physical conditions of the nebulae that should not be neglected. The heating of the dust can affect the equilibrium between the cooling and heating of the gas and, thus, the electron temperature inside the nebulae. The 
command PGRAINS, included in the last version of \cloudy, implements the presence of dust grains as described in~\cite{VanHoof04}. In all the models, the calculation was stopped when the temperature was lower than 4000 K. We also assumed that the gas has the same metallicity as the ionizing stars, covering the values 0.2Z$_\odot$ (0.008) and 0.4Z$_\odot$ (0.004) since gaseous abundances of \hii regions in this galaxy range typically between these two values (see Sect.~\ref{sec:radial_gradients}). The other elemental abundances were set in solar proportions taking as a reference the phostosphere solar values given by~\cite{Asplund05}, except in the case of nitrogen, for which we considered three different values of log(N/O), according to the mild correlation observed in Fig.~\ref{fig:ionization_ratios2}: -1.05 (classes 6 and 7), -0.95 (classes 2--5) and -0.85 (class 1). We let the age of the ionizing population, the oxygen abundance, the number of ionizing photons, some geometrical parameters (inner radius 
and filling factor) and the content of dust grains vary in order to fit the line intensity ratios of oxygen, nitrogen and sulphur to \hb, together with the \hb luminosity and the continuum at \mbox{4885 \AA{}} (obtained with the EW(\hb\onespace)). 

Once a compatible solution was found, we fixed all the fitted parameters save the dust grain content, the age of the ionizing population and the number of ionizing photons and run a grid of models varying these parameters. They are much more sensitive to line ratio variations, especially to the oxygen-to-\hb ratio. A solution in the grid was considered positive if it lied between the limits set by the shaded areas in Fig.~\ref{fig:ionization_ratios1}. Note that any line ratio can be derived just by subtracting one value from the `y'-axis to that corresponding from the `x'-axis; for instance, \mbox{log (\oiii/\hb) -- log(\oiii/\nii) = log(\nii/\hb)}. A more constrained solution for each \hii region was obtained by comparing the observables derived with the model with the extinction corrected ratios, the \hb luminosity and the EW(\hb\onespace) of each region.

Assumed intervals of the line intensity ratio for each class together with the solution for the ages are listed in Table~\ref{table:cloudy_tramos}. In general, the excitation can be explained in all cases by only one ionizing population of \mbox{3--5.5 Myr} and, unlike \starlight fitting, none around \mbox{1 Myr}. Another important output from the grid of models corresponds to the UV absorption factor due to the internal extinction within the cloud, denoted as $f_{\mathrm{d}}$, also provided in the table. That is, if \mbox{$f_{\mathrm{d}}$ = 1}, no relevant absorption by dust grains is expected and if it is larger than 1, then such a factor is missed from the observed Balmer emission light. As inferred from the table, in all cases some emission is lost due to this UV absorption. In some cases, the loss is compatible with being marginal (i.e. for classes 4--6). However, in general, about 33\% (\mbox{$f_{\mathrm{d}}$ = 1.5}) of the emitted UV photons or even more could be missed. Particularly, for class 3, the 
loss is expected to be quite high and up to more than a half of the emitted UV photons (\mbox{$f_{\mathrm{d}} > $ 2.0}) and even up to 75\% (\mbox{$f_{\mathrm{d}} \sim $ 4.0}). 

An individual model fit for each of the \hii region would be desirable. However, a few hundred iterations were needed in order to model only one class. This makes it a very long time consuming task. Nevertheless, we can complement the results obtained here with a spectrophotometric study by using a multiwavelength broad-band set of images and the spectral information provided by the Balmer emission lines. We develop this analysis in the following section. 

\subsubsection{Spectrophotometric analysis}
\label{sec:spectrophot_analysis}

\textit{XMM}-OM and SDSS imaging can provide useful information on young stellar populations. In particular, having the information of ultraviolet and \textit{u} broad-band emission makes a difference when trying to characterize ionizing stellar populations. The UV continuum is dominated by massive short-lived OB stars (\mbox{$\tau \leq$ 100 Myr}) and hence can be sensitive to star formation on time-scales of only 10 Myr or so. Likewise, as shown by~\cite{Anders04}, \textit{u}-band observations are essential for photometric investigations of young star clusters. Archive broad-band ultraviolet $UVW2$-$UVM2$ (OM) and \textit{u, i, z} (SDSS) images were retrieved.

The main goal of our spectrophotometric study is to fit the SED of the young ionizing stellar population with models using the retrieved broad-band images and the spectral information provided by the IFU data. In particular, we gathered the data set as follows.

\begin{figure}
\centering
\includegraphics[trim = 0cm -1.0cm 0cm 1cm,angle=90,clip=true,width=0.45\textwidth]{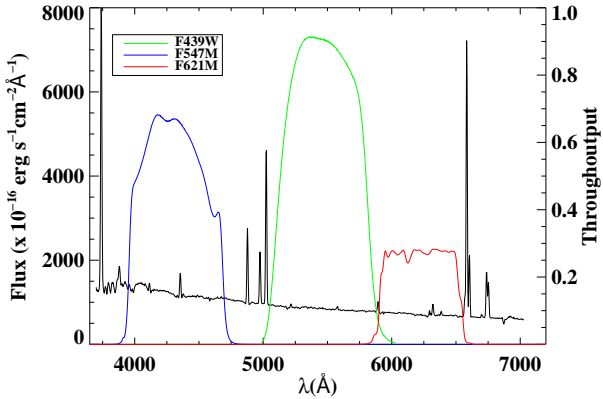}

 \caption{Convolved $HST$ filters with the IFS cubes.}
  \label{fig:hst_filters}
\end{figure}

\begin{enumerate}

 \item The five broad-band filter images just mentioned plus a set of three `simulated' broad-band images, produced with the convolution of the IFS cubes with three known $HST$ filter throughput curves: \mbox{$F439W \sim$ \textit{B} Johnson}, filter from the WFPC2 centred at \mbox{4300 \AA{}}; $F547M$, filter from the WFPC2 centred at \mbox{5479 \AA{}}; and $F621M$, filter from the WFC3 centred at \mbox{6219 \AA{}}. Fig.~\ref{fig:hst_filters} shows the integrated spectrum of the galaxy with the throughput filter curves overplotted. Note that, thanks to the star--gas decoupling analysis we avoid contamination by the nebular emission lines on the continuum broad-band images (i.e. H$\gamma$ Balmer line just in the middle of the $F439W$ filter). Altogether, with our compiled multiwavelength data, we cover homogeneously a rather large spectral range (from the near-UV to the near-IR \textit{z} band).

 \item  The \ha and \hb line intensities and the respective EW, which set important constraints on the age of the ionizing stellar populations.  The \ha\onespace/\hb ratio is also directly related to the LOS extinction. 
\end{enumerate}

With the aid of single stellar population models we can try to explain the observed photometric and spectroscopic properties of each \hii region in NGC 3310. However, we first have to take into account the non-ionizing stellar population. The results obtained with the \starlight fitting are the key to decouple the light produced by ionizing and non-ionizing stellar populations. To that end we estimated, for each region and with the knowledge of the fitted stellar populations (i.e. \starlight result), which fraction of the light at the effective wavelength of each photometric filter comes from the former or the latter. We used a conservative cut of 15 Myr to separate ionizing from non-ionizing populations. We could directly estimate this ratio with the stellar libraries used in \starlight for the $HST$ filters, since the MILES library covers a wavelength range from 3525 to 7500 \AA{}. To derive the other ratios the synthetic \popstar templates were employed, since their spectral range covers all the filter 
data set used in this study and they include the nebular continuum contribution.

We remind the reader that we performed 100 \starlight fits for each \hii region, which allowed us to estimate the uncertainties of these light ratios. In general, the light ratio for the bluest filters (i.e. \textit{UVW2,UVM2,u}) is quite low, \mbox{L$\mathrm{_{old}}$/L$\mathrm{_{young}} \leq$ 2-10\%} (i.e. practically all the light is related to ionizing population). Even if the relative uncertainties can be high (up to 60\%), they have little effect in the accuracy of the estimate of light related to young population. By contrast, the light ratio for red filters (i.e. \textit{i, z}) can be easily larger than \mbox{L$\mathrm{_{old}}$/L$\mathrm{_{young}} =~$50\%} and up to 70\%, with typical relative uncertainties of 20-30\%.

In addition, by computing these ratios we are basically subtracting the contribution of the underlying old (in the sense of non-ionizing) stellar populations to the total emitted light. Therefore, we could correct the EWs of the \hii regions by subtracting this old population from the continuum. This corrected EW was used as an input parameter for the photoionization models in the previous section.

\begin{figure*}
\includegraphics[trim = -2cm 0cm -1cm -1cm,clip=true,width=0.9\textwidth]{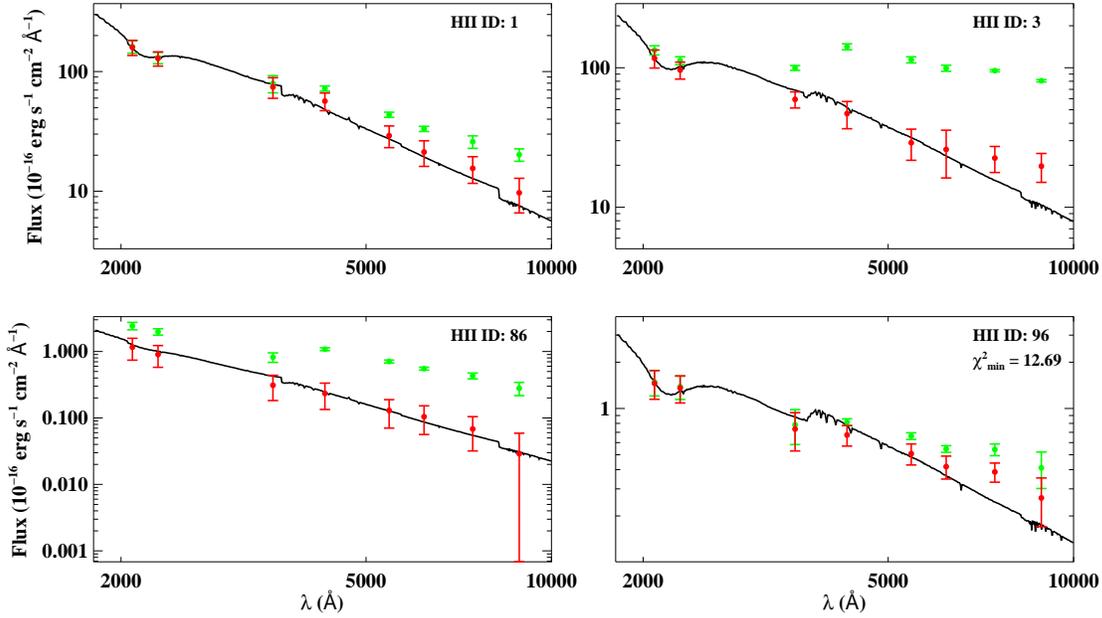}
 \caption{Synthetic best-fitting SED for four \hii regions. Photometric measurements corrected by non-ionizing population emission (not corrected) are overplotted in red (green). The correction is generally small (with some exception) for the ultraviolet data, while it is much more evident in redder filter photometric measurements. For the case where no solution is found, the value of $\chi^2_{\mathrm{min}}$ is labelled at the top right corner.}
  \label{fig:sed_fit}
\end{figure*}

The following step was to perform the photometry of the \hii regions in NGC 3310. With the `simulated' broad-band images with the $HST$ filters we directly added up all the flux contained in the spaxels that define the region. We want to use the same irregular aperture for all the images; hence, we preferred not to degrade all images to that with the worst one. Instead an IDL script was produced to add the flux of the same irregular aperture. Basically, all the images are re-sampled (preserving the flux) to have spaxels a factor of 10 smaller, so as to obtain a sub-spaxel photometry. The aperture is also re-sampled in such a way that the new aperture corresponds to the border of the previous one. Aperture coordinates are transformed to spaxel coordinates for each image and the flux within a given aperture is obtained adding up the flux of all the spaxels inside the aperture. Poisson noise is assumed, particularly important for the ultraviolet measurements, where only a few \mbox{counts s$^{-1}$} are detected 
in many regions. To account for differential spatial resolution the aperture is also shifted by 1 arcsec in four directions (north, south, east and west), and median values and dispersions are taken to compute the final flux and error for each \hii region. 

The response of the OM monitor at high count rates is not linear, an effect known as coincidence-loss ~\citep{Fordham00}. However, we have to corrected for this effect on our UV measurements because the count rates of the brightest \hii region on the individual exposures is of the order of \mbox{5 counts s$^{-1}$}, below the critical rate of \mbox{10 counts s$^{-1}$} where this effect becomes significant (10\%)  according to the ``XMM-Newton Users Handbook'', Issue 2.11, 2013 (ESA: XMM-Newton SOC). The count rate for the rest of the \hii regions is well below \mbox{5 counts s$^{-1}$}, meaning that any systematic due to not correcting for this effect is well bellow 4\%, much smaller than the typical uncertainties of the photometric measurements (see Table~\ref{table:result_chi_square}). 

Once the photometry was performed, the correction due to the underlying old population was applied. Altogether, at this stage we have the flux of the light related to the ionizing population through seven broad-band filters, the corrected EWs and the observed \ha and \hb line fluxes. We have then compared  observational and theoretical SEDs through a \mbox{$\chi^2$-fitting} procedure~\citep{Bik03} as

\begin{equation}
\chi^2(Z,\tau,A_V,m_\star) = \sum_{N} \frac{(f_{\mathrm{obs}} - f_{\mathrm{model}})^2}{\sigma^2_{\mathrm{obs}}}
\end{equation}
where $N$ denotes the number of filters and observables (photometric flux densities, fluxes and EWs of the emission lines; see Table~\ref{table:result_chi_square}) available for each \hii region; $f_{\mathrm{obs}}$ and $f_{\mathrm{model}}$ are the observed and model observables, respectively; and $\sigma_{\mathrm{obs}}$  is the weight for the fit (i.e. photometric, line flux measurements and underlying population correction uncertainties). The $\chi^2$ minimization procedure for fitting SEDs produces more satisfactory results for determining the ages, extinctions, and masses in galaxies (e.g.,~\citealt{Maoz01,Bastian05b,Diaz-Santos07}) than other widely used methods such as colour--colour diagrams.

We first assumed that the stellar population responsible for the bulk of the ionization is a single stellar population for which four parameters were to be derived: metallicity ($Z$), age ($\tau$), extinction ($A_V$) and stellar mass ($m^{\mathrm{ion}}$). We used models with three different metallicities (\mbox{0.2,0.4 and 1.0 Z$_\odot$}). The expected (minimum) $\chi^2$ value of the best fit, $\chi^2_{\mathrm{min}}$, should be equal to the number of degrees of freedom ($\nu = N - 4$). Those solutions within  $\chi^2_{\mathrm{min}} \pm \Delta\chi^2_{\mathrm{min}}$, with $\Delta\chi^2_{\mathrm{min}} = (2\nu)^{\frac{1}{2}}$, were taken to determine the range of acceptable solutions. This would be equivalent to taking the $\pm 1 \sigma$ solutions. Fig.~\ref{fig:sed_fit} illustrates the results of the SED minimization fitting technique for different \hii regions in our sample: with high (ID 1, 3) and low (ID 86, 96) S/N spectra. The change of the shape of the spectra is quite evident when the correction due to 
underlying non-ionizing population 
is applied.

In Sect.~\ref{sec:photo_models} we explored the ionization structure of the seven classes, representative of the \hii regions. As a result we could estimate characteristic age and dust absorption factor ($f_\mathrm{d}$) intervals for each class. We have complemented those results with the SED fits presented here. In particular, we computed a grid of solutions of the $\chi^2$ minimization procedure by varying $f_\mathrm{d}$ from 1.0 to 4.5, that is, changing the modelled \ha and \hb flux and the EWs, which are divided by $f_\mathrm{d}$ (when trying to recover the observed values some flux, scaled to $f_\mathrm{d}$, is lost). For a given \hii region, we then took those sets of solutions of the $\chi^2$ minimization procedure that were at the same time compatible with the age and $f_\mathrm{d}$ intervals in the class to which the region belongs. Some disagreement was encountered between the age interval derived from the $\chi^2$ fit and from the \cloudy fit for a few \hii regions. However, the maximum 
difference is less than 1 Myr. For the vast majority of the regions the model that better reproduced the observables was that with \mbox{$Z$ = 0.4Z$_\odot$}, which is in agreement with the gaseous abundances derived in Sect.~\ref{sec:abundance_properties}. Given the young nature of the ionizing population, this is expected.

The derived ages, stellar masses, internal extinctions and absorption factors are listed in Table~\ref{table:result_chi_square}. The fit was not successful for a few cases, demanding the presence of two ionizing stellar populations. A minimization analysis was performed in those cases, though large degeneracy is found on the properties of the derived populations. One of them usually ranges between 1 and 6 Myr, the other being 6--15 Myr old. Within these age ranges the mass is not generally well defined within factors of less than 5--10.  

We would like to mention that in a few cases no valid solutions were found for $f_\mathrm{d} = 1$, but for $f_\mathrm{d} > 1.0$. Thus, it may happen that the measured flux densities are related to a single stellar population but, since no solution is found, minimization techniques are immediately applied to a composite two-stellar-population model. As we have seen here, sometimes invoking a composite model is not necessary in order to reproduce the observed fluxes. 

As can be observed in the table, \hii regions are typically 2.5--5 Myr old. This is consistent with the presence of WR stars (see the spectral features in Fig.~\ref{fig:SL_examples}), given that this phase normally starts \mbox{2--3 Myr} after their birth~\citep{Meynet05}. The mass of the ionizing stars span, on the other hand, a large range, from about $10^{4}$ up to \mbox{6$\times 10^{6}$ \msun\onespace}. Note the asymmetry on the error estimates. It is also worth mentioning the typical significant absorption factors derived ($f_{\mathrm{d}}$ = 1.3--3.0). According to our combined photoionization and spectrophotometric modelling, we have seen that, in general, at least 25\% of the emitted UV photons from the OB stellar populations are absorbed by dust grains in the nebulae.

\section[]{Discussion}
\label{sec:discussion}
\subsection{Reliability on the stellar subtraction}
\label{sec:tests_SL}

\begin{figure}
\centering
\includegraphics[trim = -0.5cm 2cm 0cm -2cm,angle=90,width=0.95\columnwidth]{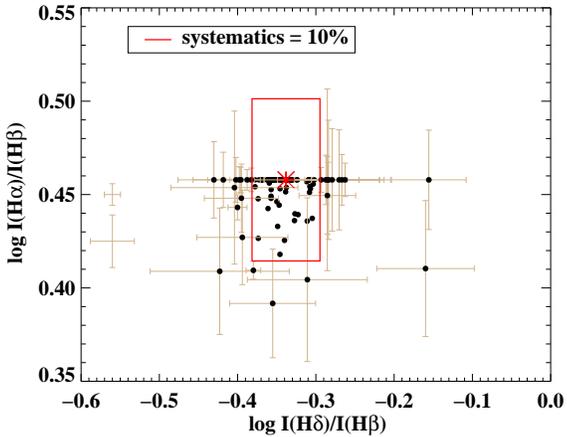}
  \caption{Balmer ratios with respect to \hb. The red asterisk marks the location on the plot of the theoretical ratio while the red box demarcates the ratio space where the values are within the expected systematics. To avoid confusion, typical error bars (median value) for \hii regions with ratios within the box are plotted on the left-hand side. The error is plotted for \hii regions with regions outside the plots.}
  \label{fig:sanity1}
 \end{figure}

In our analysis, we have been able to subtract the underlying continuum of the spectra and hence to decouple the gaseous and stellar contributions to the measured emission. For some parts of our analysis, this subtraction is critical. This is the case for the measurement of the \oiii $\lambda$4363 \AA{} line, since it is very close to \hgamma\twospace, thus likely to be affected by a significant amount of underlying absorption. Here, we pay attention to the validity of such subtraction. We have hence checked the Balmer decrement ratios obtained once the extinction is corrected for each spectrum. Had we over-subtracted continuum absorption, negative extinction values and/or inconsistent Balmer ratios (i.e. \hb to \hgamma\onespace) would have been derived. 

We present in Fig.~\ref{fig:sanity1} the derived Balmer ratios (or differences in logarithm units) and compare them with the theoretical values according to \cite{Osterbrock89} and assuming a case B recombination with \mbox{$t_\mathrm{e}$ = 10,000 K} and \mbox{$n_\mathrm{e}$ = 100 cm$^{-3}$}. Under these conditions, \mbox{\ha\onespace/\hb = 2.86} and \mbox{\hgamma\onespace/\hb = 0.459}. Only four of them lie, within uncertainties, outside the systematic box. Although there are other 17 regions with ratios that lie outside the box, within uncertainties their ratios are consistent with the theoretical values. For those few cases where the ratio was not recovered because too much subtraction was performed, an extinction of $A_V = 0$ mag was assigned.

\subsection{Biases on temperature and abundance derivations}
\label{sec:t_obs_res}

\subsubsection{Methodologies used}
\label{sec:t_methods}

\begin{figure*}
\centering
\includegraphics[angle=90,trim = -0.7cm 0cm 0cm 2cm,clip=true,width=0.95\columnwidth]{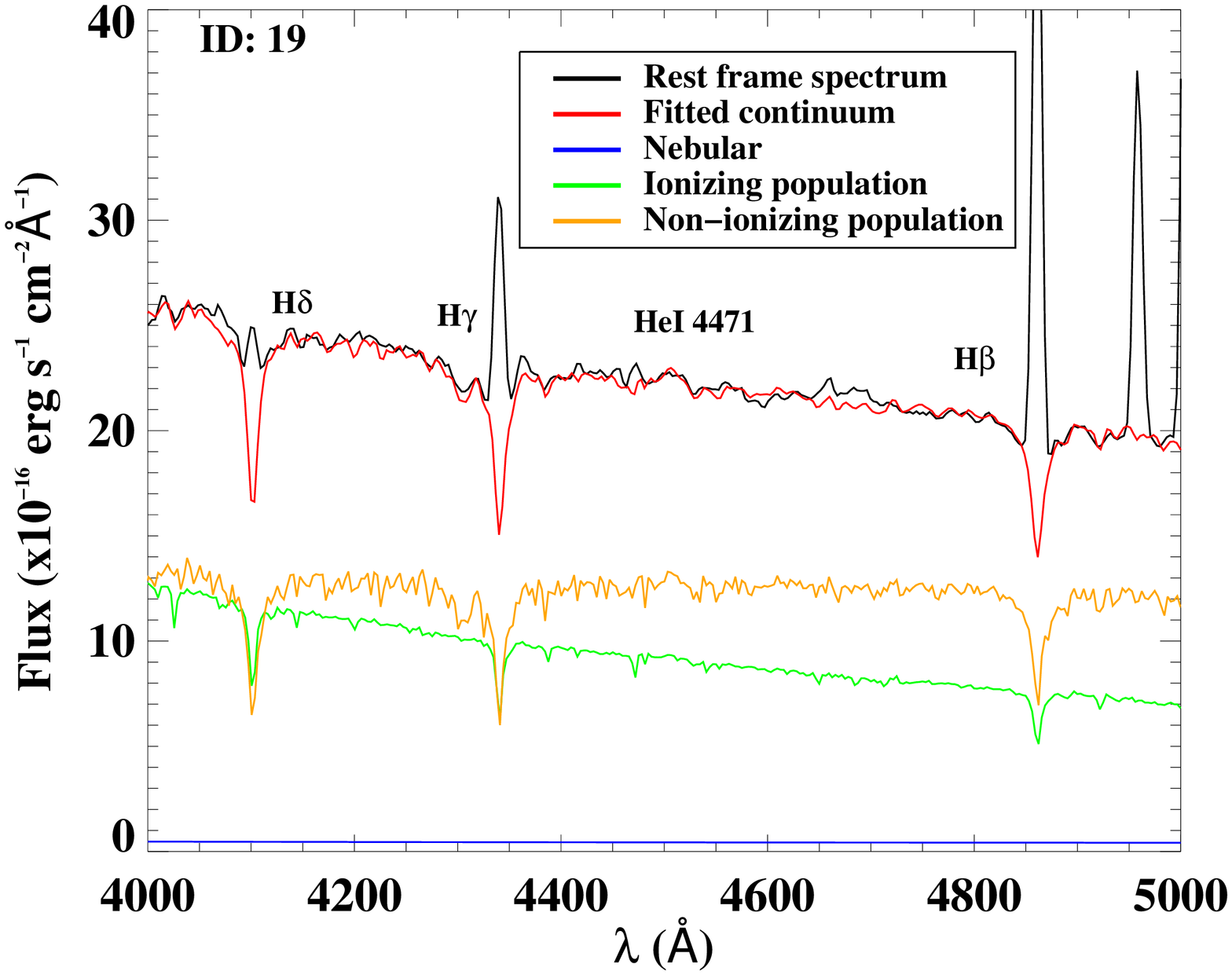}
 \includegraphics[angle=90,trim = 12cm 16.5cm 0.5cm -1cm,clip=true,width=1.0\columnwidth]{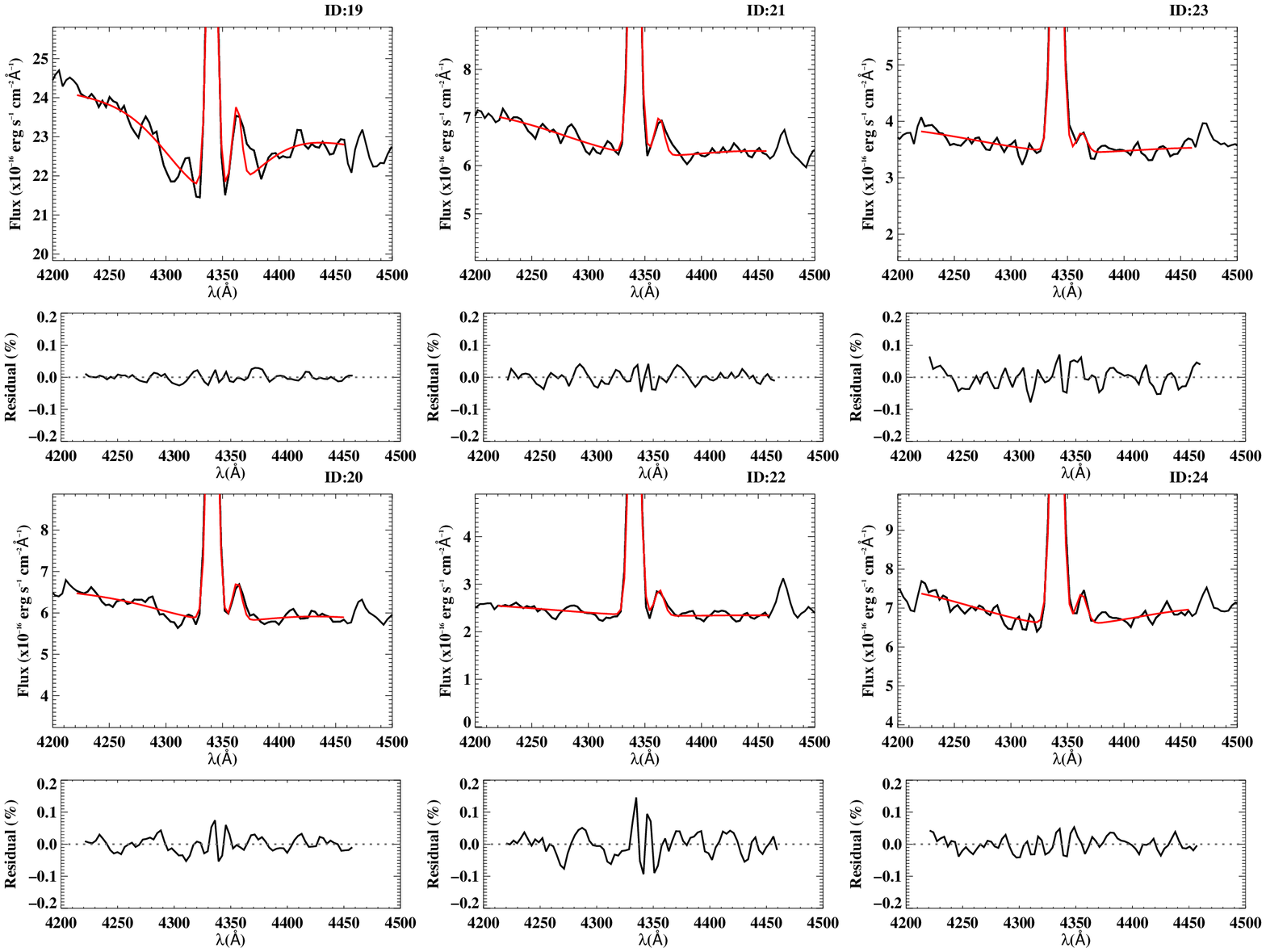}
  \caption{\textbf{Left:} rest-frame spectrum for \hii region ID 19 and the fitted components. See label for the identification of colours. The derived continuum spectrum is also plotted. Several Balmer lines (\hb,\hgamma,\hdelta) and the \hei $\lambda$4771 \AA{} line are also identified and labelled. \textbf{Right:} zoomed view centred on the \hgamma spectral range. The red line shows the resulting fit of the two Gaussians (one for \hgamma and the other for \mbox{\oiii $\lambda$4363 \AA{}}) and the negative Gaussian (for the absorption).}
   \label{fig:fit_spectrum_id19}
\end{figure*}

In Sect.~\ref{sec:radial_gradients} (Fig.~\ref{fig:metal_radial_gradients_a}) we show the abundance gradient in the disc of NGC 3310. An intrinsic dispersion of about 0.15 dex is observed along the disc for the $t_\mathrm{e}$-based derived oxygen abundances. They also show an offset of about 0.3 dex with respect to the abundances derived using strong-line calibrations. 

In principle, it could be argued that abundances obtained through strong-line calibrations do suffer from typical systematics of \mbox{0.1--0.2 dex}. However, we also have to keep in mind that accurate electron temperature determinations depend on reliable auroral line measurements, such as \mbox{\sii $\lambda$6312 \AA{}}, \mbox{\nii $\lambda$5755 \AA{}} or \mbox{\oiii $\lambda$4363 \AA{}}. The latter is very close to the \hgamma line, which in young stellar populations can show a very wide profile (see~\citealt{Diaz88} for a detailed description of this complex spectral region). Given the spectral resolution of the PINGS data, the wings of both the oxygen auroral line and \hgamma blend. In fact, as shown in Fig.~\ref{fig:fit_spectrum_id19} (left), the auroral line sets over the absorption wing of \hgamma. A prominent stellar absorption can thus critically affect the measurement of the auroral line on the observed spectrum. Furthermore, the emission of this line is rather weak in regions with 
moderate abundance \mbox{(12 + log(O/H) $\sim$ 8.0)}, and undetectable in metal-rich environments. Altogether, a reliable detection and measurement on this line can be rather awkward.

\begin{figure*}
\centering
\includegraphics[trim = 5cm 0cm 5cm 0cm,angle=90,clip=true,width=0.92\textwidth]{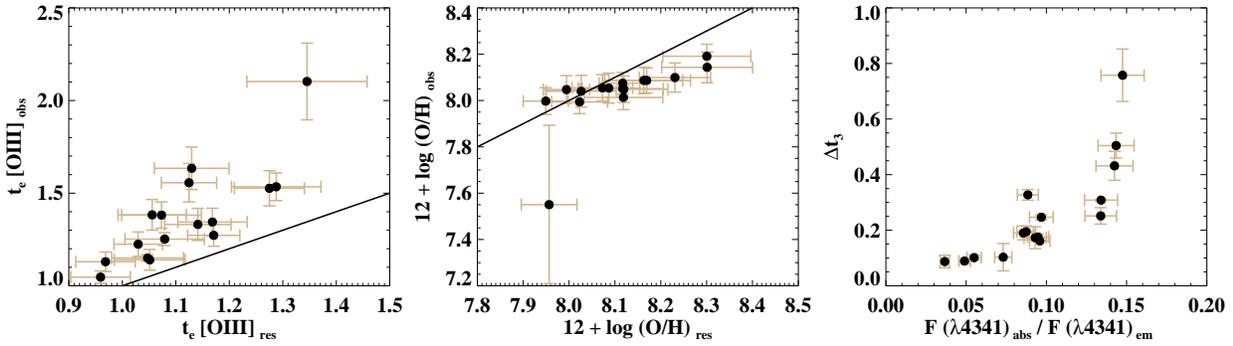}
 \caption{\textbf{Left:} comparison between the derived temperature $t_3$ using the observed and the residual spectra. \textbf{Middle}: comparison between the derived oxygen abundance using the observed and the residual spectra. In both cases the line represents the 1:1 relation. \textbf{Right}: comparison between the difference in electron temperatures ($\Delta t_3 = t_\mathrm{e} \textrm{\oiii}_\textrm{obs} - t_\mathrm{e} \textrm{\oiii}_\textrm{res}$) and the flux ratio of \hgamma in absorption to corrected \hgamma in emission.}
  \label{fig:metal_temp}
\end{figure*}

Very often, when fitting an emission line in a spectrum, a Gaussian function is assumed for the line and a straight line is used to fit the underlying continuum. For the specific case of \oiii $\lambda$4363 \AA{}, \hgamma is also included as another Gaussian function in the fit due to its proximity. We have compared our $t_\mathrm{e}$ (\oiii) ($\equiv t_3$) derivations with those that we would have obtained by fitting the auroral line directly on the `observed spectrum', without the subtraction of the continuum obtained with \starlight\onespace. Given that in most cases the  stellar absorption is noticeable in the observed spectrum, instead of a straight line, the absorption was fitted with a broad Gaussian component with negative flux. An example of this is shown in Fig.~\ref{fig:fit_spectrum_id19} (right). Note that, as illustrated in the left-hand panel, part of the flux measured for the auroral line due to the fit actually corresponds to stellar continuum.  The comparison of the electron temperature and 
the oxygen 
abundance using both methods is shown in Fig.~\ref{fig:metal_temp} (left and middle). The electron temperature is usually overestimated, which causes a general underestimation of the metallicity if the measurements are done on the observed (unsubtracted) spectra. The difference between the derived temperatures (i.e. metallicities) is correlated with the ratio between the predicted absorption of \hgamma with \starlight and the corrected \hgamma emission. It is hence directly related to the degree of absorption of the \hgamma emission-line (see Fig.~\ref{fig:metal_temp}, right). By subtracting such absorption and fitting the residual spectra we obtain higher abundance values, in better agreement with the literature. Therefore, we do not think that the methodology applied to the measurement of the \mbox{\oiii $\lambda$4363 \AA{}} line flux is behind the significant difference found between the direct and strong-line abundance estimates.

\begin{figure*}
\centering
\includegraphics[trim = 4cm 0cm 4cm 0cm,angle=90,clip=true,width=0.92\textwidth]{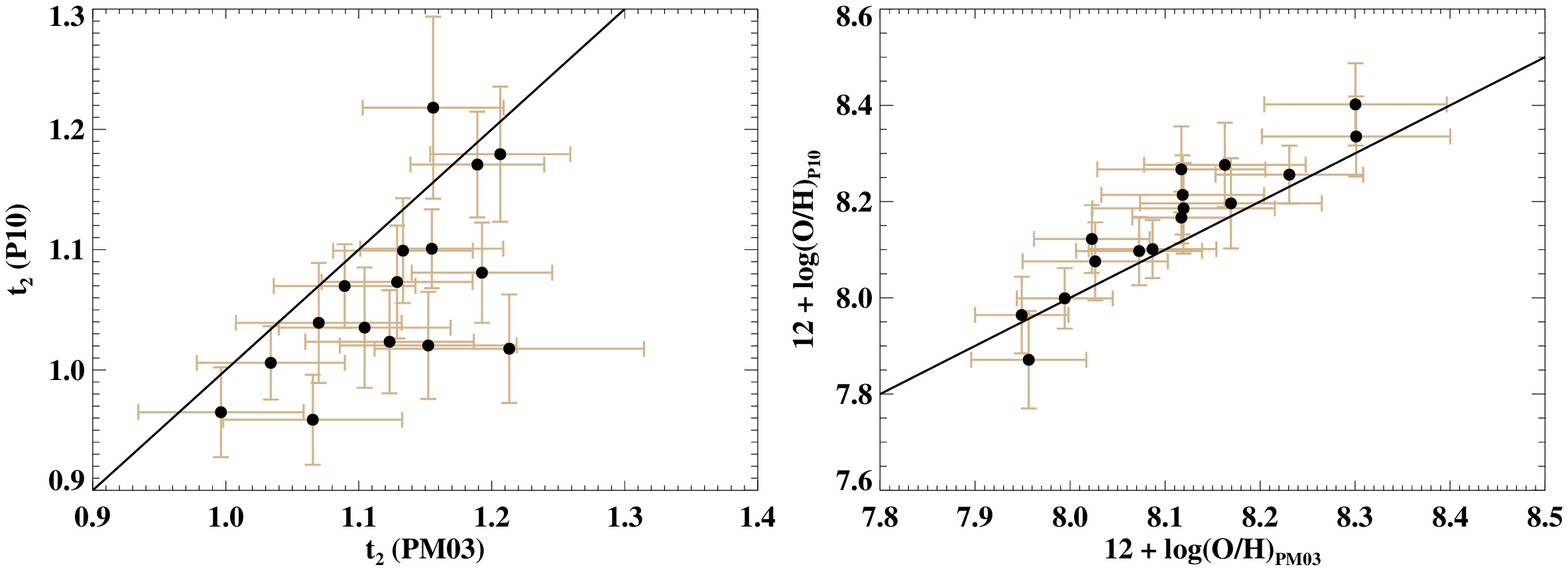}
 \caption{\textbf{Left:} comparison between the derived temperature $t_2$ using the $t_2$-$t_3$ relations provided by P\'erez-Montero \& D\'iaz (\citeyear{Perez-Montero03}; PM03) and Pilyugin et al. (\citeyear{Pilyugin10}; P10).~\textbf{Right}: relation between the derived oxygen abundance in each case. The line represents the 1:1 relation.}
  \label{fig:metal_temp2}
\end{figure*}

The use of different prescriptions used for the estimation of the $t_\mathrm{e} (\textrm{\oii})$ ($\equiv t_2$) may also contribute to the disagreement. As can be seen in Table~\ref{table:abundances}, the ion O$^+$ is generally more abundant than O$^{2+}$. That is, the abundance determination is dominated in most cases by the knowledge of $t_2$. In most cases, our direct abundance determinations rely on our derived $t_3$. Therefore, using one or other parametrizations between both electron temperatures may change our abundance estimate to a significant extent. 

There is not quite a consensus of how these temperatures are related. Several versions of the $t_2$-$t_3$ relation have been proposed, the most widely used by~\cite{Campbell86}, based on the \hii region models of~\cite{Stasinska82} is

\begin{equation}
t_2 = 0.7t_3 + 0.3
\end{equation}

Several relations have been proposed during the last decades (e.g.,~\citealt{Pagel92,Izotov97,Oey00,Pilyugin06,Pilyugin07}). Here we focus on the relations proposed by~P\'erez-Montero \& D\'iaz (\citeyear{Perez-Montero03}; hereafter PM03) and Pilyugin et al. (\citeyear{Pilyugin10}; hereafter P10). In both cases, the physical conditions were derived using the five-level atom model for O$^+$, O$^{++}$ and N$^+$ ions, using the atomic data available. We compare these two works because the $ONS$ and $O3N2$ (O/N-corrected) calibrations were anchored to direct abundance determinations using a different $t_2$-$t_3$ relation. In particular,~PM03 claim that this relation is density dependent (Eq.~\ref{eq:t2_t3}). On the other hand,~P10 propose a relation quite similar to the commonly used ones (e.g.,~\citealt{Campbell86,Pagel92}):

\begin{equation}
t_2 = 0.672t_3 + 0.314
\end{equation}

In Fig.~\ref{fig:metal_temp2}, we compare the derived electron temperature $t_2$ (left) and oxygen abundance (right) using both sets of equations. In general, if we make use of the relations proposed in P10, the derived temperatures and abundances are systematically somewhat lower and higher, respectively. Given the uncertainties, the systematics on the abundance estimate are not important or relevant for at least half of the cases, especially for abundances $\sim 8.0$ or lower. However, in a few cases, abundances can be up to 0.1--0.2 dex higher, closer to the expected abundance values obtained with the strong-line calibrations. Therefore, the use of one set of equations or another can introduce systematics in abundance determinations of typically 0.1 dex. 

\begin{figure*}
\centering
\includegraphics[trim = 4cm 0cm 4cm 0cm,angle=90,clip=true,width=0.92\textwidth]{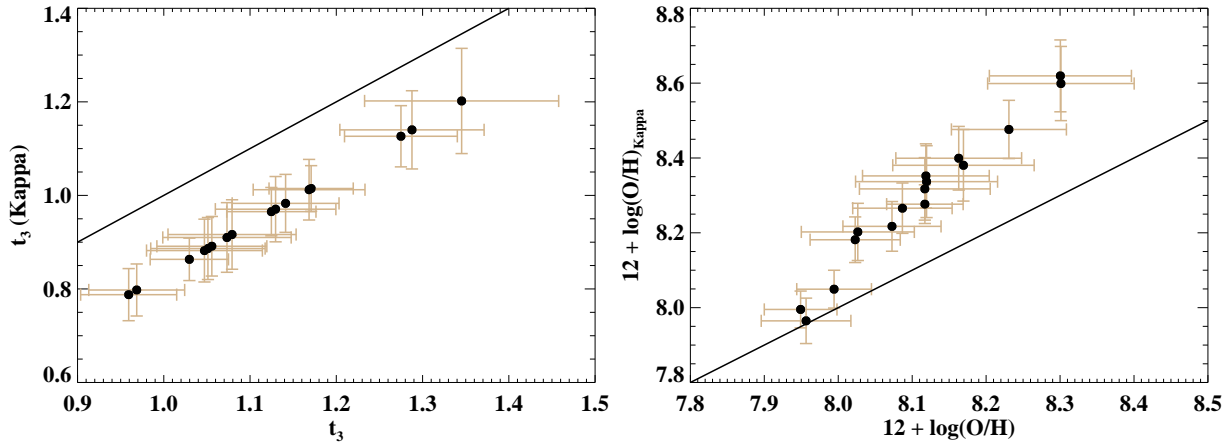}
 \caption{\textbf{Left:} comparison between the derived temperature $t_3$ using the conventional equilibrium methods (this study) and non-equilibrium methods (i.e. Kappa distribution).~\textbf{Right}: relation between the derived oxygen abundance in each case. The line represents the 1:1 relation.}
  \label{fig:metal_temp3}
\end{figure*}

Even taking into account this offset, the discrepancy between direct and strong-line abundance estimates are still significant. The compilations used in order to obtain the strong-line calibrations consist of several hundreds of \hii regions. With such a high parameter space of ionization conditions to explore (age and mass of the ionizing population, ionization parameter, metallicity, N/O, density, etc.), it would not be surprising that the direct and strong-line abundance estimates of \hii regions with different ionizing conditions as those in the compilations differed. Larger compilations or models covering as much as possible the parameter space of ionization conditions would be needed. 

\subsubsection{The abundance discrepancy problem}

Discrepancies on abundance determinations have been observed for decades. Chemical abundances determined from the optical recombination lines are systematically higher than those determined from CEL. This problem, dating back to 70 yr (\citealt{Wyse42}), was first discussed in detail by \cite{Torres-Peimbert77}, and then regularly discussed in the literature (\citealt{Liu00,Stasinska04,Garcia-Rojas06,Garcia-Rojas07,Stasinska07,Mesa-Delgado08}). This is known as the `abundance discrepancy problem'. In addition, systematic differences between abundances determined using either direct measurements of ionic temperatures, or using SEL methods are reported in the literature (e.g.,~\citealt{Kennicutt03,Bresolin07,Bresolin09,Ercolano10,Pilyugin12a,Lopez-Sanchez12}). 

Unlike the systematics discussed in Sect.~\ref{sec:t_methods}, most of the works just mentioned focus on the nature of the calibration used to define the strong-line method applied. In fact, there remains a significant offset (see~\citealt{Perez-Montero10,Dors11}) between those SEL techniques based purely upon photoionization models~\citep{McGaugh91,Kewley02,Kobulnicky04} and those based upon an empirical alignment of the strong line intensities to abundances derived in objects for which the electron temperature has been directly estimated~\citep{Bresolin04,Pilyugin05,Pilyugin12b}. In some cases the abundance determinations via direct-temperature methods are favoured, (e.g.~\citealt{Bresolin09,Ercolano10}) and the discrepancies are explained by the presence of multiple ionization sources (not taken into account in one-dimensional ionization models). In other cases, techniques based on photoionization models are preferred (e.g.,~\citealt{Stasinska05,Lopez-Sanchez12}), since they produce electron 
temperature gradients inside \hii regions, and temperature fluctuations are assumed to be the main reason for the abundance discrepancy problem.

Other physical scenarios have been proposed in order to explain this discrepancy, the most recent one exploring the fact that the electrons involved in collisional excitation and recombination processes may not be in thermal equilibrium (\citealt{Binette12,Nicholls12,Nicholls13}). Here, we can estimate how would our direct abundance estimates be affected under this assumption. However, we cannot know how our estimates using strong-line methods are affected. 

If we assume that the electrons involved in collisional excitation and recombination processes follow a non-equilibrium Kappa ($\kappa$) electron energy distribution rather than the widely assumed simple Maxwell--Boltzmann distribution, the `true' electron temperature in the ionization zone is lower than the derived value using standard methods, as shown in~\cite{Nicholls13}. They provide a means for estimating $\kappa$ with the knowledge of the ``apparent'' (i.e. derived using standard methods, like those used in this paper) \oiii~and \siii~electron temperatures (see their fig.~11). \cite{Hagele06} derived \oiii~and \siii~electron temperatures in a large compilation of \hii galaxies and giant extragalactic \hii regions. A relation between both temperatures was found as

\begin{equation}
 \mathrm{T([S{\sc III}])} = (1.19 \pm 0.08) \times \mathrm{T([O{\sc III}])}) - (0.32 \pm 0.10)
\end{equation}

Using this relation, we have made a rough estimation of \mbox{$\kappa \sim 20$} within the range of temperatures in our study, in perfect agreement with the value obtained in~\cite{Dopita13}.~\cite{Nicholls13} also provide a method to correct the electron temperature derived using standard methods, once $\kappa$ is known (see their Eq.~36). Although we are not able to compute the \siii~electron temperature, we can estimate by how much a Kappa distribution with $\kappa \sim 20$ is affecting our temperature estimates, so as to speculate by up to how much our abundance determinations can be underestimated. As Fig.~\ref{fig:metal_temp3} shows, the electron temperatures might be overestimated by typically 2000 K, which translates into an underestimation (i.e. offset) of the oxygen abundance of typically 0.2 dex and up to 0.3 dex. This range agrees well with observed offsets (e.g.~\citealt{Lopez-Sanchez12}). Note, however, that three regions still keep low-metallicity values (\mbox{12 + log(O/H) $\sim$ 8.0}) even 
if this correction is applied (Fig.~\ref{fig:metal_temp3}, right).

\subsection{Implications on the flat abundance gradient}

The subject of abundance gradients in galaxies is currently a burning issue on galaxy evolution. The steepest abundance gradients were initially seen in late-type spiral galaxies (types \mbox{Sb--Scd}). Barred galaxies were thought to present shallower gradients ~\citep{Martin94,Zaritsky94}, though more recent studies, with larger samples, have opened the debate that this may not be the case~\citep{Sanchez12b}. Metallicity gradients can also be flattened or erased in interacting galaxies and remnants from mergers.  

We have shown in Sect.~\ref{sec:radial_gradients} that the radial abundance gradient in NGC 3310 is rather flat. This is consistent with the flat gradients obtained in~\cite{Sanchez14} for merging galaxies at different levels of the interaction process. In a recent work,~\cite{Werk11} derived abundances for a handful of even more external \hii regions (up to a projected distance of almost 17 kpc or 6.7\reff). Although their abundance estimates are systematically higher than our estimates by about 0.2--0.3 dex (they use the model-based calibration of $R_{23}$), even at these distances they remain high and a flat abundance gradient is also observed. However, only a handful of \hii regions at a given azimuthal direction were observed in that study. We have observed over 100 regions and at all azimuthal directions. Therefore, combining both our and their study, we can be certain that the abundance gradient in NGC 3310 is rather flat from the very central regions to the outermost parts of the 
galaxy, well beyond 4\reff. 

This galaxy has been classified as a barred spiral. The bar and the size of the ringed structure suggest that this starburst was triggered by a bar instability~\citep{Piner95}. The likely past merger event this galaxy had with a dwarf plausibly produced the bar instability which led to the starburst. During an interaction, large amounts of gas can flow towards the central regions, carrying less enriched gas from the outskirts of the galaxy into the central regions, which can erase any metallicity gradient and dilute the central metallicity. Our results suggest that we are witnessing the consequences of such metal mixing processes.

\begin{figure}
\hspace{-0.3cm}
\includegraphics[angle=90,trim = 1cm 0cm 1cm 0cm,clip=true,width=1.03\columnwidth]{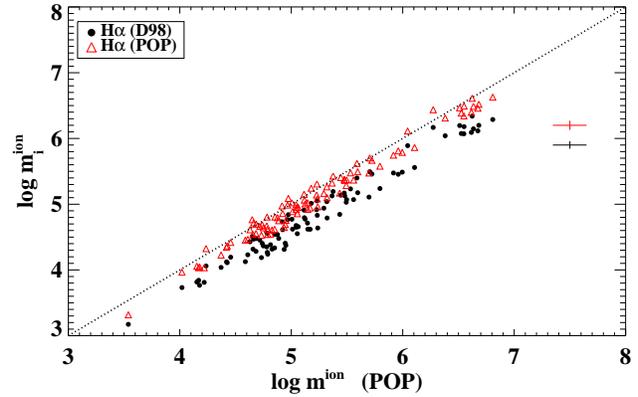}
 \caption{Comparison of the derived stellar ionizing mass using our spectrophotometric fitting (POP) with that derived using the extinction corrected \ha flux plus the corrected (from underlying non-ionizing population) \hb EW width using Diaz's~(\citeyear{Diaz98}) and our updated prescription (\ha D98 and \ha POP, respectively). Typical errors are plotted at the top right corner. Whenever the \ha flux and the EW(\hb) are used, they are corrected for the derived absorption by dust grains. The dotted line indicates a unity relation.}
  \label{fig:HII_pop2a}
\end{figure}

Werk and collaborators measured flat radial oxygen abundance gradients from the central optical bodies to the outermost regions of the galaxies in their sample. Given the different morphology of the galaxies, star-forming properties and level of disruption (13 systems, not all of them with signs of interaction) in their sample, they argued that metal transport processes in cold neutral gas rather than interactions may also play an important role in distributing the metals to the outermost parts of the galaxies (e.g., magnetorotational instabilities, thermal instability triggered self-gravitational angular momentum transport, etc.). However, given the large time-scales required (i.e. 1.5 Gyr) and that the metals generated by massive stars are generally returned to the ISM in just \mbox{$\sim$ 100 Myr}~\citep{Tenorio-Tagle96}, they concluded that the metal transport may be occurring predominantly in a hot gas component, a still unclear driver mechanism. In any case, these other mechanisms do not exclude that 
interactions (even weak interactions or minor mergers) effectively mix the chemical metal content in a galaxy.

\subsection{Star formation in NGC 3310}
\subsubsection{Different determinations of the mass of the ionizing population}

In Sect.~\ref{sec:spectrophot_analysis}, we derive ages and masses for the ionized stellar population (i.e. with ages \mbox{$\tau < 10$ Myr}). We can compare the stellar mass of the ionizing population as derived there and the values of the same masses derived from the extinction-corrected \ha fluxes, using the expression from~\cite{Diaz98}:

\begin{equation}
\label{eq:diaz98}
\mathrm{log}~m^{\mathrm{ion}} = \mathrm{log}~L(\mathrm{H}\alpha) - 0.86 \times \mathrm{log}~\mathrm{EW(H}\beta) - 32.61
\end{equation}

\begin{figure}
\centering
\includegraphics[trim = 0cm -2cm 0cm 0.5cm,clip=true,angle=90,width=1.05\columnwidth]{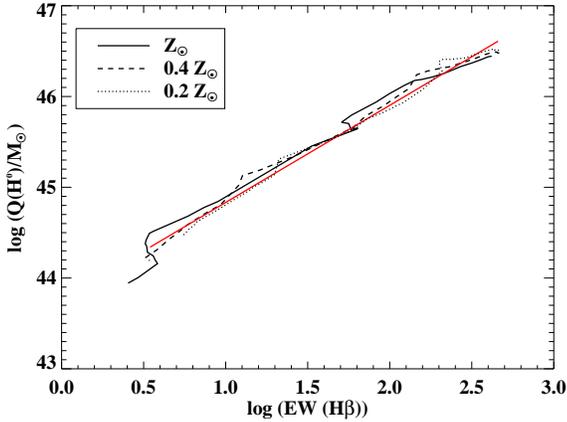}

 \caption{Relation between the number of ionizing photons per unit mass and the EW of \hb for different metallicities according to the \popstar Single Stellar Population models (Moll\'a et al.~\citeyear{Molla09},~Mart\'in-Manj\'on et al.~\citeyear{Martin-Manjon10}). Red solid line represents the best fit for the model with \mbox{$Z$ = 0.4Z$\odot$}.}
  \label{fig:models_ha}
\end{figure}

This equation takes into account the evolutionary state of the ionizing stellar cluster, and the EW used corresponds to that that would be observed in the absence of an underlying population. As can be seen in Fig.~\ref{fig:HII_pop2a} masses derived with the SED fitting are not consistent with those obtained from the \ha flux. Actually, Eq.~\ref{eq:diaz98} is based basically on models by~\cite{Garcia-Vargas95a,Garcia-Vargas95b} and \cite{Stasinska96}, all of them assuming a Salpeter IMF. We have updated this relation using \popstar, the last version of these models. We have then explored the expected correlation between the number of ionizing photons per unit mass ($Q$(H)/M$_\odot$) and the EW of \hb, since both quantities decrease with the age of the cluster (see Fig.~\ref{fig:models_ha}). A linear regression fit gives

\begin{equation}
 \mathrm{log}~Q(\mathrm{H})/\mathrm{M}_\odot = (1.07 \pm 0.02) \times \mathrm{log}~\mathrm{EW}(\mathrm{H}\beta) + (43.76 \pm 0.03)
\end{equation}
for \mbox{$Z$ = 0.4Z$_\odot$}, and

\begin{equation}
 \mathrm{log}~Q(\mathrm{H})/\mathrm{M}_\odot = (1.11 \pm 0.03) \times \mathrm{log}~\mathrm{EW}(\mathrm{H}\beta) + (43.70 \pm 0.05)
\end{equation}
for \mbox{$Z$ = Z$_\odot$}

And the mass of the ionizing cluster, corrected by the evolutionary state is (for \mbox{Z = 0.4Z$_\odot$}):

\begin{figure*}
\includegraphics[angle=90,trim = -0.5cm 0cm 0cm -1cm,clip=true,width=0.95\textwidth]{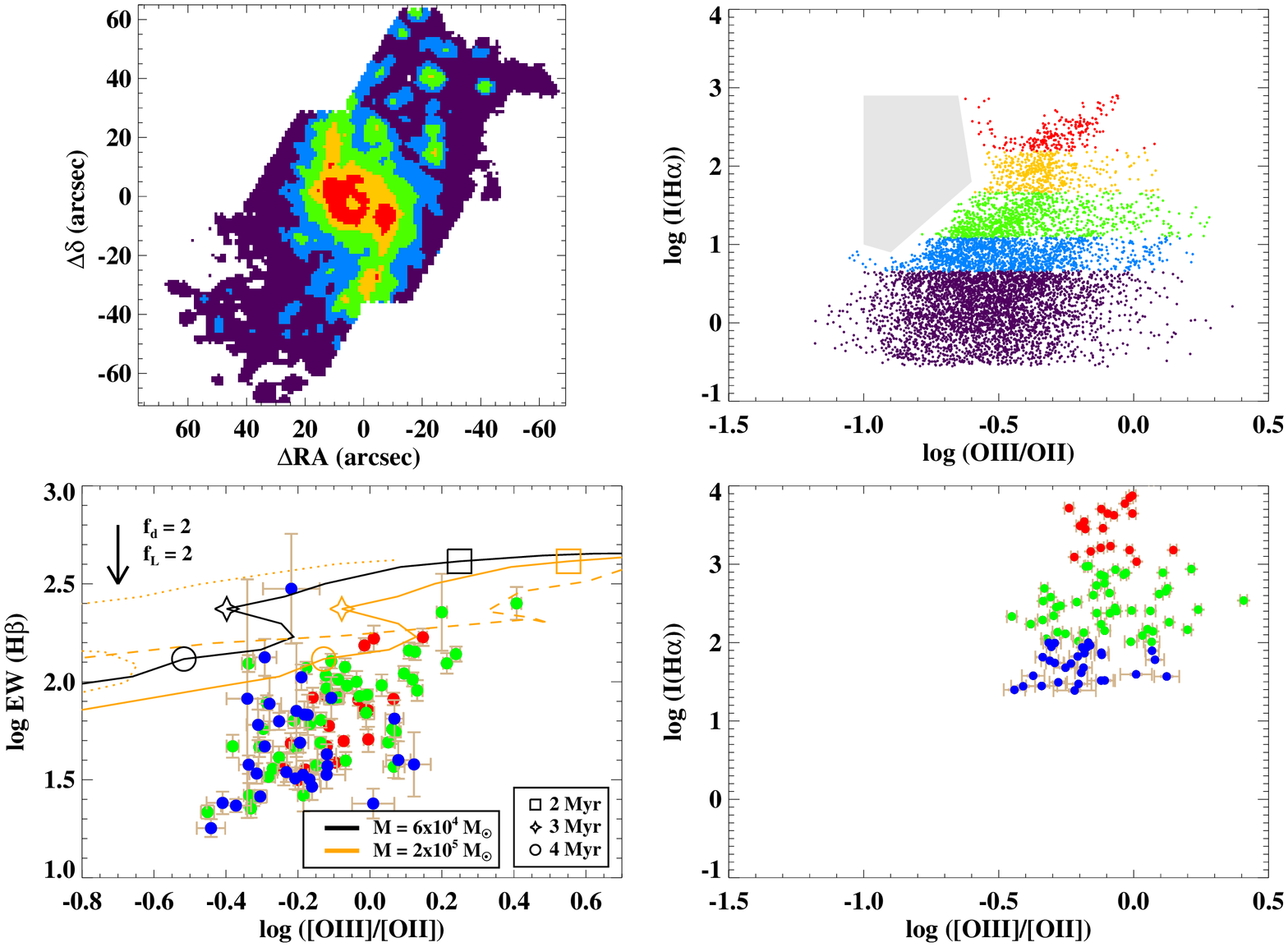}

 \caption{\textbf{Top-left:} \ha intensity map, colour coded according to the intensity intervals defined on the right-hand (top-right) panel. \textbf{Top-right:} \ha intensity (in log in arbitrary units) versus the excitation-sensitive ratio \oiii/\oii~for each spaxel. Different intensity intervals, log I(\ha), are colour coded as: $<$ 0.48 (violet), 0.48--0.90 (blue), 0.9--1.48 (green), 1.48--2.00 (orange), $>$ 2.00 (red). The grey area highlights the absence of low-excited and at the same time very luminous spaxels. \textbf{Bottom-right:} same relation as above, but for the \hii regions and different intensity intervals, log I(\ha), colour-coded as: $<$ 2.0 (blue), 2.0--3.0 (green), $>$ 4.0 (red). \textbf{Bottom-left:} EW(\hb\onespace), only corrected by the presence of underlying non-ionizing population, versus excitation relation for the \hii regions, colour-coded according to the intensity intervals defined on the right-hand (bottom-left) panel. Superimposed tracks show the evolution of 
clusters of different initial mass with time along this relation, according to \popstar models. Tracks in solid lines correspond to \mbox{Z = 0.4Z$_\odot$}, in dashed line to \mbox{Z = 0.4Z$_\odot$} and in dotted line to \mbox{Z = 0.2Z$_\odot$}. The arrow at the upper-left corner illustrates by how much the tracks should be corrected if absorption by dust grains within the nebulae ($f_\mathrm{d}$) or escape ($f_\mathrm{L}$) of half of the ionizing photons occured.}
  \label{fig:flux_cuts}
\end{figure*}

\begin{equation}
\label{eq:d98_updated}
\mathrm{log}~m^{\mathrm{ion}} = \mathrm{log}~L(\mathrm{H}\alpha) - 1.07 \times \mathrm{log}~\mathrm{EW(H}\beta) - 31.90
\end{equation}

If we use the new relation (for Z = 0.4Z$_\odot$, since the metallicity for the vast majority of \hii is around this value), then the derived mass is more similar to the adopted mass in this study (Fig.~\ref{fig:HII_pop2a}). Although within the uncertainties the mass are completely compatible, a small systematic offset of 0.1 dex is still present. Therefore, the mass obtained for the ionizing stellar population is little affected by the methodology used (SED $\chi^2$ minimization fitting or \ha and EW of \hb with the updated relation).

It is interesting to mention that, if a fraction of photons escape or dust absorption of UV photons occur, both the \ha flux and the EW measurements are affected in a similar way. Let us take $f_\mathrm{Ld}$ as the factor of photons that escape from the nebulae (leaking photons; `L') and those are absorbed by dust grains (`d') within the nebulae. Since both the \ha flux and the EW are shortened by the same factor, it is easy to work out the resulting relation for the mass of the ionizing stellar population (for $Z$ = 0.4Z$_\odot$):

\begin{flalign}
 & \mathrm{log}~m^{\mathrm{ion}} & \mspace{-20.0mu}=~& \mathrm{log}~L(\mathrm{H}\alpha_{\mathrm{obs}}) - 1.07  \times \mathrm{log}~\mathrm{EW(H}\beta_{\mathrm{obs}})~- \mspace{60.0mu}\\
 & & & - 31.90 - 0.07\times\mathrm{log}~f_{Ld} \notag 
\end{flalign}

If $f_\mathrm{Ld} = 2.0$, then the correction is just 0.02 dex. The typical range of $f_\mathrm{d}$ derived in this study (note that we have not estimated the fraction of leaking photons) is $f_\mathrm{d} = 1.3-3.5$. Actually, even if two thirds of the photons are missing (i.e. $f_\mathrm{d} = 3.0$), the correction would only amount 0.03 dex, which is lower than the typical uncertainty in the derivation of the mass.

\subsubsection{Local evolutionary tracks}

There is a clear correlation between the distance to the centre of the galaxy and the \ha luminosity measured in \mbox{NGC 3310}, the circumnuclear whereabouts being the most luminous (see emission-line maps in Fig.~\ref{fig:emission_maps}). In order to explore if there is an evolutionary effect behind this relation we have investigated the conditions of the ionization in several \ha luminosity intervals. As shown in Fig.~\ref{fig:flux_cuts} (top), the most luminous spaxels are generally also those with the hardest ionization. Even though there are a few spaxels with even higher degree of ionization, there is a clear envelope (lower part of grey area in the figure) at which no spaxel is very luminous and with low ionization. This cannot be a selection effect, because a spaxel with such conditions (being more luminous than average) should be observed. This effect, though less evident, is also observed when considering the \hii regions (Fig.~\ref{fig:flux_cuts}, bottom-right). 

The EW of \hb\onespace, once corrected by the continuum of non-ionizing populations, is a good indicator of the age of an ionizing cluster, i.e. of its evolutionary state~\citep{Dottori81}. A relation between the EW(\hb\onespace) and the degree of ionization has been observed in \hii galaxies~\citep{Hoyos06}. With our data we can reproduce a similar relation for the \hii regions of NGC 3310 (Fig.~\ref{fig:flux_cuts}, bottom-left). Thus, this effect seems to be a local one, at least at the scale of one or a few hundred pc. According to the evolutionary tracks, the higher the initial mass of the ionizing population the harder the degree of ionization in the cloud for a fixed age. Note that these tracks are also metallicity dependent. At high metallicities, for a given age and mass of the ionizing population, the degree of ionization of the gas (i.e. \oiii/\oii ratio) is lower and the EW just slightly smaller than at low metallicities. If, in general, the most luminous were more massive 
than the less 
luminous population, then the envelope we observe in the figure would be perfectly understandable. However, there is a general mismatch between observational data and models, quite evident in the figure. Given that the continuum from the underlying population has been subtracted, two reasons can be responsible for this disagreement, mainly: (i) absorption of UV photons by dust grains within the nebulae, which we have been able to roughly quantify in our analysis; (ii) escape of ionizing photons, which has been assumed to be negligible in this study. The arrow in Fig.~\ref{fig:flux_cuts} (bottom-left) exemplify by how much the tracks should be corrected by one or another. Additionally, the tracks shown in the figure only span a range up to \mbox{2$\times 10^5$ \msun}. Since the number of ionizing photons scales with the mass of the cluster, we would expect that by allowing a larger range in mass to the models the tracks would move towards harder ionizing regimes. 

With the aid of the model tracks, we can speculate on the evolutionary state of the observed \hii regions. For instance, if the mass of the ionizing population is of the order of \mbox{$10^6$ \msun} (the typical mass obtained for the most luminous ones, as reported in Table~\ref{table:result_chi_square}) and assuming that the tracks move towards harder ionizing regimes at higher masses, then two possibilities can explain the observations: (i) the most luminous population is typically older than the less luminous ionizing population; (ii) the effect of dust absorption of UV photons and/or photon leakage can be important, which would explain at the same time why no \hii regions with high luminosity and low excitation are observed (i.e. \mbox{log\oiii/\oii $>$ 0.4}).

\subsubsection{Star Formation Rates}

Although by definition, the star formation rate (SFR) of a starburst is zero when we see it several Myr later, with the knowledge of the mass and the age of the ionizing stellar population we can estimate an average SFR for each \hii region. The integrated value can give us a picture of the most recent star formation history of the galaxy (i.e. the rhythm at which the galaxy has been forming clusters during the last few Myr). We have made estimates of the SFR of the \hii regions just by computing the ratio of the mass and the age. Additionally, we have obtained another estimate by using the most widely used SFR rate calibrator in the optical, provided by Kennicutt (\citeyear{Kennicutt98}; K98). This calibrator relates the recent SFR with the \ha luminosity as

\begin{equation}
\label{eq:sfr}
 \mathrm{SFR} (\mathrm{M}_\odot \mathrm{yr}^{-1}) = 7.49 \times 10^{-42}~L\mathrm{(H}\alpha)~(\mathrm{erg s}^{-1}) 
\end{equation}

\begin{figure}
\hspace{-0.3cm}
\includegraphics[angle=90,trim = 1cm 0cm 1cm 0cm,clip=true,width=1.03\columnwidth]{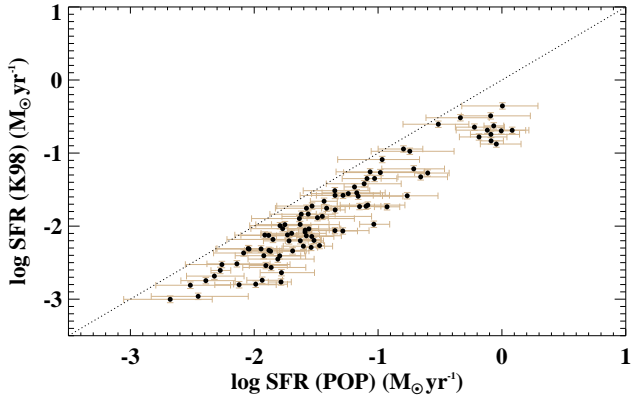}
 \caption{Comparison of the derived SFR using different methods our spectrophotometric fitting (POP) and the prescription by Kennicutt (\citeyear{Kennicutt98}; K98). Whenever the \ha flux and the EW(\hb) are used, they are corrected for the derived absorption by dust grains. The dotted line indicates a unity relation.}
  \label{fig:HII_pop2b}
\end{figure}

The estimated SFRs using Kennicutt's calibration are systematically lower than expected, even if we correct for absorption of dust grains (see Fig.~\ref{fig:HII_pop2b}). As cautioned in~\cite{Kennicutt07}, the SFR derived using Eq.~\ref{eq:sfr} for an individual \hii region, using a continuous star formation conversion relevant to entire galaxies, has limited physical meaning because the stars are younger and the region under examination is experiencing an instantaneous event when considered on any galactic evolutionary or dynamical time-scale.

The integrated SFR using each method is 4.4 (Eq.~\ref{eq:sfr}) and 13.8 (spectrophotometric SED fitting), in units of \mbox{\msun yr$^{-1}$}. These two estimates are in agreement with SFRs for starburst galaxies with moderate star formation~\citep{Kennicutt98,Leitherer00}. Given our estimated age range (\mbox{$\tau = $ 2.5 -- 5 Myr}) for the ionizing population, lower values using Kennicutt's calibration are expected.  Reported total SFR estimates are of the order of \mbox{3--8.6 \msun yr$^{-1}$}~\citep{Smith96,Werk11,Mineo12}. They are within the range of (or somewhat below than) our estimates.

\subsection{Mass growth in the disc of NGC 3310}

In Sect.~\ref{sec:spectrophot_analysis} we have shown that the ages of the ionizing population within the \hii regions span a very narrow range (\mbox{i.e.~$\tau = 2.5-5.0$ Myr};~see Table~\ref{table:result_chi_square}). With such a narrow age range and a large \hb luminosity range (i.e. log \hb$ \sim 1-3.5$;~see Fig.~\ref{fig:ionization_ratios2}), we can easily infer that there is not a correlation between the luminosity of an \hii region and the age of the ionizing population; that is, the most luminous \hii regions do not tend to be the youngest. Thus, we can assume that the luminosity of the \hii regions is roughly proportional to their masses, being their age a second-order effect.

\begin{table*}
\begin{minipage}{0.65\textwidth}
\renewcommand{\footnoterule}{}  
\begin{small}
\caption{Comparison of mass derivations and ratios with those in H10.}
\label{table:comparison_hagele2}
\begin{center}
\begin{tabular}{lccccccc}
\hline \hline
   \noalign{\smallskip}
H{\tiny II}	&	ID	&	m$^{\mathrm{ion}}$	&	m$^{\mathrm{ion}}$	&	m$^{\mathrm{LOS}}_\star$	&	m$^{dyn}_\star$	&	m$^{\mathrm{ion}}$/m$_\star$	&	m$^{\mathrm{ion}}$/m$_\star$	\\
ID	&	H10	&	&	H10	&		&	H10 	&	(\%)	&	(\%) H10	\\
(1)	&	(2)	&	(3)	&	(4)	&	(5)	&	(6)	&	(7)	&	(8)	\\
 \hline
   \noalign{\smallskip}
1+4	&	J	&	6.5$^{+1.6}_{-1.5}$	&	1.29--3.14	&	220 $\pm$ 40	&	\ldots	&	3.0$^{+1.0}_{-0.8}$	&	\ldots	\\
\noalign{\smallskip}															
3	&	N	&	6.7$^{+1.1}_{-0.9}$	&	3.49	&	531$^{+161}_{-82}$	&	74 $\pm$ 9	&	1.3 $\pm$ 0.3	&	4.7	\\
\noalign{\smallskip}															
5	&	R4	&	3.9$^{+1.7}_{-1.1}$	&	1.76	&	177$^{+106}_{-35}$	&	89 $\pm$ 3	&	2.1$^{+1.0}_{-0.7}$	&	2.0	\\
\noalign{\smallskip}															
7	&	R5+R6+S6	&	3.2$^{+1.4}_{-0.9}$	&	3.36	&	143$^{+27}_{-14}$	&	103 $\pm$ 12	&	2.1$^{+0.9}_{-0.6}$	&	3.3	\\
\noalign{\smallskip}															
8	&	R1+R2	&	1.9 $\pm$ 0.6	&	1.39	&	90$^{+42}_{-21}$	&	91:	&	2.0$^{+0.8}_{-0.6}$	&	1.5	\\
\noalign{\smallskip}															
11	&	R10+R11\footnote{Only masses for R10 are given in H10.}	&	4.4 $\pm$ 0.7	&	1.57	&	144$^{+36}_{-6}$	&	59 $\pm$ 3	&	2.9 $\pm$ 0.5	&	2.7	\\
\noalign{\smallskip}				 											
12	&	R7	&	4.7$^{+1.3}_{-1.1}$	&	0.87	&	157$^{+31}_{-11}$	&	141 $\pm$ 6	&	2.8 $\pm$ 0.7	&	0.6	\\
\hline \noalign{\smallskip}
\multicolumn{8}{@{} p{\columnwidth} @{}}{{\footnotesize \textbf{Notes.} Col (1): \hii identification ID used in this study. Col (2): \hii identification ID of the individual or group of regions identified in H10, that correspond those identified in this study. Col (3): mass of the ionizing population derived in this study. Col (4): mass of the ionizing population derived in H10. Col (5): total stellar mass along the LOS derived in this study. Col (6): dynamical mass derived in H10. Col (7): ratio of the ionizing to the total stellar mass obtained in this study. Col (8):  ratio of the ionizing to the total stellar mass (i.e. dynamical mass) obtained in H10. All masses are given in 10$^6$ \msun.}}
\end{tabular}
\end{center}
\end{small}
\end{minipage}
\end{table*}

The contribution of the ionizing population to the total stellar population for each \hii region (i.e. the young to total stellar mass ratio, m$^{\mathrm{ion}}$/m$_\star$) can be roughly estimated by taking the derived mass of the \starlight fitting as the total mass of the \hii region. Note that the total stellar mass actually represents the stellar mass along the LOS. Therefore, our estimation really represents a lower limit to the percentage. We would like to point out that the uncertainties on the derived total stellar masses (via the 100 realizations of the \starlight fitting for each region) are generally strongly asymmetric. We thus took the $1\sigma$ uncertainties as the last values included within the 68\% on the left ($\sigma^{-}$) and on the right ($\sigma^{-}$) of the mass distribution, centred on the median value.  To better estimate the uncertainties on the m$^{\mathrm{ion}}/$m$_\star$ ratio we have used the techniques developed in~\cite{Barlow03,Barlow04}. In short, when an 
experimental result is represented as $x^{+\sigma^+}_{-\sigma^{-}}$, being $\sigma^+$ and $\sigma^-$ different, a non-symmetric distribution can represent $x$. Among the functions proposed in those papers, we modelled each variable (i.e. mass)  using a ``Variable Gaussian'' parametrization. With an asymmetric distribution representing each variable we made a few tens of thousands of Monte Carlo runs for each \hii region using the {\sc idl} routine `genrand', which allows us to obtain random numbers following a given distribution (not necessarily normal) so as to obtain a reliable estimate of the mass ratio with its $\sigma^+$ and $\sigma^-$ uncertainty.

The derived m$^{\mathrm{ion}}$/m$_\star$ ratios range from 0.2 to about 7 per cent. A comparison for the circumnuclear regions in NGC 3310 identified in previous studies is shown in Table~\ref{table:comparison_hagele2}. Despite the fact that, with some exceptions, the ratios are similar, the disagreement  between ionizing population and total stellar masses separately is evident, up to factors of more than 2 for the nucleus. Masses from the ionizing population reported in H10 were computed using Eq.~\ref{eq:diaz98}. Apart from the systematic introduced by the use of the updated version presented in this study, aperture effects can contribute significantly to the differences in the reported \ha luminosities and EW(\hb\onespace), as Table~\ref{table:comparison_hagele} illustrates. In general, we have obtained higher mass values. In addition, we have generally computed higher total stellar masses because: (i) we have derived the LOS stellar mass; and (ii) H10 derived the total stellar mass of each 
cluster within an \hii region and then added up all the individual estimates for each region. In our calculations, the intracluster mass (probably from the disc) is included. Still, the ratios reported here and those obtained in H10 span a similar range, being always lower than 10\%. All these values are also similar to those reported in \hii regions and \hii galaxies observed in other studies~\citep{Alonso-Herrero01,Hagele09,Perez-Montero10}. 

\begin{figure}
\centering
\includegraphics[angle=90,trim = -1cm -2cm 1cm -1cm,clip=true,width=0.5\textwidth]{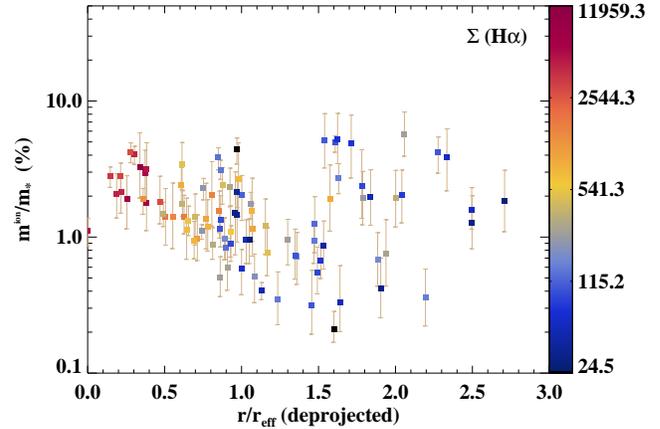}

 \caption{Radial distribution of the ratio between the stellar mass of the ionizing population and the total stellar mass along the LOS. The values are colour coded according to the \ha surface density flux in \mbox{10$^{-16}$ erg s$^{-1}$ cm$^{-2}$ \AA{}$^{-1}$ arcsec$^{-2}$}. Only data for which an estimate of the ionizing mass via the SED fitting technique is plotted.}
  \label{fig:HII_pop1}
\end{figure}

We show in Fig.~\ref{fig:HII_pop1} the radial distribution of this ratio (in terms of percentage) and its relation with the \ha flux surface density. In general, those regions with higher \ha flux surface density are located at smaller radii (with a few exceptions). The radial distribution of the young-to-total-stellar ratio is much more scattered. From the centre and up to about 1.5\reff~(where we cover the complete radial FoV), a 4.5$\sigma$ mild correlation is found, according to the Spearman's rank-order correlation test. However, if we consider the whole radial range, such correlation cease to exist in statistical terms. 

Given the narrow age range of the ionizing population we can roughly assume that any variation of the ratio between the ionizing population and the total stellar mass relates to a variation in the specific SFR (sSFR). In fact, if we compute the sSFR (just by subtracting the total mass from our estimates of the SFR) and plot it against the radial distance, we obtain a similar plot to that shown in Fig.~\ref{fig:HII_pop1}. Therefore, just by examining this figure we can have some insight on the mass growth in the galaxy disc. 

In the framework of the inside-out scenario, the SFR should be a strongly varying function of the galactocentric distance.~\cite{Munyoz-Mateos07} studied the radial profiles of sSFR for a sample of 161 nearby spiral galaxies. They found a large dispersion in the slope of these profiles with a slightly positive mean value, which they interpreted as proof of a moderate inside-out disc formation. Although they did not find any clear dependence of the sSFR gradient on the environment, they argued that transitory episodes of enhanced star formation in the inner parts of the disc can lead to a currently smaller SFR scalelength (gradual growth of the size of the disc with time) than in the past. That is, the gradient can hence be weakened or be even negative. It is well known that mergers in general can induce radial mixing processes, such as inflows of external gas on to the central regions and trigger starbursts (e.g.,~\citealt{Barnes96,Rupke10}). As mentioned before, the global starburst in NGC 3310 is likely to 
have a minor merger origin~\citep{Smith96,Kregel01,Wehner06}. The mild negative gradient of the sSFR in NGC 3310 can well be another signature of a past merger event. This suggests that the minor merger event may be playing an important role in the mass build-up on the bulge, in agreement with recent models~\citep{Hopkins10}.

\section[]{Conclusions}
\label{sec:conclusions}

We have performed an IFS analysis in the distorted spiral galaxy NGC 3310, covering up to about 3 effective radii. This represents an unprecedented simultaneous spatial and spectroscopic coverage for this galaxy, which underwent a minor merger interaction some hundred Myr ago. While major mergers are known to cause dramatic changes in the progenitor galaxies, the impact of a minor merger is still not well understood. We have thus investigated on the evolution of the stellar and chemical properties of the galaxy on account of this past event. The paper relies on the analysis of the optical spectra of about a hundred \hii regions identified along the disc and spiral arms. Chemical abundances of the gas have been obtained by analysing the continuum-subtracted emission-line spectra, different techniques have been employed, and their radial distribution has been studied. We first fitted the SED of the galaxies using the program \starlight in order to quantify the contribution of the 
underlying stellar population and perform such continuum subtraction. With the knowledge of the ionizing conditions of the \hii regions and the aid of ultraviolet and optical/NIR imaging we have characterized the properties (i.e. age, mass, SFR) of the ionizing population. The most important results of this study are summarized as follows.

\begin{enumerate}
\item All derived gaseous oxygen abundances using strong-line diagnostics, consistent with a sub-solar value \mbox{(12 + log(O/H) $\sim$ 8.2--8.4)}, are similar to those reported for circumnuclear \hii regions present in this galaxy. With a sample of over 100 \hii regions we observe a rather flat abundance gradient in the disc of NGC 3310 out to about 10 kpc away from the nucleus. We can thus confirm the evidence that the minor merger event had a substantial impact on metal mixing in the remnant.  

\item A direct ($t_\mathrm{e}$-based) oxygen abundance determination was possible in 16 \hii regions, located in the central regions of the galaxy. The derived values are somewhat lower and present more dispersion than those obtained using strong-line calibrations. With a statistically significant sample of \hii regions, we report on an offset of 0.2--0.3 dex between direct and strong-line abundance estimates. We argue several reasons behind the discrepancy: systematic uncertainties in the calibrations due to different ionization condition properties of the gas and different prescriptions used to relate electron temperatures between the ionic species. We further investigate the effect of the oversimplified assumption that the electrons in the nebulae are in thermal equilibrium on the determination of the direct abundances. Under the assumption of a non-equilibrium Kappa electron energy distribution, the adopted abundances would be increased by up to 0.2--0.3 dex.

\item With the use of single stellar population and photoionization models, we have been able to constrain the main properties of the ionizing stellar population within the \hii regions. In general, the presence of a single population is sufficient to explain the measured broad-band (from ultraviolet to $z$ band) and ionizing line-flux. The age of the ionizing population spans a narrow range of 2.5--5 Myr, whereas the mass range is quite large, from about 10$^4$ to \mbox{6$\times 10^6$ \msun\onespace}. Given the ionized gas line ratios, dust grains must be present in most of the \hii regions, causing the absorption from a few percent and up to more than two-thirds (typical absorption factors of \mbox{$f_\mathrm{d} = $1.3-3.5}) of the UV photons. 

\item We have updated prescriptions to derive the stellar mass of the ionizing population, correcting by its evolutionary state, with the knowledge of the \ha flux and the EW of \hb\onespace. The ionizing mass has little dependence on photon losses by either UV photon leakage or absorption of UV photons by dust grains within the nebulae. The typical uncertainties of the derived masses, generally within a factor of 1.5--2, are in fact much higher than any correction that should be applied were these losses even of the order of 66\%.  

\item The derived average global SFR over time-scales \mbox{$>$ 6 Myr}, 4.4 \mbox{\msun yr$^{-1}$}, is consistent with other studies made at other wavelengths. However, the average global SFR for the last few Myr (i.e. \mbox{$\lesssim$ 5 Myr}) is about a factor of 3 higher. 

\item The mass of the ionizing population represents generally less than 5\% of the estimated population along the LOS, down to about 0.2\%. Taking into account that the contribution of the old mass of the disc is not subtracted, this result is consistent with other studies that support that in circumnuclear regions the ratio of the ionizing to total stellar mass is typically of the order of a few percent, less than 10\%. The radial distribution of the specific SFR (proportional to this ratio) in the disc of NGC 3310 suggests that minor interactions indeed can play an important role in the formation and assembly of the bulge.

\end{enumerate}

\section*{Acknowledgements}

\begin{small}
We would like to thank the anonymous referee for useful comments that have helped improve the quality and clarify the content of the paper. This work has been partially supported by projects AYA2010-21887-C04-03 and AYA2010-21887-C04-01 of the Spanish National Plan for Astronomy and Astrophysics, and by the project AstroMadrid, funded by the Comunidad de Madrid government under grant CAM S2009/ESP-1496, partially using funds from the EU FEDER programme. 
FFRO acknowledges the Mexican National Council for Science and Technology (CONACYT) for financial support under the programme Estancias Posdoctorales y Sab{\'a}ticas al Extranjero para la Consolidaci{\'o}n de Grupos de Investigaci{\'o}n, 2010-2012. 
This research has made use of the NASA/IPAC Extragalactic Database which is operated by the Jet Propulsion Laboratory, California Institute of Technology, under contract with the National Aeronautics and Space Administration. We have made use of observations obtained with \mbox{$XMM-Newton$}, an ESA science mission with instruments and contributions directly funded by ESA Member States and NASA.
\end{small}

\renewcommand{\item}{\itemold}

\bibliographystyle{mn2e}

\bibliography{my_bib.bib}{}

\newpage

\begin{onecolumn}

\appendix

\begin{footnotesize}
\begin{landscape}
\section{Catalogue tables}
\begin{longtable}{@{\hspace{0.20cm}}l@{\hspace{0.20cm}}c@{\hspace{0.20cm}}c@{\hspace{0.20cm}}c@{\hspace{0.20cm}}c@{\hspace{0.20cm}}c@{\hspace{0.20cm}}c@{\hspace{0.20cm}}c@{\hspace{0.20cm}}c@{\hspace{0.20cm}}c@{\hspace{0.20cm}}c@{\hspace{0.20cm}}c@{\hspace{0.20cm}}c@{\hspace{0.20cm}}c@{\hspace{0.20cm}}c@{\hspace{0.20cm}}c@{\hspace{0.20cm}}}
\caption{Catalogue of \hii regions and reddeing corrected derived line intensities}\label{table:hii_catalogue}\\
\hline \hline
   \noalign{\smallskip}
\multicolumn{1}{c}{\hii ID} \vline &
\multicolumn{1}{c}{$\Delta$ AR}  &
\multicolumn{1}{c}{$\Delta \delta$}  &
\multicolumn{1}{c}{$r_\mathrm{d}$}\vline  &
\multicolumn{1}{c}{\oii~3727,3729} &
\multicolumn{1}{c}{H$\gamma$}  &
\multicolumn{1}{c}{H$\beta$}  &
\multicolumn{1}{c}{\oiii~4959}  &
\multicolumn{1}{c}{\oiii~5007} &
\multicolumn{1}{c}{\nii~6548}  &
\multicolumn{1}{c}{H$\alpha$}  &
\multicolumn{1}{c}{\nii~6584}  &
\multicolumn{1}{c}{\sii~6717}  &
\multicolumn{1}{c}{\sii~6731} \vline &
\multicolumn{1}{c}{log I(H$\beta$)}  &
\multicolumn{1}{c}{c (H$\beta$)}  \\

\multicolumn{1}{c}{(1)} \vline &
\multicolumn{3}{c}{(2--4)} \vline &
\multicolumn{10}{c}{(5--14)} \vline &
\multicolumn{1}{c}{(15)} &
\multicolumn{1}{c}{(16)} \\

\noalign{\smallskip}
 \hline
\endfirsthead

\hline \hline
   \noalign{\smallskip}
\multicolumn{1}{c}{\hii ID} \vline &
\multicolumn{1}{c}{$\Delta$ AR}  &
\multicolumn{1}{c}{$\Delta \delta$}  &
\multicolumn{1}{c}{$r_\mathrm{d}$}\vline  &
\multicolumn{1}{c}{\oii~3727,3729} &
\multicolumn{1}{c}{H$\gamma$}  &
\multicolumn{1}{c}{H$\beta$}  &
\multicolumn{1}{c}{\oiii~4959}  &
\multicolumn{1}{c}{\oiii~5007} &
\multicolumn{1}{c}{\nii~6548}  &
\multicolumn{1}{c}{H$\alpha$}  &
\multicolumn{1}{c}{\nii~6584}  &
\multicolumn{1}{c}{\sii~6717}  &
\multicolumn{1}{c}{\sii~6731} \vline &
\multicolumn{1}{c}{log I(H$\beta$)}  &
\multicolumn{1}{c}{c (H$\beta$)}  \\

\multicolumn{1}{c}{(1)} \vline &
\multicolumn{3}{c}{(2--4)} \vline &
\multicolumn{10}{c}{(5--14)} \vline &
\multicolumn{1}{c}{(15)} &
\multicolumn{1}{c}{(16)}  \\

\noalign{\smallskip}
 \hline
\endhead

 \hline\multicolumn{16}{r}{Continued on next page}\\
\endfoot
\endlastfoot
   \noalign{\smallskip}
1 & 9.09 & -4.86 & 11.88 & 280.2 $\pm$ 5.1 & 44.7 $\pm$ 0.5 & 100.0 $\pm$ 0.5 & 81.3 $\pm$ 1.7 & 244.5 $\pm$ 2.6 & 16.7 $\pm$ 0.2 & 287.0 $\pm$ 2.9 & 48.3 $\pm$ 1.7 & 31.2 $\pm$ 0.3 & 23.7 $\pm$ 0.2 & -12.38 & 0.13 $\pm$ 0.03 \\
2 & -7.08 & 0.27 & 8.92 & 294.8 $\pm$ 6.0 & 44.4 $\pm$ 0.4 & 100.0 $\pm$ 0.4 & 72.4 $\pm$ 2.0 & 219.2 $\pm$ 2.8 & 20.7 $\pm$ 0.2 & 287.0 $\pm$ 2.1 & 60.0 $\pm$ 1.0 & 37.9 $\pm$ 0.3 & 28.7 $\pm$ 0.2 & -12.58 & 0.16 $\pm$ 0.03 \\
3 & 0.00 & 0.00 & 0.00 & 283.1 $\pm$ 9.9 & 50.9 $\pm$ 1.5 & 100.0 $\pm$ 0.9 & 39.6 $\pm$ 1.1 & 123.6 $\pm$ 2.3 & 41.4 $\pm$ 0.7 & 287.0 $\pm$ 4.4 & 119.8 $\pm$ 1.9 & 48.7 $\pm$ 0.9 & 39.2 $\pm$ 0.8 & -12.74 & 0.14 $\pm$ 0.03 \\
4 & 7.91 & -9.67 & 13.07 & 313.1 $\pm$ 6.6 & 46.6 $\pm$ 0.5 & 100.0 $\pm$ 0.6 & 71.9 $\pm$ 1.6 & 218.7 $\pm$ 2.5 & 18.1 $\pm$ 0.2 & 287.0 $\pm$ 2.9 & 52.3 $\pm$ 1.5 & 38.9 $\pm$ 0.3 & 29.6 $\pm$ 0.3 & -12.68 & 0.07 $\pm$ 0.03 \\
5 & -4.71 & 3.69 & 6.60 & 279.0 $\pm$ 7.0 & 46.5 $\pm$ 0.5 & 100.0 $\pm$ 0.6 & 67.2 $\pm$ 1.7 & 208.9 $\pm$ 2.5 & 22.9 $\pm$ 0.3 & 287.0 $\pm$ 2.5 & 66.3 $\pm$ 1.2 & 37.1 $\pm$ 0.4 & 28.0 $\pm$ 0.4 & -12.81 & 0.08 $\pm$ 0.03 \\
6 & -5.85 & -6.50 & 10.64 & 281.5 $\pm$ 6.6 & 45.0 $\pm$ 0.6 & 100.0 $\pm$ 0.5 & 55.9 $\pm$ 1.4 & 169.6 $\pm$ 2.0 & 23.2 $\pm$ 0.3 & 287.0 $\pm$ 2.2 & 67.4 $\pm$ 1.1 & 45.2 $\pm$ 0.4 & 33.7 $\pm$ 0.3 & -12.81 & 0.05 $\pm$ 0.03 \\
7 & -2.81 & 7.00 & 7.56 & 284.1 $\pm$ 6.8 & 47.4 $\pm$ 0.5 & 100.0 $\pm$ 0.5 & 59.1 $\pm$ 1.4 & 181.0 $\pm$ 2.0 & 21.5 $\pm$ 0.2 & 287.0 $\pm$ 2.4 & 62.4 $\pm$ 1.2 & 40.4 $\pm$ 0.3 & 29.6 $\pm$ 0.3 & -12.83 & 0.09 $\pm$ 0.03 \\
8 & -10.52 & 4.11 & 13.38 & 310.0 $\pm$ 8.0 & 46.1 $\pm$ 0.6 & 100.0 $\pm$ 0.5 & 73.7 $\pm$ 1.7 & 225.7 $\pm$ 2.5 & 19.6 $\pm$ 0.2 & 287.0 $\pm$ 2.2 & 56.6 $\pm$ 1.1 & 40.2 $\pm$ 0.4 & 30.3 $\pm$ 0.3 & -12.61 & 0.28 $\pm$ 0.02 \\
\end{longtable}
\end{landscape}
\end{footnotesize}

\begin{footnotesize}
\begin{landscape}
\begin{table*}
\hspace{6cm}
\begin{minipage}{\textheight}
\caption{Catalogue of external fibres and reddening corrected derived line intensities}\label{table:ext_fibers_catalogue}
\hspace{-6cm}
\begin{tabular}{@{\hspace{0.25cm}}l@{\hspace{0.25cm}}c@{\hspace{0.25cm}}c@{\hspace{0.25cm}}c@{\hspace{0.25cm}}c@{\hspace{0.25cm}}c@{\hspace{0.25cm}}c@{\hspace{0.25cm}}c@{\hspace{0.25cm}}c@{\hspace{0.25cm}}c@{\hspace{0.25cm}}c@{\hspace{0.25cm}}c@{\hspace{0.25cm}}c@{\hspace{0.25cm}}c@{\hspace{0.25cm}}c@{\hspace{0.25cm}}}
\hline \hline
   \noalign{\smallskip}
\multicolumn{1}{c}{Ext ID} \vline &
\multicolumn{1}{c}{$\Delta$ AR}  &
\multicolumn{1}{c}{$\Delta \delta$}  &
\multicolumn{1}{c}{$r_\mathrm{d}$}\vline  &
\multicolumn{1}{c}{\oii~3727,3729} &
\multicolumn{1}{c}{H$\beta$}  &
\multicolumn{1}{c}{\oiii~4959}  &
\multicolumn{1}{c}{\oiii~5007} &
\multicolumn{1}{c}{\nii~6548}  &
\multicolumn{1}{c}{H$\alpha$}  &
\multicolumn{1}{c}{\nii~6584}  &
\multicolumn{1}{c}{\sii~6717}  &
\multicolumn{1}{c}{\sii~6731} \vline &
\multicolumn{1}{c}{log I(H$\beta$)}  &
\multicolumn{1}{c}{c (H$\beta$)}  \\

\multicolumn{1}{c}{(1)} \vline &
\multicolumn{3}{c}{(2--4)} \vline &
\multicolumn{9}{c}{(5--13)} \vline &
\multicolumn{1}{c}{(14)} &
\multicolumn{1}{c}{(15)} \\

 \hline
   \noalign{\smallskip}
1& 44.80 & -114.40 & 123.09 & 225.0 $\pm$ 17.1 & 100.0 $\pm$ 1.2 & 111.9 $\pm$ 2.4 & 316.3 $\pm$ 4.8 & 11.9 $\pm$ 1.4 & 287.0 $\pm$ 4.5 & 34.5 $\pm$ 1.6 & 25.4 $\pm$ 1.4 & 17.1 $\pm$ 1.3 & -14.18 & 0.23 $\pm$ 0.01 \\
2& 44.80 & -112.80 & 121.63 & 646.3 $\pm$ 72.2 & 100.0 $\pm$ 3.1 & 76.8 $\pm$ 5.3 & 235.3 $\pm$ 10.2 & 22.8 $\pm$ 2.4 & 287.0 $\pm$ 10.5 & 65.8 $\pm$ 3.7 & 50.8 $\pm$ 3.0 & 38.7 $\pm$ 2.7 & -14.30 & 0.33 $\pm$ 0.02 \\
3& 44.80 & -107.20 & 116.56 & 277.2 $\pm$ 60.2 & 100.0 $\pm$ 3.4 & 41.2 $\pm$ 4.8 & 133.9 $\pm$ 7.9 & 19.7 $\pm$ 4.8 & 255.3 $\pm$ 12.8 & 57.1 $\pm$ 6.1 & 58.7 $\pm$ 5.7 & 44.6 $\pm$ 5.2 & -14.94 & 0.00 \\
4& -21.80 & -71.00 & 81.56 & 307.5 $\pm$ 4.2 & 100.0 $\pm$ 0.6 & 97.9 $\pm$ 1.1 & 304.0 $\pm$ 2.3 & 12.5 $\pm$ 0.7 & 287.0 $\pm$ 2.3 & 36.3 $\pm$ 0.8 & 28.7 $\pm$ 0.7 & 21.0 $\pm$ 0.7 & -13.74 & 0.17 $\pm$ 0.01 \\
\end{tabular}
\end{minipage}
\end{table*}
\end{landscape}
\end{footnotesize}

\begin{footnotesize}
\begin{landscape}
\begin{longtable}{@{\hspace{0.30cm}}l@{\hspace{0.30cm}}c@{\hspace{0.30cm}}c@{\hspace{0.30cm}}c@{\hspace{0.30cm}}c@{\hspace{0.30cm}}c@{\hspace{0.30cm}}c@{\hspace{0.30cm}}c@{\hspace{0.30cm}}c@{\hspace{0.30cm}}c@{\hspace{0.30cm}}c@{\hspace{0.30cm}}c@{\hspace{0.30cm}}c@{\hspace{0.30cm}}c@{\hspace{0.30cm}}c@{\hspace{0.30cm}}}
\caption{Properties of the \hii regions derived via $\chi^2$ minimization + \cloudy analsyis}\label{table:result_chi_square}\\
\hline \hline
   \noalign{\smallskip}
\multicolumn{1}{c}{\hii} \vline &
\multicolumn{1}{c}{$UVW2$}  &
\multicolumn{1}{c}{$UVM2$}  &
\multicolumn{1}{c}{$u$}  &
\multicolumn{1}{c}{$F439W$}  &
\multicolumn{1}{c}{$F547M$}  &
\multicolumn{1}{c}{$F621M$}  &
\multicolumn{1}{c}{$i$}  &
\multicolumn{1}{c}{$z$} \vline &
\multicolumn{1}{c}{EW (\ha)}  &
\multicolumn{1}{c}{EW (\hb)} \vline &
\multicolumn{1}{c}{log ($\tau$)}  &
\multicolumn{1}{c}{log ($m^ {\mathrm{ion}}$)}  &
\multicolumn{1}{c}{$A_V$} &
\multicolumn{1}{c}{$f_\mathrm{d}$} \\

\multicolumn{1}{c}{ID} \vline &
\multicolumn{8}{c}{10$^{-16}$ erg s$^{-1}$ cm$^{-2}$ \AA{}$^{-1}$} \vline &
\multicolumn{1}{c}{\AA{}} &
\multicolumn{1}{c}{\AA{}}\vline &
\multicolumn{1}{c}{yr} &
\multicolumn{1}{c}{\msun} &
\multicolumn{1}{c}{mag} &
\multicolumn{1}{c}{} \\   

\multicolumn{1}{c}{(1)} \vline &
\multicolumn{8}{c}{(2--9)} \vline &
\multicolumn{1}{c}{(10)}  &
\multicolumn{1}{c}{(11)} \vline &
\multicolumn{1}{c}{(12)}  &
\multicolumn{1}{c}{(13)}  &
\multicolumn{1}{c}{(14)} &
\multicolumn{1}{c}{(15)} \\

\noalign{\smallskip}
 \hline
\endfirsthead

\hline \hline
   \noalign{\smallskip}
\multicolumn{1}{c}{\hii} \vline &
\multicolumn{1}{c}{$UVW2$}  &
\multicolumn{1}{c}{$UVM2$}  &
\multicolumn{1}{c}{$u$}  &
\multicolumn{1}{c}{$F439W$}  &
\multicolumn{1}{c}{$F547M$}  &
\multicolumn{1}{c}{$F621M$}  &
\multicolumn{1}{c}{$i$}  &
\multicolumn{1}{c}{$z$} \vline &
\multicolumn{1}{c}{EW (\ha)}  &
\multicolumn{1}{c}{EW (\hb)} \vline &
\multicolumn{1}{c}{log ($\tau$)}  &
\multicolumn{1}{c}{log ($m^ {\mathrm{ion}}$)}  &
\multicolumn{1}{c}{$A_V$} &
\multicolumn{1}{c}{$f_\mathrm{d}$} \\

\multicolumn{1}{c}{ID} \vline &
\multicolumn{8}{c}{10$^{-16}$ erg s$^{-1}$ cm$^{-2}$ \AA{}$^{-1}$} \vline &
\multicolumn{1}{c}{\AA{}} &
\multicolumn{1}{c}{\AA{}}\vline &
\multicolumn{1}{c}{yr} &
\multicolumn{1}{c}{\msun} &
\multicolumn{1}{c}{mag} &
\multicolumn{1}{c}{} \\   

\multicolumn{1}{c}{(1)} \vline &
\multicolumn{8}{c}{(2--9)} \vline &
\multicolumn{1}{c}{(10)}  &
\multicolumn{1}{c}{(11)} \vline &
\multicolumn{1}{c}{(12)}  &
\multicolumn{1}{c}{(13)}  &
\multicolumn{1}{c}{(14)} &
\multicolumn{1}{c}{(15)} \\

\noalign{\smallskip}
 \hline
\endhead

\noalign{\smallskip}
 \hline\multicolumn{15}{r}{Continued on next page}\\
\endfoot
\endlastfoot
   \noalign{\smallskip}
1 & 159.4  $\pm$ 23.1 & 128.6  $\pm$ 17.5 & 74.3  $\pm$ 14.6 & 56.6  $\pm$ 9.6 & 29.1  $\pm$ 6.0 & 21.2  $\pm$ 5.1 & 15.5  $\pm$ 3.9 & 9.7  $\pm$ 3.1 & 598  $\pm$ 71 & 81  $\pm$ 5 & 6.61 $^{+0.07}_{-0.06}$ & 6.61 $^{+0.26}_{-0.22}$ & 0.49 $^{+0.25}_{-0.19}$ & 1.5 $\pm$ 0.35 \\
\noalign{\smallskip}
2 & 84.2  $\pm$ 8.5 & 70.0  $\pm$ 6.5 & 49.7  $\pm$ 5.8 & 34.5  $\pm$ 6.3 & 19.9  $\pm$ 4.5 & 16.2  $\pm$ 4.6 & 10.8  $\pm$ 3.1 & 7.9  $\pm$ 2.8 & 463  $\pm$ 85 & 71  $\pm$ 7 & 6.61 $^{+0.04}_{-0.08}$ & 6.53 $^{+0.18}_{-0.24}$ & 0.64 $^{+0.20}_{-0.17}$ & 1.7 $\pm$ 0.30 \\
\noalign{\smallskip}
3 & 116.7  $\pm$ 17.2 & 96.3  $\pm$ 13.5 & 59.3  $\pm$ 7.8 & 46.9  $\pm$ 10.4 & 29.0  $\pm$ 7.3 & 25.9  $\pm$ 9.7 & 22.5  $\pm$ 4.8 & 19.7  $\pm$ 4.6 & 207  $\pm$ 61 & 35  $\pm$ 7 & 6.72 $^{+0.01}_{-0.02}$ & 6.81 $^{+0.07}_{-0.09}$ & 0.63 $^{+0.11}_{-0.10}$ & 1.6 $\pm$ 0.20 \\
\noalign{\smallskip}
4 & 119.1  $\pm$ 22.8 & 99.3  $\pm$ 18.0 & 69.4  $\pm$ 13.8 & 32.8  $\pm$ 5.2 & 16.8  $\pm$ 3.3 & 12.4  $\pm$ 2.9 & 12.4  $\pm$ 3.5 & 7.8  $\pm$ 2.9 & 564  $\pm$ 67 & 80  $\pm$ 5 & 6.60 $^{+0.03}_{-0.06}$ & 6.38 $^{+0.12}_{-0.18}$ & 0.39 $^{+0.21}_{-0.18}$ & 1.6 $\pm$ 0.25 \\
\noalign{\smallskip}
5 & 80.7  $\pm$ 10.6 & 65.3  $\pm$ 8.3 & 40.6  $\pm$ 5.7 & 33.4  $\pm$ 6.5 & 20.5  $\pm$ 4.8 & 17.2  $\pm$ 5.1 & 10.1  $\pm$ 2.9 & 7.7  $\pm$ 2.7 & 289  $\pm$ 75 & 50  $\pm$ 7 & 6.70 $^{+0.03}_{-0.06}$ & 6.61 $^{+0.13}_{-0.17}$ & 0.64 $^{+0.13}_{-0.14}$ & 1.6 $\pm$ 0.45 \\
\noalign{\smallskip}
6 & 165.1  $\pm$ 10.8 & 134.4  $\pm$ 7.2 & 73.0  $\pm$ 3.9 & 54.4  $\pm$ 8.4 & 28.5  $\pm$ 5.3 & 21.1  $\pm$ 5.1 & 15.0  $\pm$ 2.0 & 9.5  $\pm$ 1.7 & 268  $\pm$ 21 & 38  $\pm$ 1 & 6.69 $\pm$ 0.03  & 6.68 $^{+0.06}_{-0.07}$ & 0.46 $\pm$ 0.06  & 1.9 $\pm$ 0.25 \\
\end{longtable}
\label{lastpage}
\end{landscape}
\end{footnotesize}

\end{onecolumn}

\end{document}